\makeatletter

\@ifundefined{@par@version@dash}{%
\def\@parse@version#1{\@parse@version@0#1}
\def\@parse@version@#1/#2/#3#4#5\@nil{%
\@parse@version@dash#1-#2-#3#4\@nil}
\def\@parse@version@dash#1-#2-#3#4#5\@nil{%
  \if\relax#2\relax\else#1\fi#2#3#4 }
}{}
\makeatother
\documentclass[
superscriptaddress,
 amsmath,amssymb,
 aps,
 showpacs,
prc
]{revtex4-2} 
\usepackage[breaklinks=true,colorlinks=true,linkcolor=blue]{hyperref}
\usepackage{xcolor} %for highlighting text in color
\usepackage{graphicx}% Include figure files
\usepackage{bmpsize}% Use natural size of figure
\usepackage{dcolumn}% Align table columns on decimal point
\usepackage[normalem]{ulem} %for \sout{} (strikeout)
\usepackage{xfrac}
\usepackage{bm}% bold math
\usepackage{xcolor, soul}
\sethlcolor{yellow}
\newcommand{\delectron}{\mbox{$d_\text{e}$}}
\newcommand{\dproton}{\mbox{$d_\text{p}$}}
\newcommand{\dneutron}{\mbox{$d_\text{n}$}}

\newcommand{\dmuon}{\mbox{$d_\text{$\mu$}$}}
\newcommand{\ecm}{\mbox{$e \cdot \text{cm}$}}

\begin{document}

\title{On Possibilities of High Precision Fundamental Physics Experiments \\ in Spin-Transparent Storage Rings of Low Energy Polarized Electron Beams}

\author{R.~Suleiman} 
\email{Corresponding author: suleiman@jlab.org}
\affiliation{Thomas Jefferson National Accelerator Facility, Newport News, Virginia 23606 USA}
\author{V.~S.~Morozov}
\affiliation{Oak Ridge National Laboratory, Oak Ridge, Tennessee 37831, USA}
\author{Ya.~S.~Derbenev}
\affiliation{Thomas Jefferson National Accelerator Facility, Newport News, Virginia 23606 USA}

%\collaboration{The SDM Collaboration}
\date{\today}

\begin{abstract} % Non-structured version
We present a new design of highly specialized small storage rings for low energy polarized electron beams. The new design is based on the transparent spin methodology that cancels the spin precession due to the magnetic dipole moment at any energy while allowing for spin precession induced by the fundamental physics of interest to accumulate. The buildup of the vertical component of beam polarization can be measured using standard Mott Polarimetry that is optimal at low electron energy. Systematic uncertainties are suppressed using counter-rotating bunched beams with various polarization orientations. These rings can be used to directly measure the permanent electric dipole moment of the electron, relevant to CP violation and matter-antimatter asymmetry in the universe, and to search for dark energy and ultra-light dark matter. 
\end{abstract}

\maketitle

\section{\label{sec:introduction}Introduction}

The electric dipole moment (EDM) is very sensitive to physics beyond the Standard Model and new sources of Charge-conjugation and Parity (CP) violation~\cite{RevModPhys.91.015001,Fortson:2003,doi:10.1142/S0218301317300028}. Such CP violation, beyond what is present in the weak interaction, could signal the presence of new physics and explain the puzzle of the matter-antimatter asymmetry in the universe. Searching for permanent EDMs has attracted a lot of effort over the last 70 years and there are many
ongoing experiments searching for a nonzero atomic or neutron EDMs. However, there are no direct measurements of the electron or proton EDMs. The EDM upper limit of the electron (~\delectron{} $< 4.1 \times 10^{-30}$~\ecm{}, 90\% C.L.) has been extracted from a measurement using HfF$^+$ ion~\cite{roussy2022new} while the proton limit (~\dproton{} $< 2.0 \times 10^{-25}$~\ecm{},  95\% C.L.) was obtained using $^{199}\mathrm{Hg}$ atom~\cite{PhysRevLett.116.161601}. Direct measurements of the EDM upper limits only exist for the neutron (~\dneutron{} $< 3.6 \times 10^{-26}$~\ecm{}, 95\% C.L.)~\cite{PhysRevLett.124.081803} and the muon~\cite{PhysRevD.80.052008} where muon limit (~\dmuon{} $< 1.8 \times 10^{-19}$~\ecm{}, 95\% C.L.) was measured at Brookhaven National Lab in conjunction with the anomalous magnetic dipole moment, $g-2 (\equiv 2G)$.

Any measurement of EDM relies on measuring the spin precession rate in an electric field in the particle rest frame:
\begin{equation}
  \frac{d{\vec S}}{d\tau} =  {\vec \mu} \times {\vec B}_{\rm rest} +  {\vec d} \times {\vec E}_{\rm rest} , \label{eq:T-BMT_restFrame}  
\end{equation}
   
\noindent where the magnetic dipole moment (MDM) and EDM are defined as ${\vec \mu}=(G+1)(q/mc){\vec S}$ and ${\vec d}=(\eta/2) (q/mc) {\vec S}$, respectively, where $q$ and $m$ are the particle charge and mass, and $c$ is the speed of light. However, for charged particles, such a measurement cannot be made while keeping the particle at rest since electric field leads to acceleration. Therefore, to be able to apply electric field and keep the charged particle trapped, a storage ring must be used. For a charged particle moving in electric and magnetic fields, we have the following generalized Thomas-BMT equation of spin precession in the lab's Cartesian frame~\cite{Fukuyama2013}:

\begin{equation}
  \frac{d{\vec S}}{dt} = \left( {\vec \omega}_{\rm MDM} + {\vec \omega}_{\rm EDM} \right) \times {\vec S}, \label{eq:eq:T-BMT_labFrame}  
\end{equation}

\noindent with:

\begin{equation}
   {\vec \omega}_{\rm MDM} = -\frac{q}{mc} \left( \frac{1+G}{\gamma} {\vec B}_{\rm rest} + \frac{1}{\gamma+1} {\vec \beta} \times {\vec E}_{\rm rest} \right)\; {\rm and} \label{eq:eq:T-BMT_restFrame-MDM}  
\end{equation}

\begin{equation}
  {\vec \omega}_{\rm EDM} = -\frac{\eta}{2} \frac{q}{mc} {\vec E}_{\rm rest}, \label{eq:T-BMT_labFrame-EDM}  
\end{equation}

\noindent where ${\vec v} \equiv {\vec \beta} c$ and $\gamma$ are the particle velocity and Lorentz energy factor, respectively. Here, as usual, symbol ${\vec S}$ represents the spin as a vector in the particle rest frame; appearance of the term $\propto {\vec \beta} \times {\vec E}_{\rm rest}$ is due to relativistic transformation of the spin direction with velocity rotation (Thomas precession~\cite{doi:10.1080/14786440108564170}). In terms of electric and magnetic fields in the lab's frame, we get~\cite{Fukuyama2013}: 

\begin{equation}
  {\vec \omega}_{\rm MDM} = -\frac{q}{mc} \left[ \left( \frac{1}{\gamma} + G \right) {\vec B}_\bot + \frac{1}{\gamma} (1+G) {\vec B_\parallel} + \left( \frac{1}{\gamma + 1} + G \right) {\vec E} \times {\vec \beta} \right] \; {\rm and} \label{eq:T-BMT_FieldslabFrame-MDM}  
\end{equation}

\begin{equation}
  {\vec \omega}_{\rm EDM} = -\frac{\eta q}{2mc} \left( \frac{1}{\gamma} {\vec E}_\parallel + {\vec E_\bot} + {\vec \beta} \times {\vec B} \right). \label{eq:eq:T-BMT_FieldslabFrame-EDM}  
\end{equation}

The general principle of an EDM measurement in a ring is based on compensation of the MDM spin rotation either instantaneous or over a turn while allowing for turn-by-turn accumulation of the EDM precession. Observation of a polarization rotation then indicates presence of the EDM. This measurement requires account of and compensation for systematic effects associated with field errors, beam emittances and background magnetic fields.  

A flat reference orbit can generally be designed employing vertical magnetic $B_y$ and radial electric $E_x$ fields. In a coordinate system connected to the particle velocity on the design orbit (the Frenet-Serret frame), the MDM spin precession about the vertical axis has an angular frequency $\omega_{y,{\rm MDM}}$ of:

\begin{equation}
  \omega_{y,{\rm MDM}} = -\frac{q}{mc} \left( G B_y - \frac{1 - \gamma^2 \beta^2 G}{\gamma^2 \beta} E_x \right). \label{eq:T-BMT_labFrame-ME}  
\end{equation}

Considering Eq.~\ref{eq:T-BMT_labFrame-ME}, two experimental approaches were developed to measure the EDM in a storage ring:
\begin{enumerate}
\item{}	An all electric ring with $B_y=0$ and $\gamma^2 = 1 + 1/G$, {\em i.e.}, ``magic energy'' case. This works only for $G>0$ and at a very specific energy. A recent European collaboration~\cite{abusaif2019storage} has proposed an experiment to measure the proton ($G=1.79$) EDM with a sensitivity of $10^{-29}$~\ecm{} using polarized proton beams at the magic energy of 232.8~MeV in an all-electric precision storage ring of about 500~m in circumference. To reduce systematic uncertainties, clock-wise (CW) and counter clock-wise (CCW) beams will be circulating concurrently.

An interesting advanced modification of the magic energy approach is a hybrid ring using electrical bending and magnetic focusing. The magnetic focusing is used to compensate for background magnetic fields while still allowing for the simultaneous storage of CW and CCW beams. With such a ring (about 800~m in circumference), a proposal has been developed at Brookhaven National Lab to measure the EDM of the proton~\cite{PhysRevLett.93.052001,doi:10.1063/1.4967465,PhysRevAccelBeams.22.034001,PhysRevD.105.032001}.

\item{}	A combined electric/magnetic ring with $GB_y = \frac{1 - \gamma^2 \beta^2 G}{\gamma^2 \beta} E_x$. It was considered to use the Cooler Synchrotron (COSY) at the Forschungszentrum J\"ulich, Germany, as such a ring~\cite{PhysRevLett.117.054801} to measure the deuteron ($G=-0.143$) EDM at 1.0 GeV/c.
\end{enumerate}

Notably, the above proposals aim to measure the EDM of the proton and deuteron but not of the electron. In fact, there is no proposal to measure the electron ($G=0.00116$) EDM at the magic energy (15.01 MeV, $\gamma=29.38$) because there is no viable polarimetry at that energy.

This paper presents a method to measure the EDM of charged particles in small storage rings with a beam energy range of a few MeV or less. Our approach to EDM measurement is based on the use of a Spin-Transparent (ST) Figure-8 orbit symmetry. The ST method was developed for magnetic synchrotrons. It was extensively studied while designing polarized electron, proton, deuteron, and ${^3}$He beams for the Electron-Ion Collider~\cite{PhysRevLett.124.194801,Kondratenko:2019ubq}. In our proposed EDM measurement technique, this method was extended to storage rings designed using purely electric fields to allow for two counter-rotating (CR) electron beams A and B (CRA and CRB) to circulate concurrently. Applied to an EDM search, the ST method offers new mechanisms for suppression of destructive effects associated with energy spread and beam emittances. 
    
For electrons, we develop a Figure-8 all-electric ring design with beam energy below 1 MeV modulated along the orbit. At this low electron energy, Mott polarimetry reaches its peak efficiency. The concept of a small Figure-8 spin-transparent storage ring for electron EDM search and its preliminary optics design are presented in Section~\ref{sec:concept}. Section~\ref{sec:mott} describes a Mott polarimeter, as part of the electron EDM ring, used to measure accumulation of the vertical polarization. Summary of our theoretical considerations can be found in Appendix~\ref{app:theory} with full details to be published in a future dedicated arXiv paper. Appendix~\ref{app:long-paper} lists the subjects to be covered in that future paper.

The development of a concept for EDM search using a polarized charged particle beam stored in a ring requires a careful consideration of many issues associated with long-term beam polarization stability and lifetime. One must formulate relevant requirements on the beam intensity and quality. Other issues are associated with polarization control and measurement. Our preliminary estimates are provided below in the paper. We build on the experience and innovations accumulated by other research groups planning to measure EDM with storage rings~\cite{abusaif2019storage,doi:10.1063/1.4967465,PhysRevAccelBeams.22.034001}. The key features of our approach are discussed in the paper and can be summarized as follows:
\begin{enumerate}
\item{} electron, proton, and deuteron EDM search experiments can be done in small low-energy rings,
\item{} the ST configuration of the storage ring makes the spin tune energy-independent and insensitive to the beam emittances,
\item{} Mott polarimetry is efficient for electrons in the considered energy range,
\item{} low energy polarimetry can be used for protons and deuterons,
\item{} small orbit size allows for efficient stochastic cooling, and  
\item{} the Spin Echo approach may further advance the precision of spin control.
\end{enumerate}

We also note that, as in the magic-energy-based design~\cite{PhysRevD.103.055010}, the proposed small storage rings also allow studies of the nature of Dark Matter and Dark Energy. A Figure-8 electron storage ring can be used to look for spin precession induced by dark energy and ultra-light (axion) dark matter. The field gradient of the low-mass axion couples to the spin of transversely-polarized electrons stored in the ring with a sensitivity proportional to the relativistic beam velocity, $\beta$, and the beam spin coherence time (SCT).

\section{\label{sec:concept}New Proposed Concept}
The new concept relies on an ST ring design with Figure-8 being a natural example of such a ring. An ST ring offers new techniques for suppression of systematic effects and allows one to search for the EDM at any energy. In particular, the integrated effect of the whole ring on the magnetic moment is naturally canceled to zeroth order at any energy due to the ST property of the ring. Techniques for compensating the first and second order effects have also been developed~\cite{PhysRevLett.124.194801}. Thus, an ST ring can serve as an ideal base for high-precision spin experiments. Moreover, it relies only on conventional accelerator technologies.

With the MDM spin rotation canceled, observation of a spin rotation then indicates presence of the EDM. This is the general principle of an EDM measurement in a storage ring. This measurement requires compensation and account of the systematic effects associated with the MDM. For example, one must eliminate the MDM spin effect due to closed orbit excursion and beam emittances. Closed orbit excursion is always present in a real ring but, in an ST ring, its spin effect can be precisely measured and corrected. Such a correction can be done in a systematic way using only fields that are weak compared to the design fields. This technique has been developed and verified by simulations while designing polarized Figure-8 rings of the Jefferson Lab Electron-Ion Collider~\cite{PhysRevLett.124.194801}. The next-order spin effects arise due to the beam emittances and can be suppressed through ring optics design.

\subsection{Benefits of Proposed Concept}

This new concept of a small ST storage ring has the following features beneficial to EDM search: 
\begin{itemize}
    \item long SCT, 
    \item energy-independent spin tune, 
    \item bunched and un-bunched (coasting) beams, 
    \item any energy including low energies, 
    \item no synchrotron radiation, 
    \item minimum safety issues – no radiation and activation, 
    \item straightforward polarimetry, 
    \item two CR beams, 
    \item small easily-manageable size, 
    \item good control of systematic effects and imperfections including background magnetic fields, 
    \item low cost, and
    \item the possibility of using it as a test-bed for larger-scale experiments.
\end{itemize}
More details about these unique benefits will be provided throughout this paper.

\subsection{ST Ring Design}

The EDM effect in a storage ring is a tiny signal on top of a relatively large MDM spin precession. Therefore, the most promising approach to extracting the EDM contribution is a scheme where the MDM signal is suppressed with a very high accuracy while the EDM signal remains. Thus, the basis of such an EDM measurement scheme is detection of a spin rotation with the MDM signal compensated. This compensation can be done locally like in the case of a magic-energy ring or globally like in an ST ring where the MDM signal is compensated over a single turn. 

One advantage of the ST configuration is that MDM suppression is a natural consequence of the ring topology occurring regardless of the beam energy. It immediately follows from this property that there is no spin decoherence due to the beam energy spread. The main problem with directly using an ST ring for this application is that the EDM signal in such a ring with only transverse magnetic and/or electric fields also vanishes. 

We describe a Figure-8 ring configuration with different beam energies in different ring sections. The net bend at each of the energies is still zero preserving the MDM suppression. The different energy sections are connected by static longitudinal electric field sections. These fields have no effect on the MDM spin precession but break the degeneracy of the EDM spin precession. The level of the expected EDM signal is discussed below.

One of the main challenges for measuring the EDM in an MDM-suppressed system is precise control of systematic effects. One must confidently differentiate a real EDM signal from a false one caused by residual MDM effects. Perhaps, the most powerful technique for suppression of the systematic effects is to use two CR beams in a direction-symmetric ring. Since any bending magnetic field breaks the symmetry between the CR beams, we choose an all-electric ring. The strength of a CR beam system for systematic effect suppression will be discussed in more details at the end of this paper.

We employ static longitudinal electric field sections for acceleration and deceleration. Unlike warm RF cavities, they allow for beam energy recovery thus minimizing the ring's energy consumption. In fact, the power requirements of warm RF cavities make them impractical at the beam currents of interest. On the other hand, static longitudinal fields do not provide beam bunching, which is needed for time separation of the CR bunches. Time separation of the bunches is in turn needed for their simultaneous independent stochastic cooling. As will be shown later, stochastic cooling is necessary to maintain the bunch intensities sufficient from statistical-accuracy point of view. Thus, bunching allows for simultaneous storage of high-intensity CR bunches. As stated above, this is a critical tool for suppression of many systematic effects. Therefore, we introduce a dedicated RF cavity to bunch the beams. Since it operates at zero voltage crossing ({\em i.e.}, will not change central energy of the bunch), its voltage and power requirements are quite modest as will be shown later. The RF electric field of the bunching cavity is accompanied by oscillating magnetic field, which interacts with the particle's MDM. However, its effect averages to zero over time due to synchrotron oscillations.  

\begin{figure}[!htbp]
\includegraphics[width=0.75\columnwidth]{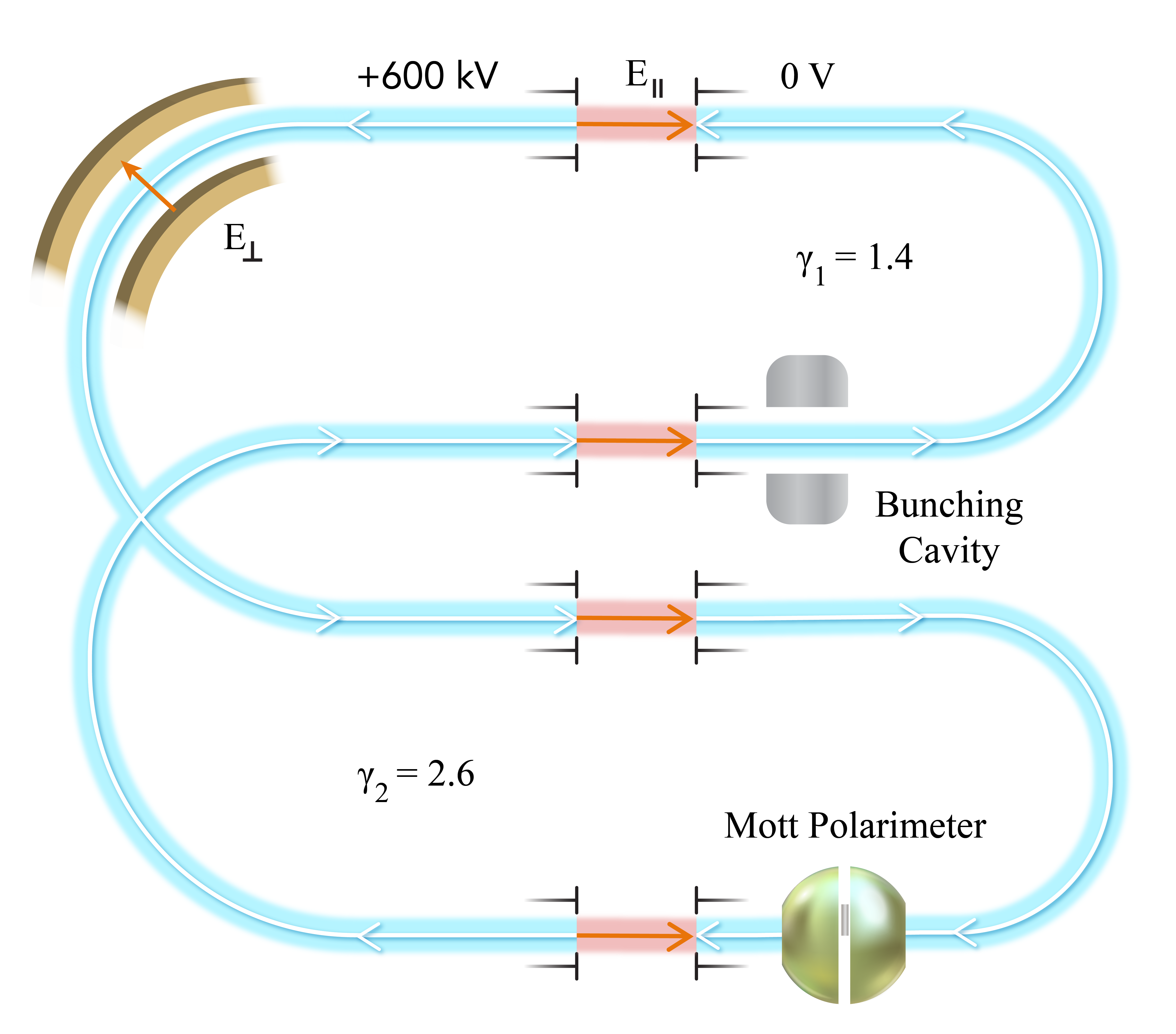}
\caption{ 
Layout of a two-energy ST storage ring for measuring the electron EDM (figure not drawn to scale). Only one of the two CR electron beams is shown. 
The ring uses only static electric fields ($E_\parallel$ and $E_\perp$) except for a single RF bunching cavity. The high-energy arcs are
floating at high voltage of 600 kV. Both vertical and horizontal polarization components can be simultaneously measured using a Mott polarimeter.
\label{fig:ring_schematics}
}
\end{figure}

To measure the electron EDM, the simplest two-energy ST ring schematic conforming to the above principles is shown in Fig.~\ref{fig:ring_schematics}. Partly because of being simple, it is also quite compact. Compactness is important for maximizing the rate of EDM signal accumulation. The ring in Fig.~\ref{fig:ring_schematics} consists of two low-energy and two high-energy arcs. The beam directions in the two arcs of each energy are opposite making the net bending angle zero. The low- and high-energy arcs are connected to potentially common capacitor plates providing acceleration and deceleration of 600 kV. A bunching RF cavity is located in a low-energy section to minimize its voltage requirements.

The ring in Fig.~\ref{fig:ring_schematics} consists of arc sections where the electric field remains transverse to the reference electron velocity and of straight sections where the electric field is longitudinal to the beam direction. We consider the spin dynamics in these two cases separately. Bunching RF fields do not contribute to this picture because their average first-order spin effect is zero. Using the Thomas-BMT equation extended to include the EDM, the spin precession per unit length in the transverse electric fields of the arc sections is given by

\begin{equation}
\label{eq:A}
  \vec{\omega}^{E_\perp}_{\rm lab} = -\frac{q}{mv} \left[  \underbrace{ \frac{\eta E_\perp}{2c} \hat{E}}_\text{EDM} - \underbrace{ \left(G +  \frac{1}{\gamma+1} \right) \frac{vE_\perp}{c^2} \hat{v} \times \hat{E}}_\text{MDM} \right],
\end{equation}
where $q$ is the particle charge, $m$ is its mass, $v$ is the particle speed, $\eta$ is the electric dipole factor defined below, $E_\perp$ is the field strength, $c$ is the speed of light, $\hat{E}$ is a unit vector along the electric field direction, $G$ is the magnetic anomaly, $\gamma$ is the Lorenz energy factor, and $\hat{v}$ is a unit vector along the particle velocity. As indicated in Eq.~\ref{eq:A}, its first and second terms describe the respective contributions of EDM and MDM to the spin rotation. The electric dipole factor $\eta$ is introduced to specify the proportionality factor between the electric dipole moment $\vec{d}$ and the particle spin $\vec{S}$ as
\begin{equation}
\label{eq:B}
  \vec{d} = \frac{\eta}{2} \frac{q}{mc} \vec{S}.
\end{equation}

It is convenient to express Eq.~\ref{eq:A} in terms of the precession rate $d\theta/ds$ of the particle's orbital angle $\theta$ per unit orbit length $s$
\begin{equation}
\label{eq:C}
  \vec{\omega}^{E_\perp}_{\rm lab} = -\frac{\eta}{2} \frac{\gamma v}{c} \frac{d\theta}{ds} \hat{E} + \left(1- \frac{1-\gamma^2 \beta^2 G}{\gamma} \right) \frac{d\theta}{ds} \hat{y} \;,
\end{equation}
where $\hat{y}$ is the axis perpendicular to the ring plane (vertical). Equation~\ref{eq:C} was obtained using the relation
\begin{equation}
\label{eq:D}
  E = \frac{m \gamma v^2}{q} \frac{d\theta}{ds}\;.
\end{equation}

We next convert Eq.~\ref{eq:C} to the accelerator reference frame where the $\hat{z}$ axis points along the particle velocity and the $\hat{x}$ axis forms a right-handed coordinate system with the vertical $\hat{y}$ axis and $\hat{z}$:
\begin{equation}
\label{eq:E}
  \vec{\omega}^{E_\perp}_{\rm acc} = -\frac{\eta \gamma \beta}{2} \frac{d\theta}{ds} \hat{x} - \frac{1-\gamma^2 \beta^2 G}{\gamma} \frac{d\theta}{ds} \hat{y} \;.
\end{equation}
The magic energy approach relies on cancellation of the second term in Eq.~\ref{eq:E} by going to the magic energy defined by the $1-\gamma^2 \beta^2 G=0$ condition in an all-electric ring. In a conventional circular ring, the EDM signal then accumulates. Some of the issues with this approach are a relatively large ring size defined by the magic energy and control of the spin coherence. In a Figure-8 ring design, where the net orbital rotation per turn is zero, the net spin rotation also automatically cancels every turn at any energy avoiding spin decoherence to first order. However, in a single-energy Figure-8 ring, the EDM signal also averages to zero. We solve this problem by adding sections of static longitudinal electric fields, as illustrated in Fig.~\ref{fig:ring_schematics}. They have no effect on the MDM spin rotation but, at the same time, break the ring symmetry for the EDM spin rotation. Only EDM can cause spin precession in longitudinal electric field $E_\parallel$. The spin rotation angle per unit length is then given by
\begin{equation}
\label{eq:F}
  \vec{\omega}^{E_\parallel}_{\rm acc} = -\frac{\eta q E_\parallel}{2mcv \gamma} \hat{z}\;.
\end{equation}

We now analyze the expected EDM sensitivity of such a ring and compare it to that of a magic-energy ring. Since the ring in Fig.~\ref{fig:ring_schematics} is a sequence of electrostatic arcs and straights, one can first obtain the spin precession angle and axis for each of these sections separately. Using Eq.~\ref{eq:E}, the spin precession in an arc is
\begin{equation}
\label{eq:G}
  \vec{\psi}_{\rm arc} = \int_{\rm arc}{\vec{\omega}^{E_\perp}_{\rm acc} dz} = - \left( \frac{\eta \gamma \beta}{2} \hat{x} +  \frac{1-\gamma^2 \beta^2 G}{\gamma} \hat{y} \right) \theta_{\rm arc} \;,
\end{equation}
where $\theta_{\rm arc}$ is the arc's orbital bending angle. For the spin precession in a straight, Eq.~\ref{eq:F} gives
\begin{equation}
\label{eq:H}
  \vec{\psi}_{\rm str} = \int_{\rm str}{\vec{\omega}^{E_\parallel}_{\rm acc} dz} = -\frac{\eta}{2} \ln \frac{\gamma_1 + \sqrt{\gamma_1^2 - 1}}{\gamma_2 + \sqrt{\gamma_2^2 - 1}} \hat{z} \;,
\end{equation}
where $\gamma_1$ and $\gamma_2$ are the particle's Lorentz energy factors at the entrance and exit of the straight, respectively.
One can next use Eqs.~\ref{eq:G} and~\ref{eq:H} to obtain and sequentially multiply the spin rotation matrices of the individual sections. One thus builds a spin rotation matrix of the entire ring. It can be analyzed to extract the spin rotation axis and angle over a single turn. This approach, however, yields a rather cumbersome expression even for the simple layout shown in Fig.~\ref{fig:ring_schematics}. It makes it difficult to understand the functional dependencies and relative importance of the different design parameters. Such a direct calculation is still useful to benchmark an approximate but highly accurate calculation described below.

It is instructive to treat the EDM signal as a perturbation to the MDM spin motion. This is of course valid to a high degree of accuracy because the EDM spin effect is many orders of magnitude smaller than the MDM one. Such an analysis technique is used, for example, to analyze the spin effect of ring imperfections and calculate a spin resonance strength. This technique has been demonstrated to yield accurate results~\cite{PhysRevLett.124.194801}.

According to this technique, the EDM effect is considered as a perturbation in the so-called natural spin reference frame ($\hat{e}_1$, $\hat{e}_2$, $\hat{e}_3$). The initial orientation of the spin reference frame can be chosen arbitrarily but, at the start point, it is convenient to align it with the accelerator reference frame:
\begin{equation}
\label{eq:I}
  \hat{e}_1(s=0) = \hat{x}, \;\;\;\;\;\; \hat{e}_2(s=0) = \hat{y}, \;\;\;\;\;\; \hat{e}_3(s=0) = \hat{z} \;.
\end{equation}
The natural frame follows the MDM spin motion around the ring, {\em i.e.}, the MDM component of the spin remains fixed in such a frame. Since the ring is spin transparent, the natural frame is one-turn periodic and returns to the orientation given by Eq.~\ref{eq:I} on every pass. Evolution of the natural-frame unit vectors in the accelerator reference frame is described by
\begin{equation}
\label{eq:J}
  \frac{d\vec{e}}{d\theta} = - \frac{1-\gamma^2 \beta^2 G}{\gamma} [\hat{y} \times \vec{e}] \;.
\end{equation}

In the perturbative approximation, the components of the spin rotation per particle turn due to EDM are given in the spin frame by the following integral over the ring circumference 
\begin{equation}
\label{eq:K}
  (\psi^{\rm OT}_{\rm EDM})_i = \oint{\vec{\omega}_{\rm EDM}(s) \hat{e}_i(s) ds} \;,
\end{equation}
where OT stands for One Turn and $\vec{\omega}_{\rm EDM}(s)$ is the spin precession due to EDM alone per unit orbital length. The EDM components $\vec{\omega}^{E_\perp}_{\rm EDM}$ and $\vec{\omega}^{E_\parallel}_{\rm EDM}$ of the total spin precession rates follow from Eqs.~\ref{eq:E} and~\ref{eq:F} for transverse and longitudinal electric fields, respectively. They are given by
\begin{equation}
\label{eq:L}
  \vec{\omega}^{E_\perp}_{\rm EDM} = - \frac{\eta \gamma \beta}{2} \frac{d\theta}{ds} \hat{x} \; {\rm and}
\end{equation}
\begin{equation}
\label{eq:M}
  \vec{\omega}^{E_\parallel}_{\rm EDM} = - \frac{\eta q E_\parallel}{2mcv\gamma} \hat{z} \;.
\end{equation}

By substituting Eqs.~\ref{eq:J},~\ref{eq:L}, and~\ref{eq:M} into Eq.~\ref{eq:K} and completing the integral for the ring configuration shown in Fig.~\ref{fig:ring_schematics}, we arrive at the following expressions for the spin rotation components:
\begin{equation}
\label{eq:N}
  (\psi^{\rm OT}_{\rm EDM})_1 = -2 \eta \left[ \frac{\gamma_2^2 \beta_2}{1-\gamma_2^2 \beta_2^2 G} - \frac{\gamma_1^2 \beta_1}{1-\gamma_1^2 \beta_1^2 G} - \ln \frac{\gamma_2 + \sqrt{\gamma_2^2 - 1}}{\gamma_1 + \sqrt{\gamma_1^2 - 1}} \right] \sin\left( \frac{\omega_M^1}{2} \pi\right) \sin\left( \frac{\omega_M^2}{2} \pi\right)  \sin\left( \frac{\omega_M^2 - \omega_M^1}{2} \pi\right),
\end{equation}
\begin{equation} 
\label{eq:O}
  (\psi^{\rm OT}_{\rm EDM})_2 = 0 ,
\end{equation}
\begin{equation}
\label{eq:P}
  (\psi^{\rm OT}_{\rm EDM})_3 = 2 \eta \left[ \frac{\gamma_2^2 \beta_2}{1-\gamma_2^2 \beta_2^2 G} - \frac{\gamma_1^2 \beta_1}{1-\gamma_1^2 \beta_1^2 G} - \ln \frac{\gamma_2 + \sqrt{\gamma_2^2 - 1}}{\gamma_1 + \sqrt{\gamma_1^2 - 1}} \right] \sin\left( \frac{\omega_M^1}{2} \pi\right) \sin\left( \frac{\omega_M^2}{2} \pi\right)  \cos\left( \frac{\omega_M^2 - \omega_M^1}{2} \pi\right).
\end{equation}
The quantity $\omega_M^n$ in Eqs.~\ref{eq:N} and~\ref{eq:P} stands for
\begin{equation} 
\label{eq:Q}
  \omega_M^n \equiv - \frac{1+G-\gamma_n^2 G}{\gamma_n} \;,
\end{equation}
where $n=1$, 2 specifies the electron Lorentz energy factor $\gamma_n$ in the two different energy regions. The physical meaning of $\omega_M^n$ is the spin precession due to MDM per unit orbital angle in transverse electric field. 

The magnitude of the spin rotation due to EDM in one turn can be obtained from Eqs.~\ref{eq:N}--\ref{eq:P} as:
\begin{equation}
\label{eq:R}
  \left| \psi^{\rm OT}_{\rm EDM} \right| = \left|2 \eta \left[ \frac{\gamma_2^2 \beta_2}{1-\gamma_2^2 \beta_2^2 G} - \frac{\gamma_1^2 \beta_1}{1-\gamma_1^2 \beta_1^2 G} - \ln \frac{\gamma_2 + \sqrt{\gamma_2^2 - 1}}{\gamma_1 + \sqrt{\gamma_1^2 - 1}} \right] \sin\left( \frac{\omega_M^1}{2} \pi\right) \sin\left( \frac{\omega_M^2}{2} \pi\right) \right| \;,
\end{equation}

We assume that, at the start of integration in Eq.~\ref{eq:K}, the spin frame is aligned with the accelerator and lab frames. Since the spin frame returns to its initial orientation after each turn around the orbit, it always coincides with the accelerator and lab reference frames at the start point. Therefore, the spin rotation components given by Eqs.~\ref{eq:N}--\ref{eq:P} and~\ref{eq:R} are the same in all three reference frames.
 
\begin{figure}[!htbp]
\includegraphics[width=0.75\columnwidth]{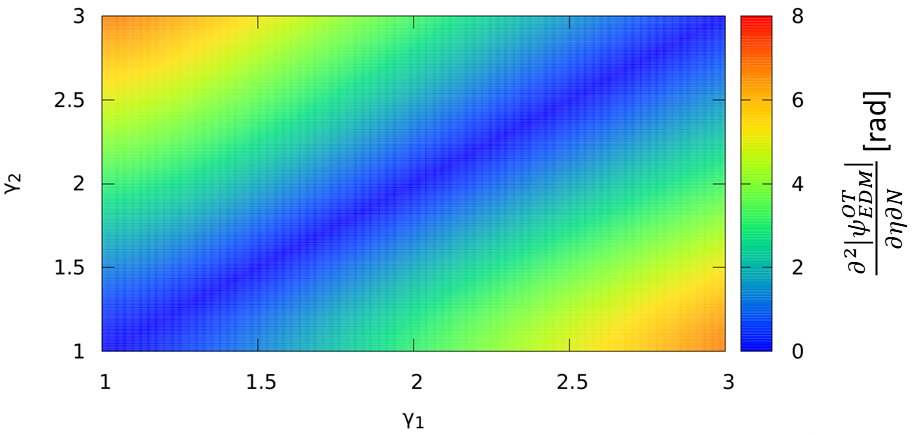}
\caption{ 
Magnitude of the EDM spin rotation in the two-energy ST ring per unit $\eta$ per turn $\partial^2 \left| \psi_{\rm EDM}^{\rm OT} \right|/(\partial \eta \partial N)$ as a function of $\gamma_1$ and $\gamma_2$.
\label{fig:EDMperTurn}
}
\end{figure}

The closed expressions in Eqs.~\ref{eq:N}--\ref{eq:P} and~\ref{eq:R} make it straightforward to analyze the parametric dependence of the EDM spin rotation. One can see that the EDM spin rotation axis lies in the horizontal plane. Besides the fundamental constant $G$, the EDM spin rotation in Eq.~\ref{eq:R} only depends on the choice of the electron beam energies $\gamma_1$ and $\gamma_2$. The magnitude of the EDM spin rotation per unit $\eta$ per turn $N$, $\partial^2 \left| \psi_{\rm EDM}^{\rm OT} \right|/(\partial \eta \partial N)$, is plotted in Fig.~\ref{fig:EDMperTurn} as a function of $\gamma_1$ and $\gamma_2$. As expected, the EDM signal goes to zero on the diagonal of Fig.~\ref{fig:EDMperTurn} where $\gamma_1=\gamma_2$ and the spin dynamics reduces to that in a single-energy ST ring. Reaching a substantial EDM signal requires a significant difference between $\gamma_1$ and $\gamma_2$. However, while increasing the energy difference provides a greater EDM spin rotation per single turn, it is also associated with increase in the ring size and therefore increase in the revolution time. Since the figure of merit is the EDM signal build up per unit time, one must optimize the signal per turn and the revolution time together to maximize their ratio.

In the discussion below, we select conservative energies and their difference corresponding to $\gamma_1=1.4$ and $\gamma_2=2.6$, {\em i.e.}, the electrostatic potential difference over the accelerating/decelerating gaps corresponds to 600~kV.
 
\begin{figure}[!htbp]
\includegraphics[width=0.75\columnwidth]{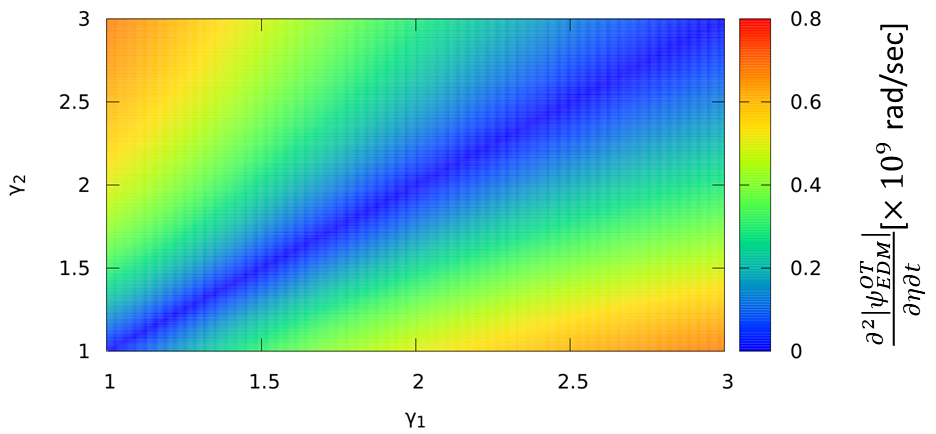}
\caption{ 
Magnitude of the EDM spin rotation in the two-energy ST ring per unit $\eta$ per unit time $\partial^2 \left| \psi_{\rm EDM}^{\rm OT} \right|/(\partial \eta \partial t)$ as a function of $\gamma_1$ and $\gamma_2$.
\label{fig:EDMperTime}
}
\end{figure}

For the choice of $\gamma_1=1.4$ and $\gamma_2=2.6$, the EDM precession is $\partial^2 \left| \psi_{\rm EDM}^{\rm OT} \right|/(\partial \eta \partial N)=4.24$ rad. For comparison, let us consider the EDM precession $\partial^2 \left| \psi_{\rm EDM,ME}^{\rm OT} \right|/(\partial \eta \partial N)$ that one gets in a racetrack at the magic energy (ME). One can readily obtain its expression from Eq.~\ref{eq:E}:
\begin{equation}
\label{eq:S}
  \frac{\partial^2 \left| \psi_{\rm EDM,ME}^{\rm OT} \right|}{\partial \eta \partial N} = \frac{\pi}{\sqrt{G}} = 92.24 \; {\rm rad}. 
\end{equation}
To compare the efficiencies of the two-energy ST and magic-energy concepts, we need to evaluate their achievable rates of the EDM spin rotation per unit time. This in turn requires estimates of their typical revolution times. Let us assume bending and accelerating/decelerating electric fields of $|E|=10$ MV/m. We also assume that these fields constitute only a half of the ring circumference, {\em i.e.}, a packing factor of 0.5. The rest of the ring is occupied by other machine components and experimental apparatus. The electron revolution time in the two-energy ST ring of Fig.~\ref{fig:ring_schematics} is
\begin{equation}
\label{eq:T}
  T= 2 \frac{mc^2/e}{c |E|} \left( 2 \pi \gamma_1 \beta_1 + 2 \pi \gamma_2 \beta_2 + 4 \left[\sqrt{\gamma_2^2-1} - \sqrt{\gamma_1^2-1} \right] \right).
\end{equation}
The EDM spin rotation per unit time is
\begin{equation}
\label{eq:U}
  \frac{\partial^2 \left| \psi_{\rm EDM}^{\rm OT} \right|}{\partial \eta \partial t} = \frac{1}{T} \frac{\partial^2 \left| \psi_{\rm EDM}^{\rm OT} \right|}{\partial \eta \partial N}.
\end{equation}
The EDM signal in terms of the EDM spin rotation per unit $\eta$ per unit time $\partial^2 \left| \psi_{\rm EDM}^{\rm OT} \right|/(\partial \eta \partial t)$ is plotted in Fig.~\ref{fig:EDMperTime} as a function of $\gamma_1$ and $\gamma_2$. For $\gamma_1=1.4$ and $\gamma_2=2.6$, the EDM precession rate is ${\partial^2 \left| \psi_{\rm EDM}^{\rm OT} \right|}/{(\partial \eta \partial t)} = 0.46 \times 10^{9}$ rad/s. 

Under the same field strength and ring composition assumptions as above, the revolution time around a racetrack ring at the magic energy is given by
\begin{equation}
\label{eq:V}
  T_{\rm ME} = 2 \frac{mc^2/e}{c |E|} \frac{2\pi}{\sqrt{G}}.
\end{equation}
With the revolution time of Eq.~\ref{eq:V}, the EDM spin precession rate in a magic-energy ring is ${\partial^2 \left| \psi_{\rm EDM,ME}^{\rm OT} \right|}/{(\partial \eta \partial t)} = 1.47\times10^{9}$~rad/s. Thus, the large difference in the EDM spin rotation per turn between the spin-transparent and magic-energy rings is largely balanced out by the fact that the spin-transparent ring can be much more compact due to the low electron energies.

Assuming an electron EDM of~\delectron{} $=10^{-29}$~\ecm{} corresponding to $\eta=1.04\times10^{-18}$, the expected EDM spin precession rates are about 1.53 and 0.48~nrad/s for the magic-energy and two-energy ST rings, respectively. Note that these are rough estimates. The specific spin-transparent ring design described below results in ${\partial^2 \left| \psi_{\rm EDM}^{\rm OT} \right|}/{(\partial \eta \partial t)} = 0.29 \times 10^{9}$ rad/s and ${\partial \left| \psi_{\rm EDM}^{\rm OT} \right|}/{\partial t} = 0.30$~nrad/s, which are in a reasonable agreement with the estimates. This level of signal should be experimentally measurable given adequate suppression of systematic effects. Table~\ref{tab:ring-comp} summarizes the EDM spin rotation parameters in the two scenarios. For the ST ring, Table~\ref{tab:ring-comp} shows the more conservative parameters for the illustration design discussed later.   

Note that this comparison is a purely hypothetical exercise, since there are currently no proposals for measuring the electron EDM in a storage ring at the magic energy. The main difficulties with applying the magic-energy approach to electrons are lack of suitable polarimetry and complications due to synchrotron radiation.

\setlength{\tabcolsep}{12pt}
\begin{table*}[!htb] \caption{Comparison of the EDM spin rotations in the ME and two-energy ST rings. }
 \label{tab:ring-comp}
    \centering
    \begin{tabular}{l|c|c|c|c}
\hline \hline
 & & & &  \\
Scheme           & Energy  & $\partial^2 \left| \psi_{\rm EDM}^{\rm OT} \right|/(\partial \eta \partial N)$ &  $\partial^2 \left| \psi_{\rm EDM}^{\rm OT} \right|/(\partial \eta \partial t)$ &  ${\partial \left| \psi_{\rm EDM}^{\rm OT} \right|}/{\partial t}$ \\ 
                 &  configuration  & (rad)        & ($\times 10^9$~rad/s) & (nrad/s) \\ 
                 &  $(\gamma)$                      &              &                       & assuming \\ 
                 &                        &              &                       & $\eta=$ \\ 
                 &                        &              &                       & $1.04\cdot 10^{-18}$ \\ 
 & & & &  \\ \hline 
 & & & &  \\
 ME & 29.38            & 92.24        & 1.47                     & 1.53 \\ 
 & & & &  \\ \hline
 & & & &  \\
 ST & (1.4, 2.6)  & 4.24         & 0.29                     & 0.30 \\ 
 & & & &  \\ \hline \hline
    \end{tabular}
\end{table*}

\subsection{Two-Energy ST Electron EDM Ring: Optics Design}

An EDM ring optics design must provide high efficiency in terms of the EDM spin rotation rate, long spin coherence time, adequate momentum acceptance and dynamic aperture, low emittance growth rates due to Intra-Beam Scattering (IBS), and reasonable stored beam size. We choose the so-called Bates design~\cite{FLANZ1985325} for the EDM ring arcs. Each arc consists of five alternating bends located symmetrically about its center as $\pm | \theta_1 | \wedge \mp | \theta_2 | \wedge  \pm 2| \theta_3 | \wedge  \mp | \theta_2 | \wedge \pm | \theta_1 |$ where $\theta_i$ are the three different angular parameters defining the arc's geometry. We use half-circular arcs so that, for each arc, 
\begin{equation}
\label{eq:bates_arc}
  \left | 2\sum_{i = 1}^3 \theta_i \right | = 180^\circ.
\end{equation}
Such an arc can be made very compact with most of its length occupied by bending elements, which is important for maximizing the EDM spin rotation rate. The inefficiency of its alternating end bends not contributing to the EDM signal is largely made up for by its compactness. The transverse optics design is based on the weak-focusing principle providing robustness and allowing for large beam size and momentum spread. Horizontal focusing comes purely from the bending curvature while vertical focusing is provided by weak electric field gradient. 

Due to the change of the bending direction between the adjacent arcs, each arc has to be achromatic. The dispersion is suppressed in each of the high-energy Bates bends by adjusting $\theta_i^{high}$ while keeping Eq.~\ref{eq:bates_arc} satisfied. In the high-energy arcs, we keep the absolute bending radii $(|\rho_i^{high}|)$ of the different bending sections and therefore their horizontal focusing strengths $(\propto 1/(\rho_i^{high})^2)$ fixed. Such a design is relatively simple from the engineering point of view by having only a few adjustment knobs. By not having large beam envelope variations, the Bates design is also efficient in terms of having minimal IBS growth rates. IBS is the main mechanism for degradation of intense electron beams at low energy. 

\begin{figure}[!htbp]
\includegraphics[width=0.75\columnwidth]{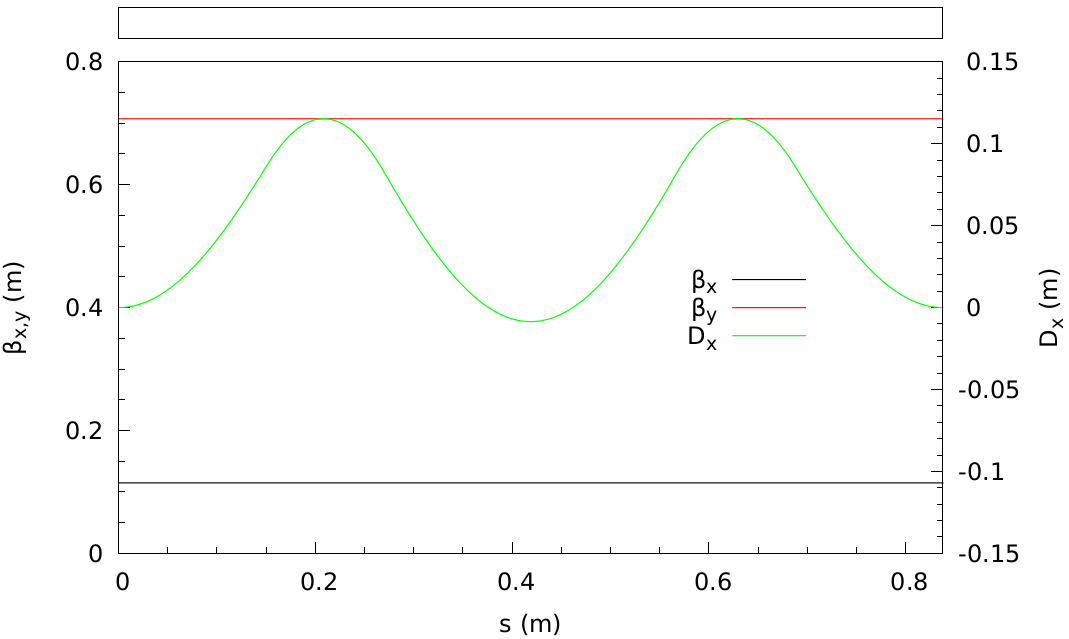}
\caption{ 
Optics of the high-energy arc showing the Twiss $\beta$ functions in both planes ($\beta_{x,y}$) and the horizontal dispersion ($D_x$) along the arc.
\label{fig:OpticsHiArc}
}
\end{figure}

The optics of a high-energy Bates arc is shown in Fig.~\ref{fig:OpticsHiArc}. The cell is designed using combined-function electron lenses. To minimize the arc length and therefore maximize the EDM rotation per unit time, we set the bending electric field to its practical limit of 10~MV/m. The vertical focusing electric field gradient $(\partial E_y/\partial y)_{high}$ is set to 2.3~MV/m$^2$. Even though our concept is based on a weak-focusing design, its high curvature $1/\rho$ provides relatively strong horizontal focusing $\propto 1/\rho^2$. Balanced by vertical focusing, it results in small Twiss $\beta_{x,y}$ functions in both planes and in a small horizontal dispersion $D_x$. This helps us keep the maximum beam size under control. 

When transitioning between the low and high-energy sections, the beam geometric emittances $\varepsilon_{x/y}$ scale inversely proportionally to the beam momentum with the normalized emittances $\varepsilon_{x/y}^N$ staying constant:
\begin{equation}
\label{eq:emit_sc}
  (\varepsilon_{x/y}^N)_{low} = (\varepsilon_{x/y}^N)_{high} = (\beta\gamma)_{low} (\varepsilon_{x/y})_{low} 
  = (\beta\gamma)_{high} (\varepsilon_{x/y})_{high}.
\end{equation}
Assuming for simplicity that the Twiss $\alpha$ functions are zero at the entrance and exit of each arc and noting that the beam size does not change significantly during passage through a longitudinal-field section, Eq.~\ref{eq:emit_sc} leads to the following optical matching requirement on the Twiss $\beta$ functions at the low- and high-energy ends of each longitudinal-field section
\begin{equation}
\label{eq:X}
  \frac{(\beta_{x/y})_{low}}{(\beta_{x/y})_{high}} = \frac{p_{low}}{p_{high}} \;.
\end{equation}

The condition of Eq.~\ref{eq:X} can in principle be met by scaling the dimensions and the absolute electric field strengths of the low-energy arcs by factors of $p_{low}/p_{high}$ and $\beta_{low}/\beta_{high}$, respectively. While the resulting shortening of the low-energy arc lengths benefits the EDM rotation rate, there is still a limit on how small the bending radius $\rho$ can be. For obvious geometric reasons, it cannot be smaller than the horizontal beam pipe half-aperture size $a_x$. Even though it may be feasible to approach this limit with an electrostatic design, we adhere to a more conventional picture in our example design by requiring $\rho$ to be much greater than $a_x$. $a_x$ in turn must be much greater than the rms horizontal beam size $\sigma_x$ to provide a sufficient beam stay clear and maintain a long beam lifetime. These requirements can be summarized as:
\begin{equation}
\label{eq:row_low_lim}
  \rho \gg a_x \gg \sigma_x.
\end{equation}
The bending radius of $\gamma=2.6$ electrons in 10~MV/m transverse electric field is
\begin{equation}
\label{eq:W}
  \rho^{high} = \frac{m\gamma v^2}{|qE|} = \frac{mc^2(\gamma^2-1)}{|qE|\gamma} \simeq 11.3 \; {\rm cm}.
\end{equation}
As it will be shown below, the expected $\sigma_x$ of a stochastically-cooled beam is about 6~mm. We meet the requirement of Eq.~\ref{eq:row_low_lim} by considering $a_x$ of 3~cm. However, this does not leave much room for reducing $\rho$ below that of the high-energy arcs. Therefore, we keep the bending radii $(\rho_i^{low})$ of the different bending sections of the low-energy arcs nearly the same as $\rho_i^{high} = \rho^{high}$ of the high-energy arcs. Unlike $\rho_i^{high}$, $\rho_i^{low}$ are allowed to vary independently from each other and are only slightly adjusted to match $(\beta_x)_{low}$ while keeping the low-energy arcs achromatic. The smallest $\rho_i^{low}$ is still reasonably large at about 8.3~cm. The bending angles $\theta_i^{low}$ of the low-energy arc sections are the same as $\theta_i^{high}$. These changes have no effect on $(\beta_y)_{low}$, which only depends on the constant value of $(\partial E_y/\partial y)_{low}$ in this design. The condition of Eq.~\ref{eq:X} on $(\beta_y)_{low}$ is met by scaling $(\partial E_y/\partial y)_{low}$ by a factor of $\gamma_{high}/\gamma_{low}$ to 4.2~MV/m$^2$.

\begin{figure}[!htbp]
\includegraphics[width=0.75\columnwidth]{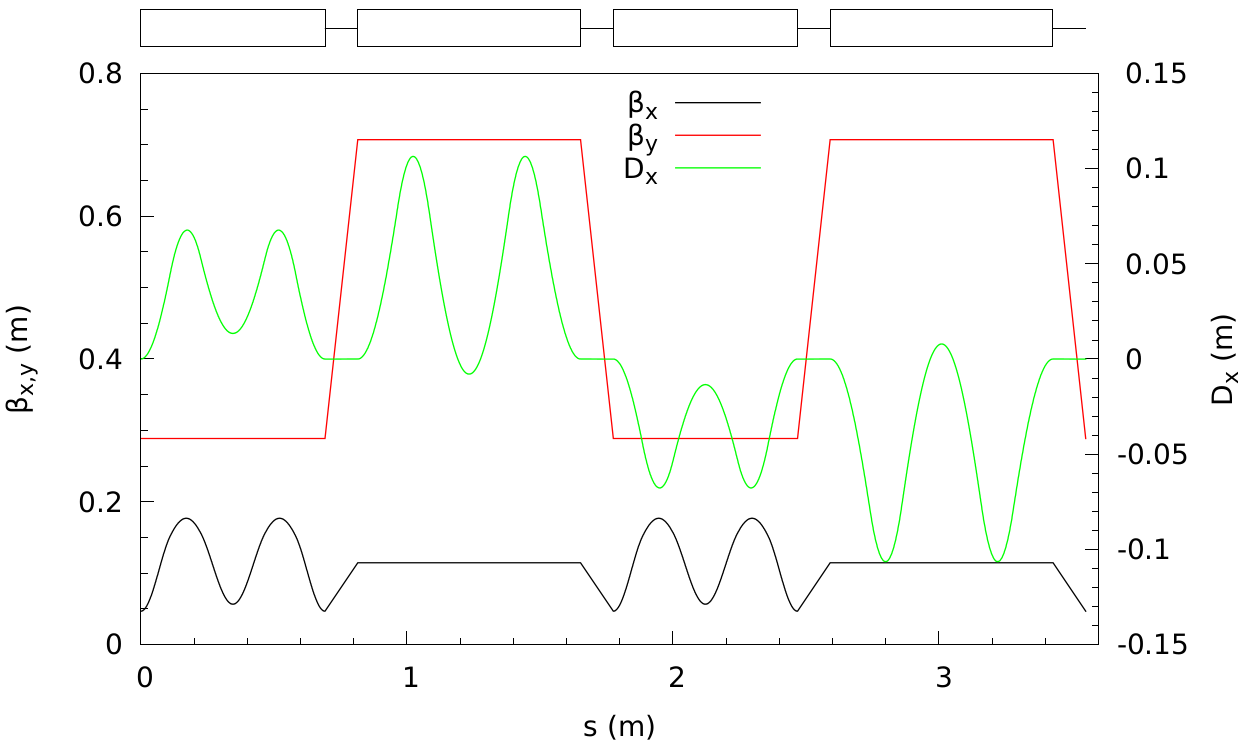}
\caption{ 
Complete optics of the EDM two-energy ST ring showing the Twiss $\beta$ functions in both planes ($\beta_{x,y}$) and the horizontal dispersion ($D_x$) along the ring.
\label{fig:OpticsEntireRing}
}
\end{figure}

\begin{figure}[!htbp]
\includegraphics[width=0.75\columnwidth]{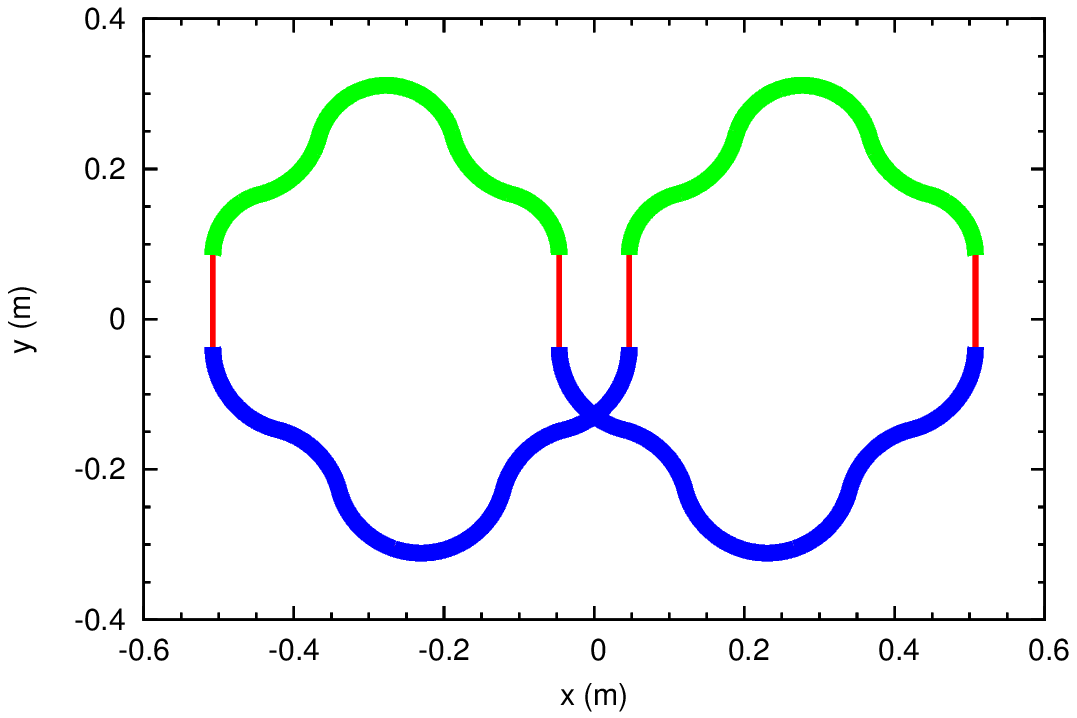}
\caption{ 
Footprint of the EDM ST ring drawn to scale showing the high-energy arcs in blue, the low-energy arcs in green, and the longitudinal static electric field sections in red.
\label{fig:RingFootprint}
}
\end{figure}

Figure~\ref{fig:OpticsEntireRing} shows the optics of the entire ring accounting for the change in the Twiss $\beta$ functions at the energy change regions. The corresponding footprint of the ring is shown in Fig.~\ref{fig:RingFootprint}. The main parameters of the ring design are summarized in Table~\ref{tab:ring-param}.
 
\begin{table*}[!htb] \caption{Geometrical and optical ring parameters. }
\label{tab:ring-param}
\centering
\begin{tabular}{lll}
\hline \hline
%\vspace{0.8cm}
Quantity                 & \multicolumn{2}{l}{Value} \\ \hline
Ring circumference	     & \multicolumn{2}{l}{3.55~m} \\ 
Circulation frequency $f_c$	 & \multicolumn{2}{l}{68.5~MHz} \\
Horizontal/vertical $Q_x/Q_y$ betatron tunes & \multicolumn{2}{l}{4.(53)/1.(14)} \\
Straight section length	 & \multicolumn{2}{l}{12.3~cm} \\ 
Beam pipe aperture $\pm a_x$  & \multicolumn{2}{l}{$\pm 3$~cm} \\ 
Longitudinal $E_{acc/dec}$ field & \multicolumn{2}{l}{5~MV/m} \\ 
$M_{56}$ of accelerating/decelerating straight & \multicolumn{2}{l}{22.3/54.8~mm} \\
$M_{66}$ of accelerating/decelerating straight & \multicolumn{2}{l}{0.31/3.23} \\ \\ 
Section                      & Low-energy                 & High-energy \\ \hline
Relativistic $\gamma$    & 1.4 & 2.6 \\
Arc bending angles $\theta_i$ & $\pm 75.0^\circ$, $\mp 61.0^\circ$, $\pm 76.0^\circ$ & $\pm 75.0^\circ$, $\mp 61.0^\circ$, $\pm 76.0^\circ$ \\
Arc bending radii $\rho_i$ & $\pm 8.3$, $\mp 10.8$, $\pm 9.2$~cm & $\pm 11.3$, $\mp 11.3$, $\pm 11.3$~cm \\
Arc bending $(E_x)_i$ field & $\mp 4.2$, $\pm 3.3$, $\mp 3.8$~MV/m & $\mp 10$, $\pm 10$, $\mp 10$~MV/m \\
Maximum Twiss $\beta_x/\beta_y$ & 17.7/28.9~cm & 11.5/70.7~cm \\ 
Maximum horizontal dispersion $D_x$ & 9.7~cm & 11.5~cm \\ 
Horizontal/vertical $\xi_x/\xi_y$ chromaticities per $180^\circ$ arc & -0.37/-1.93 & -0.63/-4.14 \\
$M_{56}$ per $180^\circ$ arc & 18.9~cm & 10.2~cm \\
Ring slip factor $\eta$ & -0.314 & -0.769 \\
\hline \hline
\end{tabular}
\end{table*}

\subsection{Longitudinal Dynamics}
As discussed above, beam bunching is necessary to distinguish and stochastically cool simultaneously-stored CR electron beams and as a means of efficient suppression of systematic effects. Therefore, a bunching RF cavity is inserted into the ring. In a matched equilibrium state with weak longitudinal focusing, each bunch has nearly constant length and rms absolute momentum spread around the ring. Since the slip factor $\eta$ is defined in terms of the relative momentum spread, it has different values in the low- and high-energy sections of the ring as shown in Table~\ref{tab:ring-param}. 

We choose a harmonic number of five providing sufficient bucket length for relatively intense bunches. Four of the buckets are filled while the fifth one is kept empty for the purposes of injection, extraction and ion cloud clearing. Considering densely populated buckets, we assume one quarter of the bucket length to be the rms bunch size. The equilibrium momentum spread at that bunch size is determined by a balance of the IBS diffusion and stochastic cooling as discussed in Section~\ref{sec:ibs-cooling}. The resulting beam parameters with and without cooling are given in Table~\ref{tab:beam-param}. The case without cooling is much less preferable because of large emittances and/or low achievable beam intensities. Therefore, we focus on the longitudinal dynamics parameters of a cooled bunched beam. 

The above bunch length, the momentum spread of a cooled beam provided in Table~\ref{tab:beam-param}, and the ring's slip factor specified in Table~\ref{tab:ring-param} determine the bunching cavity's voltage according to
\begin{equation}
\label{eq:bunching_volt}
  eV_{rf} = -\frac{\beta h \eta}{2\pi f_{rf}^2} \frac{p^2c^4}{E} \left( \frac{\sigma_{\Delta p/p}}{\sigma_z}\right)^2.
\end{equation}
Equation~\ref{eq:bunching_volt} gives a lower voltage when the bunching cavity is placed in the low-energy section rather than in the high-energy one. The voltage required in the former case is given in Table~\ref{tab:long-dyn} along with other key parameters relevant to the longitudinal beam dynamics in the two-energy ring.

\begin{table*}[!htb] \caption{Longitudinal dynamics parameters for a stochastically cooled beam. The bunching cavity is placed in the low-energy section.}
\label{tab:long-dyn}
\centering
\begin{tabular}{ll}
\hline \hline
Quantity                 & Value \\ \hline
Harmonic number $h$ & 5 \\
RF frequency $f_{rf}$ & 342.6~MHz \\
Bunch length $\sigma_z$ & 17.8~cm \\ 
Bunching voltage amplitude $V_{rf}$ & 1.67~kV \\
Synchrotron tune $\nu_{syn}$ & 0.029 \\
\hline \hline
\end{tabular}
\end{table*}

\begin{table*}[!htb] \caption{Beam parameters without and with stochastic cooling. }
\label{tab:beam-param}
\centering
\begin{tabular}{lllll}
\hline \hline
%\vspace{0.8cm}
Quantity & \multicolumn{4}{l}{Value} \\ \hline
Number of bunches & \multicolumn{4}{l}{4 CRA and 4 CRB} \\
Charge per bunch $Q_b$ & \multicolumn{4}{l}{2~nC at injection immediately reduced to 1~nC at store} \\ 
Beam current $I$ & \multicolumn{4}{l}{274~mA CRA and 274~mA CRB} \\
Stochastic cooling & \multicolumn{2}{l}{Off} & \multicolumn{2}{l}{On} \\
IBS growth times $\tau_x^{IBS}/\tau_y^{IBS}/\tau_z^{IBS}$ & \multicolumn{2}{l}{$10^4/10^4/10^4$~s} & \multicolumn{2}{l}{$40/40/4$~s} \\ \\
Section & Low-energy & High-energy & Low-energy & High-energy \\ \hline
rms momentum spread $\sigma_{\Delta p/p}$ & 0.22 & 0.091 & 0.034 & 0.014 \\
rms geometric $\varepsilon_x/\varepsilon_y$ emittances & 0.88/1.23~mm & 0.36/0.50~mm & 0.14/0.11~mm & 0.056/0.045~mm \\
Maximum rms beam size $\sigma_x^{max}/\sigma_y^{max}$ & 25/19~mm & 12/19~mm & 5.9/5.6~mm & 3.0/5.6~mm \\
\hline \hline
\end{tabular}
\end{table*}

\section{Beam Physics Issues and Limitations}

\subsection{IBS and Stochastic Cooling}
\label{sec:ibs-cooling}
Searching for EDM signal requires detection of minute changes in the beam's polarization. Therefore, the polarimeter must provide high-precision measurement of the polarization during an experiment. With systematic effects suppressed as discussed in Section~\ref{sec:sys-uncert}, the measurement precision is determined by the statistical error. Thus, one must store a sufficiently large number of particles in the ring. The statistical considerations for the polarimetry suggest that the following procedure. A 2~nC electron charge is initially injected into each RF bucket except for the empty one in each direction of circulation. The initial beam polarization is then immediately measured with good statistics reducing the bunch charge down to 1~nC ($N_e=6.24\times10^9$ of stored electrons). The reduced-charge bunches are stored for an extended period of time and their polarization is monitored for an EDM signal. The resulting statistics provides an adequate sensitivity level as discussed below.

With such a high bunch charge at such low energies, the greatest concern for the beam lifetime comes from IBS. The emittance growth rates due to IBS are directly proportional to the electron phase-space density and are inversely proportional to the second power of the beam energy. Consequently, the beam emittances and therefore the beam sizes must be sufficiently large to keep the growth rates at an acceptable level. We use the optics shown in Fig.~\ref{fig:OpticsEntireRing} and the beam and ring parameters listed in Tables~\ref{tab:ring-param}, \ref{tab:long-dyn}, and \ref{tab:beam-param} to estimate the emittance limitations imposed by IBS. 

Let us first consider the case without any beam cooling. We apply the Conte-Martini~\cite{Conte:1985sm} IBS model implemented in MAD-X~\cite{MAD-X} to calculate the emittance growth rates. MAD-X does not directly allow for calculating multiple-energy storage rings. Thus, we treat the two different-energy sections separately and wrap a generic optimizer around MAD-X calculations. We account for damping/anti-damping, optics scaling, and the difference in the geometric size of and in the amount of charge contained in each energy section. The optimizer solves the non-linearly-coupled Conte-Martini equations by adjusting the transverse emittance sizes $\varepsilon_{x/y}$ and the beam's relative momentum spread $\sigma_\delta$ until their results for the $\tau_{x/y/z}^{\rm IBS}$ IBS growth times in all three dimensions reach a level of $10^4$~s, {\em i.e.}, 
\begin{equation}
\label{eq:XII}
  \tau_x^{\rm IBS} = \tau_y^{\rm IBS} = \tau_z^{\rm IBS} = 10^4 \; {\rm s}. 
\end{equation}
The resulting emittances and relative momentum spread satisfying Eq.~\ref{eq:XII} are listed in Table~\ref{tab:beam-param}. Table~\ref{tab:beam-param} also gives the maximum rms beam sizes corresponding to these emittances and the optics parameters provided in Table~\ref{tab:ring-param}. Clearly, at the desired intensity, the uncooled beam sizes in Table~\ref{tab:beam-param} are not compatible with the maximum beam pipe aperture that can be used in a ring of such a small size.

We next consider the case when the beam is cooled stochastically. In an equilibrium, IBS is balanced by cooling, {\em i.e.}, the IBS growth rates are equal to the cooling rates. Therefore, we first need to make an assumption about the cooling performance. 

We assume that cooling is primarily longitudinal with about 10\% of the total cooling decrement coupled into the transverse dimensions. In addition to distributing the cooling decrements, our model accounts for coupling of the IBS rates. A typical stochastic cooling time can be estimated~\cite{MARRINER200411} as 
\begin{equation}
\label{eq:XV}
  \tau_z^{\rm cool} \sim \frac{N_e}{2W} , 
\end{equation}
where $N_e$ is the number of electrons and $W$ is the cooling system's bandwidth. For $N_e$ of $6.24\times 10^9$ and the state of the art $W$ of 10~GHz, Eq.~\ref{eq:XV} gives $\tau_z^{\rm cool}$ of 0.3~s. 

Equation~\ref{eq:XV} provides only an approximate estimate for the lower limit of the cooling time. The longitudinal dynamics parameters of our example EDM ring listed in Table~\ref{tab:long-dyn} allow for a more accurate estimate of $\tau_z$. Including both wanted and unwanted mixing effects as well as amplifier noise and assuming optimal gain, it is given by~\cite{Mohl1983hmt}
\begin{equation}
\label{eq:tau_z_acc}
  \tau_z^{\rm cool} \simeq \frac{N_e}{W} \frac{M+U}{(1-\tilde{M}^{-2})^2}, 
\end{equation}
where $U$ is the noise-to-signal power ratio and $M$ and $\tilde{M}$ are the wanted and unwanted mixing factors, respectively. For particles with the maximum longitudinal offsets $z_{max} \ge \sigma_z$, the synchrotron tune shift $\Delta \nu_{syn} > 0.15 \nu_{syn}$~\cite{Lee2018} and
\begin{equation}
\label{eq:want_mix_fact}
  M < \frac{c}{8 W \sigma_z \Delta \nu_{syn}} = \frac{c}{1.2 W \sigma_z \nu_{syn}} . 
\end{equation}

It is reasonable to assume $U=1$ and $\tilde{M}^{-1}=M^{-1}/2$. Then, for the parameters given in Table~\ref{tab:long-dyn}, Eqs.~\ref{eq:want_mix_fact} and \ref{eq:tau_z_acc} result in $M \lesssim 4.9$ and $\tau_z\lesssim 3.7$~s, respectively. Therefore, we assume a longitudinal cooling time $\tau_z^{\rm cool}$ of 4~s. Since longitudinal IBS and cooling balance each other in equilibrium, we need to find the beam parameters corresponding to the IBS times of 
\begin{equation}
\label{eq:XVI}
  \tau_x^{\rm IBS} = \tau_y^{\rm IBS} = 10 \tau_z^{\rm cool} = 40~{\rm s}, \;\;  \tau_z^{\rm IBS} = \tau_z^{\rm cool} = 4~{\rm s}, 
\end{equation}
where we assume that 10\% of the longitudinal cooling decrement is coupled into each of the transverse dimensions. We use the same optimization procedure as described above for the uncooled case to find the transverse emittances and momentum spread satisfying Eq.~\ref{eq:XVI}. They are listed in Table~\ref{tab:beam-param} along with the corresponding maximum beam sizes in the low and high-energy sections of the ring. 

\begin{figure}[!htbp]
\includegraphics[width=0.75\columnwidth]{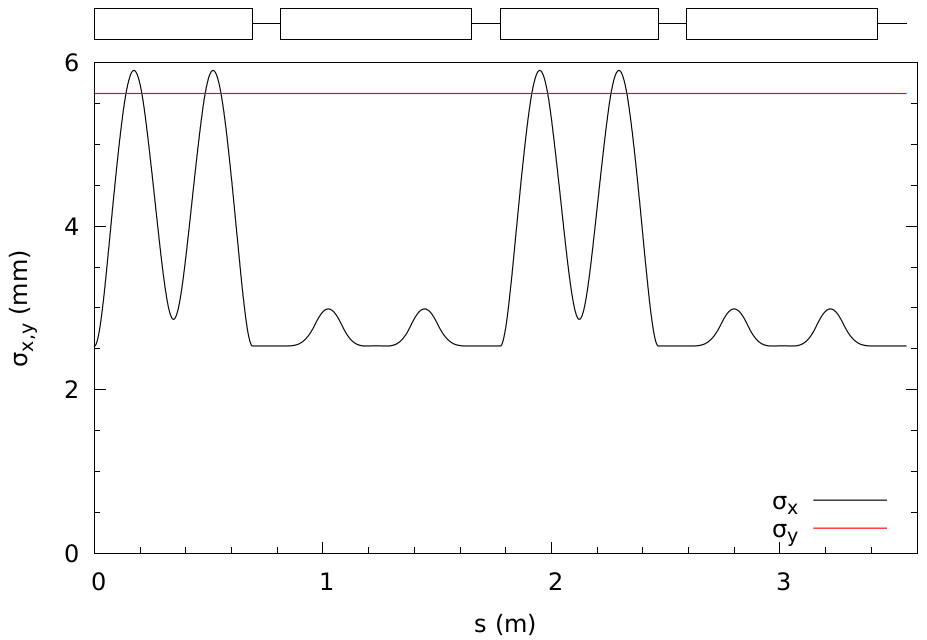}
\caption{ 
Beam sizes along the EDM ring when stochastic cooling is applied.
\label{fig:BeamSizeCooling}
}
\end{figure}

Figure~\ref{fig:BeamSizeCooling} plots the rms transverse beam sizes around the ring for the cooled beam parameters in Table~\ref{tab:beam-param} and the optics of Fig.~\ref{fig:OpticsEntireRing}. The beam sizes with stochastic cooling in Fig.~\ref{fig:BeamSizeCooling} are certainly more manageable and are more straightforward to realize in practice than those without cooling whose maximum values are summarized in Table~\ref{tab:beam-param}.

\subsection{Space Charge}
Another limitation on the amount of stored charge comes from the betatron tune shifts $\Delta \nu_{x/y}^{\rm SC}$ due to space charge fields. Beam stability requires that $|\Delta \nu_{x/y}^{\rm SC}|$ do not exceed the space charge threshold $|\Delta \nu_{x/y}^{\rm thr}|$ of about 0.3 (or even higher~\cite{PhysRevAccelBeams.24.044001,4441116}). As with IBS, the space charge effect is inversely proportional to the second power of the beam energy thus growing stronger with reduction in energy and being of a concern in our low-energy case.

The direct space-charge tune shift in a bunched beam is given by
\begin{equation}
\label{eq:XIX}
  |\Delta \nu_{x/y}^{\rm SC}| = \frac{r_e}{4\pi \varepsilon_{x/y}^N} \frac{N_b}{\sqrt{2\pi}\sigma_z} \oint{\frac{ds}{\beta \gamma^2}} = \frac{r_e}{4\pi \varepsilon_{x/y}^N} \frac{N_b}{\sqrt{2\pi}\sigma_z} \left( \frac{l_1}{\beta_1 \gamma_1^2} + \frac{l_2}{\beta_2 \gamma_2^2} + 4  \frac{mc^2}{e | E_\parallel | } \left[ \tan^{-1} ( \beta_2\gamma_2 ) - \tan^{-1} ( \beta_1 \gamma_1 ) \right] \right),
\end{equation}

\noindent where $L$ is the ring circumference, $l_1$ and $l_2$ are the lengths of the $\gamma_1$ and $\gamma_2$ energy sections, respectively, and $E_\parallel$ is the accelerating/decelerating electric field strength.

Using the parameter values from Tables~\ref{tab:ring-param}, \ref{tab:long-dyn}, and \ref{tab:beam-param} in Eq.~\ref{eq:XIX}, we get
\begin{equation}
\label{eq:XX}
  |\Delta \nu_{x/y}^{\rm SC}| = 3.4/4.2 \times 10^{-2} \ll |\Delta \nu_{x/y}^{\rm thr}| \sim 0.3.
\end{equation}
Thus, we conclude that space charge is not an issue in our case even in the cooled beam scenario.

More importantly, each stored beam experiences the field of the CR beam. Its local effect is a factor of $\gamma^2(1+\beta^2)$ stronger than the self-field interaction. The resulting tune shift is a factor of about 6 greater than that of a single beam: 
\begin{equation}
\label{eq:bb_tune_shift}
|\Delta \nu_{x/y}^{\rm BB}| = 0.2/0.25 \lesssim |\Delta \nu_{x/y}^{\rm thr}| \sim 0.3.
\end{equation}
Fortunately, it is still consistent with the typical threshold of 0.3. We also investigated Coulomb instabilities but these are overcome
by Landau damping.

\subsection{Incoherent (Single Electron) Synchrotron Radiation}
For electrons with $\gamma=2.6$, the power radiated by a single electron in free-space is estimated to be about 0.3~keV/s and for $\gamma=29.38$ (the magic energy case) the synchrotron radiation is about 35 keV/s per single electron. For the new ST ring described here, at this low electron energy combined with the proposed vacuum chamber design, some shielding effect is expected that suppresses the emission of synchrotron radiation~\cite{Murphy:1996yt,Warnock:1990} such that the bunching cavity can easily compensate for this energy loss. In contrast, there is no such shielding effect in the magic energy ring and synchrotron radiation is another major drawback when compared to the low energy ST ring. 

\subsection{Residual Gas and Beam Lifetime}

Beam lifetime or beam store time should be at least as long as the SCT. The vacuum level in such small ring can be as low as $1\times10^{-12}$~Torr (dominated by H$_2$ molecules) and can be achieved with combination of extreme low out-gassing due to very high-temperature bake-outs and very deliberate pumping using ion-pumps and non-evaporable getters~\cite{doi:10.1116/1.5010154}. The total energy loss of the electron beam due to interaction with residual gas molecules after 1~day store time is estimated to be about 1.1~keV and, as in the case of synchrotron radiation, the bunching cavity will compensate for any energy loss. The lifetime due to residual gas scattering is estimated to be 12 days. 

\subsection{CR Beams Interaction and Beam Lifetime}

The expected lifetime due to the CR beam-beam interaction is estimated to be more than 1~day. No energy loss is expected here since the identical beams are moving in opposite directions - no energy exchange in the scattering but will affect the emittance because of the angular distribution.

\section{Spin Stability and Precision Control: Spin Coherence Time}

\subsection{Spin Coherence Time}
SCT is the time beam stays polarized in the storage ring.  It is important to have a long SCT since this is the time available to accumulate and observe the EDM signal. 
After beam has been stored one unit of SCT, the second beam polarization measurement is performed.  For the second polarization measurement, the second half of the beam is extracted on the polarimeter target.
The proton EDM proposals at magic energy are limited by SCT of about 1000 s~\cite{PhysRevLett.117.054801}, where depolarization is caused by the spread in MDM spin precession mainly due to energy spread and being slightly off exact magic energy, while the beam lifetime is much longer. Since in our case the spin tune is energy independent, the energy spread does not contribute to depolarization in the first order. The main limitation comes from the spin tune spread due to the beam emittances. 

The spin motion in an ST ring is governed by a zero-integer spin resonance. The main parameter describing a spin resonance is its strength defined as the number of spin precessions per unit turn. The zero-integer spin resonance strength consists of coherent and incoherent parts. The coherent part is determined by the ring imperfections and is the same for all particles. It does not cause depolarization. It can be precisely compensated using weak correcting elements and its residual effects can be further suppressed by comparing polarization dynamics of the CR beams. The limit on SCT is determined primarily by the incoherent part of the spin resonance $\varepsilon_{inc}$, which causes the spins of particles with different betatron and synchrotron amplitudes precess at slightly different rates. There are also other effects, such as spin diffusion due to stochastic cooling, that need further study.
The ST theory~\cite{Filatov2020} still needs to be extended to electric rings to provide an analytic expectation for $\varepsilon_{inc}$ in the proposed ring described above. 
For now, we only provide an upper limit on $\varepsilon_{inc}$ based on the desired coherence time $\tau_{coh}$. Consider a particle with its motion invariants equal to the rms emittances of the bunch. Let us define $\tau_{coh}$ as the amount of time it takes the spin of such a particle to deviate by $60^\circ$ from its original direction. 
The spin component of this particle along the initial spin direction is then equal to $\cos 60^\circ = 1/2$ of its original value. 
Since the number of turns it takes the spin to complete such a rotation is $N_{coh}=1/(6\varepsilon_{inc})$, $\varepsilon_{inc}$ must satisfy
\begin{equation}
\label{eq:sct}
  \varepsilon_{inc} < \frac{1}{6 \tau_{coh} f_c}.
\end{equation}
For $\tau_{coh}$ of $10^5$~s (about 1~day) and $f_c$ of Table~\ref{tab:ring-param}, Eq.~\ref{eq:sct} gives $\varepsilon_{inc} < 2.4 \cdot 10^{-14}$. Just for scale, such levels of $\varepsilon_{inc}$ are reachable in magnetic ST rings even without applying any special suppression measures~\cite{Kondratenko2017}. 

\section{Dark Matter and Dark Energy Searches}

As in the ME based design, the low energy ST storage ring can also be used to search for spin precession induced by dark energy and ultra-light (axion) dark matter (DE/DM). 
The axion field gradient couples to the spin of transversely polarized electrons stored in the ring with a sensitivity proportional to the relativistic beam velocity, $\beta$, and SCT~\cite{PhysRevD.103.055010}. 
Since the proposed ST ring has a small size, one ring can be used to store longitudinally polarized electrons to measure EDM while a second ring can be used to store transversely polarized electrons for DE/DM search. 
Unlike the EDM ring, for the DE/DM ring, one can use an identical ring to EDM, or Figure-8 electric ring without longitudinal electric fields. 
For this search, the spin rotates around the electron's velocity and the main systematic uncertainty is longitudinal background magnetic field that rotates the spin of the CRA and CRB beams in the same direction.
However, DE/DM field will rotate the spin in opposite directions, resulting in cancellation of this background effect when combining CRA and CRB data.

Another method to probe dark matter is to measure oscillation in the EDM signal in the presence of an axion field~\cite{PhysRevD.104.096006, PhysRevD.99.083002,Pretz2020}.

\section{\label{sec:mott}Polarized Electron Source, Mott Polarimetry and Expected EDM Limit}

This section presents some details about the other hardware needed in an experiment to measure the electron EDM. Besides the ring discussed earlier, the polarized electron source and the Mott polarimeter are the other major equipment. 

\subsection{Polarized Electron Source}

\begin{figure}[!htbp]
\includegraphics[width=1.00\columnwidth]{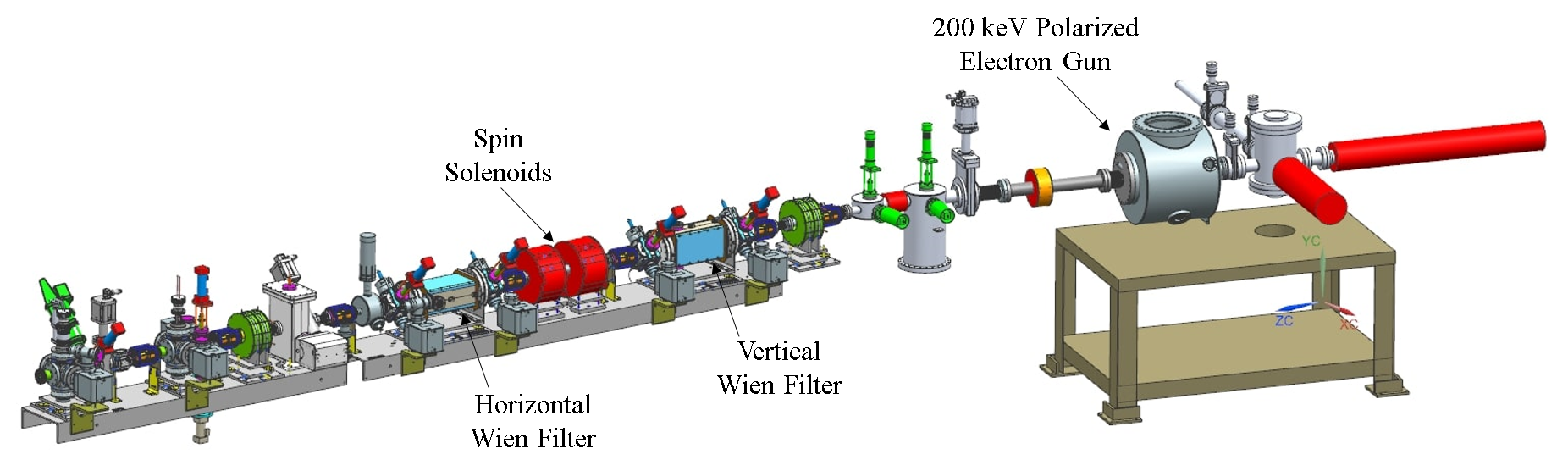}
\caption{\label{fig:source_schematics}
Jefferson Lab polarized electron gun and 3D spin manipulator. }
\end{figure}

Figure~\ref{fig:source_schematics} shows the main parts of the polarized electron source.  The polarized electron beam is generated when a 780~nm laser light is incident on a photocathode. The linearly polarized laser light first passes through a Pockels Cell that produces either right-handed or left-handed circularly polarized light. The photocathode is a super-lattice strained GaAs that produces 90\% longitudinally polarized electrons with either positive or negative helicity depending on the circular polarization of the laser light. The switching between positive and negative helicity is accomplished by changing the voltage applied to the Pockels Cell. The GaAs wafer is first activated (work function is lowered) in a preparation chamber by adding Cs and NF$_3$. After the activation, the photocathode is moved to the dc HV chamber where it is inserted in a cathode electrode that can be biased to -200~kV. The emitted electrons are accelerated in the cathode-anode gap and will have the time structure of the laser light. Nowadays, the polarized electron source can deliver beam polarization of 0.90 on a regular basis. More details about the polarized electron source can be found here~\cite{PhysRevSTAB.13.010101,ADDERLEY2023167710, Adderley:2011ri}.

The beamline is equipped with a 3D spin manipulator that can rotate the spin of the electrons in any direction. The first part of the spin manipulator is a Wien filter that rotates the spin vertically followed by a set of two solenoids that can rotate the spin in the transverse direction. Finally, a second Wien Filter is capable of rotating the spin in the horizontal plane. A similar spin manipulator has been in used reliably at Jefferson Lab over the last twelve years~\cite{Adderley:2011ri}.

\subsection{Electron Mott Polarimetry}

The build-up of the vertical component of the electron beam polarization due to spin precession from longitudinal to vertical caused by EDM (or from radial to vertical caused by DE/DM) can be measured using the conventional Mott polarimeter. The polarimeter can provide continuous and non-invasive measurement and simultaneously determines both the horizontal and the vertical asymmetries. The horizontal asymmetry is sensitive to the vertical polarization component and thus physics of interest while the vertical asymmetry will be used to monitor the horizontal polarization component.

\begin{figure}[!htbp]
\includegraphics[width=0.75\columnwidth]{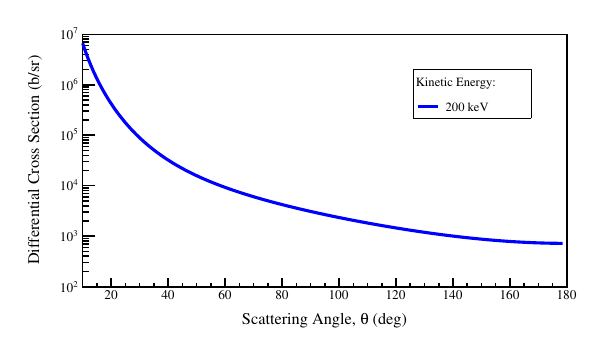}
\caption{\label{fig:mott_dcs}
Differential cross section of Mott scattering from $^{238}\mathrm{U}$ nucleus.}
\end{figure}

\begin{figure}[!htbp]
\includegraphics[width=0.75\columnwidth]{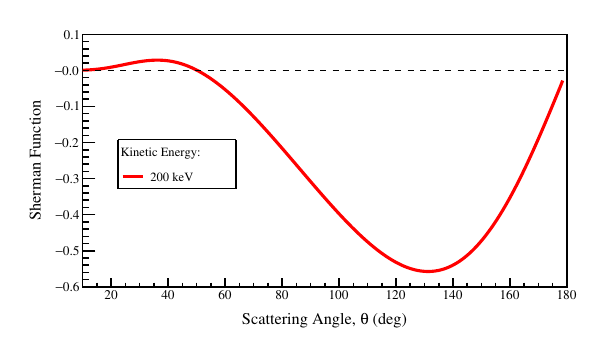}
\caption{\label{fig:mott_sf}
Sherman Function (Analyzing Power) of Mott scattering from a single $^{238}\mathrm{U}$ atom.}
\end{figure}

\begin{figure}[!htbp]
\includegraphics[width=0.75\columnwidth]{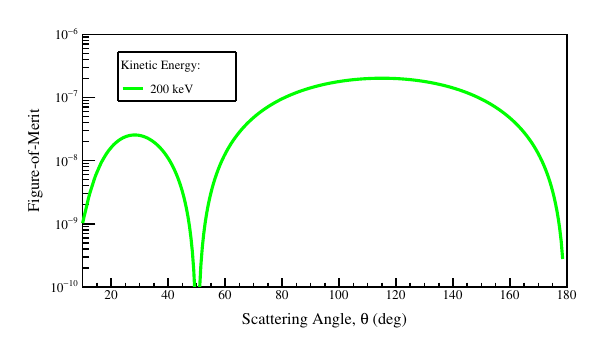}
\caption{\label{fig:mott_fom}
Polarimeter Figure-of-Merit of Mott scattering from 100 nm $^{238}\mathrm{U}$ foil into 1 msr solid angle.}
\end{figure}

Mott polarimetry based on elastic scattering of transversely polarized electrons from high-Z atomic nuclei is the standard technique for measuring electron beam polarization in the energy range from 20~keV to a few MeV~\cite{doi:10.1063/1.1143371,doi:10.1063/1.3556593, doi:10.1142/S0218301318300047, PhysRevC.102.015501}. Figure~\ref{fig:mott_dcs} shows the differential cross section of 200~keV kinetic energy electrons ($\gamma=1.4$) from $^{238}\mathrm{U}$ nucleus~\cite{Roca_Maza_2017,Roca_Maza}. The analyzing power of Mott scattering from a single atom, known as the Sherman Function, is shown in Fig.~\ref{fig:mott_sf} and can be as large as 0.56 at $130^\circ$ scattering angle. The Figure-of-merit, defined as: $fom = \epsilon A_y^2$, is plotted in Fig.~\ref{fig:mott_fom} for a 100~nm foil and a 1 msr detector solid angle. Here, $\epsilon$ is the polarimeter efficiency defined as the ratio of detected electrons over incident electrons and $A_y$ is the analyzing power (Sherman Function).    
 
\begin{figure}[!htbp]
\includegraphics[width=0.75\columnwidth]{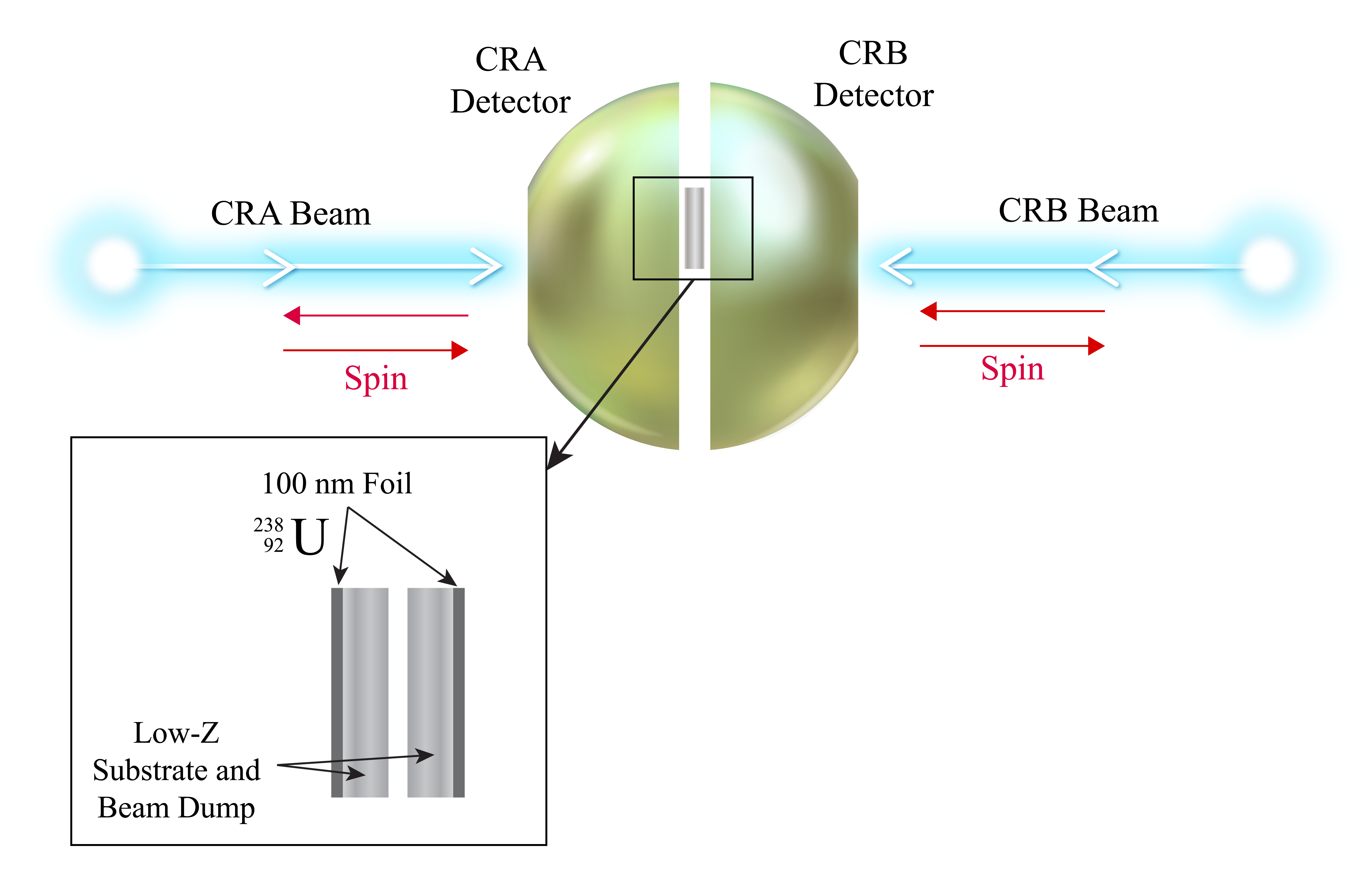}
\caption{\label{fig:pol_schematics}
Schematic of the Mott Polarimeter. The detectors will cover the scattering angles from $90^\circ$ to $160^\circ$ with full azimuthal coverage. The target is made of two 100~nm $^{238}\mathrm{U}$ foils with a low-Z substrate as a beam dump. }
\end{figure}

The schematic of a Mott polarimeter to measure electron beam transverse polarization is shown in Fig.~\ref{fig:pol_schematics}. A target made of two 100~nm $^{238}\mathrm{U}$ foil with a low-Z substrate can be inserted to intercept part of the beam. The substrate will serve also as a beam dump.

The detectors of the scattered electrons will cover the scattering angles from $90^\circ$ to $160^\circ$ with full azimuthal coverage. Over the detector acceptance, the relevant quantities to Mott polarimetry are:
\begin{itemize}
\item{}	Detector solid angle acceptance: $\int_{-\pi/2}^{\pi/2} {\int_{90^\circ}^{160^\circ} {sin\theta d\theta} d\phi} = 3.0$ steradian
\item{}	Integrated total cross section: $ {\int_{-\pi/2}^{\pi/2} {\int_{90^\circ}^{160^\circ} {cs(\theta) sin\theta} d\theta} d\phi} = 4876$ b
\item{}	Polarimeter efficiency: $\epsilon = \frac{\rho_U d_{\rm foil} N_A}{M_U} \int_{-\pi/2}^{\pi/2} {\int_{90^\circ}^{160^\circ} {cs(\theta) sin\theta d\theta} d\phi} = 0.0024$
\item{}	Polarimeter average analyzing power: $A_y = - \frac{\int_{-\pi/2}^{\pi/2} {\int_{90^\circ}^{160^\circ} {A_y(\theta) cs(\theta) sin\theta d\theta} d\phi}}{\int_{-\pi/2}^{\pi/2} {\int_{90^\circ}^{160^\circ} {cs(\theta) sin\theta d\theta} d\phi}} = 0.45$
\item{}	Polarimeter Figure-of-merit: $fom= \frac{\rho_U d_{\rm foil} N_A}{M_U} \int_{-\pi/2}^{\pi/2} {\int_{90^\circ}^{160^\circ} {cs(\theta) A_y^2(\theta) sin\theta d\theta} d\phi} = 4.9 \times10^{-4}$
\item{}	Mott measurement statistical uncertainty of vertical polarization due to EDM spin precession: $\delta P=\frac{1}{\sqrt{2P^2  Q / e fom}}= 0.029\%$ ($Q=2$~nC per bunch, two bunches CRA or CRB)
\end{itemize}

\subsection{Expected Electron EDM Statistical Limit}

Table~\ref{tab:pol_par} lists the parameters relevant for a polarization measurement. The initial electron beam polarization is measured by inserting the Mott target to intercept half the beam at $t=0$. The final electron beam polarization is performed at the end of the storage time at $t=$ SCT. For this polarization measurement scheme, the statistical uncertainty of the EDM measurement per fill can be calculated as~\cite{PhysRevD.104.096006, abusaif2019storage}:

\begin{equation}
  \sigma_{\rm EDM} = \sqrt{8} \frac{d_e}{\sqrt{Q/e \; \epsilon \;} \; A_y \; P \; \Omega_{\rm EDM} \; {\rm SCT} }  \; . \label{eq:EDM_limit}  
\end{equation}

With the numerical values given in Table~\ref{tab:pol_par}, the statistical uncertainty per fill on the EDM measurement is about $2.5 \times 10^{-28}$~\ecm{}. After five years of data taking, the projected statistical limit is about $5.8 \times 10^{-30}$~\ecm{} with the expectation that further optimization and improvements will lower this limit. 

\begin{table*}[!htb] \caption{Parameters of Mott Polarimetry. }
 \label{tab:pol_par}
    \centering
    \begin{tabular}{lcl}
\hline \hline
%\vspace{0.8cm}
Electrons per fill                & $N_e=Q/e$         & $5.0 \times 10^{10} (2.5 \times 10^{10}$ CRA, $2.5 \times 10^{10}$ CRB) \\
Polarimeter efficiency            & $\epsilon$    & 0.0024 \\
Analyzing power 	              & $A_y$         & 0.45 \\
Longitudinal polarization         & $P$           & 0.90 \\
Precession rate              & $\;\;\;\;\;\;\;\;$ $\Omega_{\rm EDM}$ $\;\;\;\;\;\;\;\;$ & 0.30~nrad/s (calculated assuming~\delectron{} $= 1 \times 10^{-29}$~\ecm{}) \\
Spin coherence time               & SCT           & 1~day (86400~s) \\
   \hline \hline
    \end{tabular}
\end{table*}

One possible scheme to operate the EDM storage ring and perform Mott measurements is as follows:
\begin{itemize}
\item{}	Ring will be filled with four polarized electron bunches, each 2~nC, in one direction and another four bunches, each 2~nC, in opposite direction.
\item{}	Four possible filling schemes for CRA:
\begin{enumerate}
\item{}	First bunch with positive helicity (CRA,+h), second bunch with negative helicity (CRA,-h), or
\item{}	First bunch with negative helicity (CRA,-h), second bunch with positive helicity (CRA,+h).
\item{}	Third bunch with in-radial polarization (CRA,+r), fourth bunch with out-radial polarization (CRA,-r), or
\item{}	Third bunch with out-radial polarization (CRA,-r), fourth bunch with in-radial polarization (CRA,+r). Or,
\item{}	Third bunch with up-vertical polarization (CRA,+v), fourth bunch with down-vertical polarization (CRA,-v), or
\item{}	Third bunch with down-vertical polarization (CRA,-v), fourth bunch with up-vertical polarization (CRA,+v).
\end{enumerate}
\item{}	Four possible filling schemes for CRB - similar to CRA.
\item{} Use half the charge of each bunch to measure the initial bunch polarization at $t=0$. Now, each bunch is 1~nC.
\item{}	Store for period of SCT = 1~day.
\item{} The longitudinally polarized bunches will be used to measure EDM while the radial and vertical polarized bunches will be used to measure the systematic uncertainties related to background electromagnetic fields.
\item{}	EDM will accumulate a positive vertical polarization component for (CRA,+h) and (CRB,-h) and negative vertical component for (CRA,-h) and (CRB,+h).
\item{}	Measure left-right and up-down asymmetries at the end of store time, $t=$ SCT by extracting the bunches to intercept the Mott target.
\item{}	Measure scattered electron asymmetries of each beam in its back-angle detector. Forward scattered electrons will be stopped in the substrate. Target ladder will be electrically isolated to measure intercepted charge from each beam.
\item{}	Expected systematic uncertainties of Mott Polarimeter of $\delta P \approx 1$ $\mu$rad. This is similar to the expected value for proton polarimetry at COSY~\cite{BRANTJES201249}.
\item{}	Mott measurement statistical uncertainty of vertical polarization due to EDM spin precession per fill (CRA and CRB) is $\delta P \approx 0.020\%$.
\item{}	After total running of 1825 fills (or five years), achieve measurement of vertical polarization due to EDM spin precession to statistical uncertainty of $\delta P \approx 4.7 \times 10^{-6}$ or 4.7~$\mu$rad.
\end{itemize}

\section{\label{sec:sys-uncert}Systematic Uncertainties}

Since the spin rotation due to EDM is much smaller than that of the MDM ($\eta/G \approx 10^{-15}$), MDM spin rotations limit the smallest EDM that can be measured and introduce systematic uncertainties~\cite{PhysRevD.105.032001, PhysRevAccelBeams.25.064001}. In the TS ring, the MDM spin rotations average to zero over a single turn. However, fringe and environmental electromagnetic fields and errors in the construction and alignment of the ring elements may introduce non-zero MDM spin rotations. One approach to further suppress residual MDM effects is the use of state-of-the-art shielding of environmental fields where the small size of the storage ring makes shielding very practical. Another approach relies on the fact that the EDM is time-reversal violating while the majority of the problems are time-reversal conserving. The time-reversal of the measurement technique can be implemented by the use of two CR beams. A third approach to suppress systematic uncertainties is to reverse the direction of the electron spin and combine data, for example collected during EDM measurements, as: $\frac{ {\rm (CRA,+h)} - {\rm (CRA,-h)} + {\rm (CRB,-h)} - {\rm (CRB,+h)}}{4}$. Finally, a fourth approach will use two of the four bunches (with either radial or vertical polarization, as was discussed in the previous section) in each direction of the ring to control background electromagnetic fields. Not all systematic effects cancel with either beam velocity or spin reversals, some systematic effects mimic the EDM signal and these will require more scrutiny~\cite{PhysRevD.105.032001}.

In the EDM measurement, the main systematic uncertainty is a radial background magnetic field that generates a vertical spin rotation identical to EDM, {\em i.e.}, it mimics EDM (no cancellations between contributions to CRA and CRB beams with either helicities). Any field left over after extensive shielding may be monitored by observing any vertical separation of the CRA and CRB beams. The separation is very small and state-of-the-art shielding of environmental fields is our first approach; this is made easier with such small ring. A recent innovation to reduce this effect envisioned a ring with electrical bending but with magnetic focusing (hybrid ring), albeit now vertical electric field is the main systematic uncertainty~\cite{PhysRevAccelBeams.22.034001}.

The coherent spin effect of the time-varying electric
and magnetic fields of stochastic cooling averages down to zero. There may still be a depolarizing incoherent spin effect in higher orders of the
fields that may limit SCT. It needs to be studied but it
should not in principle interfere with the measurement of
the coherent EDM spin effect.

For the DE/DM search, the spin rotates around the electron's velocity and the main systematic uncertainty is longitudinal background magnetic field that rotates the spin of the CRA and CRB beams in the same direction. However, DE/DM interaction will rotate the spin in opposite directions, resulting in cancellation of this background effect when combining CRA and CRB data.

\section{Conclusions}

We described a new method to directly measure the electron EDM in a small spin-transparent ring in energy range below 1 MeV. Mott polarimetry is proposed to detect the EDM or DE/DM related spin precession and is being recognized exclusively efficient in this energy range of electrons. All the envisioned issues and potential restrictions pointed earlier in this paper are manageable and detailed studies and evaluations will be presented in a future paper. The elicited limitations and required compensation or protection measures (including application of stochastic cooling) are not found to be crucial for the required effectiveness of this approach. The presented approach has the following advantages: low energy, a room size facility, spin-achromatic beam transport, ease precision and operation control, attainability of the ultra-high vacuum, and conventional polarimetry. Additionally, a similar ring to the one discussed above can be used to measure the positron EDM. Generating and accumulating polarized positron bunches have been extensively studied in relation to other projects~\cite{FangleiLin} and and was found to be very achievable.
 
Techniques of compensation for spin de-coherency due to beam emittances are under a consideration.  In particular, an intriguing possibility of implementing the Spin Echo technique~\cite{sym13030398} in electric ring facility is under study. The described avenue looks immediately extendable to polarized proton and deuteron beams of same energy and size range or close to that, thus becoming also a viable and feasible, chromaticity-free alternative to the magic energy method.

After all, the described approaches and ideas in organizing the spin-transparent beam and polarimetry are thought applicable, after proper modifications, to other exciting opportunities as high precision search for manifestation of the Dark Matter and Dark Energy of the Universe.

\begin{acknowledgments}
We appreciate very helpful discussions with M.~Blaskiewicz, 
A.~Hutton, Yu.~N.~Filatov, A.~M.~Kondratenko, M.~A.~Kondratenko,
V.~Lebedev, R.~Li, M.~Poelker, and T.~Satogata. This material is
based upon work supported by the U.S. Department of Energy, Office of Science, Office of Nuclear Physics under contract DE-AC05-06OR23177. This manuscript has been authored in part by
UT-Battelle, LLC, under contract DE-AC05-00OR22725 with
the US Department of Energy (DOE). The publisher acknowledges the US government license to provide public access under the DOE Public Access Plan (http://energy.gov/downloads/doe-public-access-plan).
\end{acknowledgments}

%%\newpage

\appendix

\section{\label{app:theory}Summary of Theoretical Considerations}

We can make a few statements resulting from our theoretical analysis omitting their mathematical proof (which is too detailed for this short arXiv paper but will be provided in future longer arXiv paper):

\begin{enumerate}

\item{} Accumulation of the stationary EDM precession requires presence of either longitudinal and/or transverse electric ﬁelds in the lab frame.
\item{} The resulting EDM precession eﬀect can be expressed via integrals over the sections with the transverse ﬁelds.
\item{} To accumulate the EDM precession in transverse electric ﬁeld, transverse magnetic ﬁeld must also be present with exception of the magic energy case.
\item{} In the Figure-8 conﬁguration, one has to provide an effective correlation between the radial electric ﬁeld and the MDM driven spin rotation. 
\item{} For small $\gamma G$ (electrons), MDM phase advance can include a few $360^\circ$ spin rotations in the arcs of our special Figure-8 synchrotron design.
\item{} With magnetic and electric bends, Figure-8 based approach can be efficiently used for low energy protons and deuterons with appropriate optimization of the spin phase/ﬁelds correlation. Note that one may consider a hadron energy as low as a few MeV coming out of an RFQ.
\item{} When suitable for optimization, spin can make additional full rotations in electric or magnetic loops introduced in arcs.
\item{} There are two theorems proved for an arbitrary (ﬂat or non-ﬂat) closed orbit:
\begin{enumerate}
\item{} 1st order average strength of storage ring EDM is zero in magnetic rings 
\item{} Similar, 1st order average strength of storage ring EDM is zero in rings with purely transverse electric ﬁeld – with only exclusion of magic energy point.
\end{enumerate}

\item{} So besides magic energy storage ring EDM, there are the following three diﬀerent types of a closed orbit design to set storage ring:

\begin{enumerate}
\item{} rings with closed orbit design on interleaving magnetic and electric bends, 
\item{} rings with magnetic bends interleaving with longitudinal (accelerating/decelerating) electric ﬁeld, and 
\item{} rings with electric bends interleaving with longitudinal (accelerating/decelerating) electric ﬁeld.
\end{enumerate}

\item{} Typical energy range consistent with reasonable range of acceleration/deceleration (static or RF) is about a few MeV; static looks preferred, but RF is considerable as well. It should be noted, however, that EDM precession average rate in modulated energy electric rings is proportional to factor $\gamma-1$, so the rate efficiency of this scheme is low compared to a magic energy ring for protons, or compared to a constant low energy electro-magnetic ring for protons or deuterons.
\item{} Figure-8 can be employed (unlikely to the magic energy case) for all possible low energy storage configurations.
\item{} Two CR beams can be stored in modulated energy all electric ring – that may be crucial for EDM detection.

\end{enumerate}

\newpage

\section{\label{app:long-paper}Subjects of Future arXiv Paper}

These are the subjects to be covered in the future longer arXiv paper:
 
\begin{verbatim}

I. Introduction

II. General properties of MDM and EDM spin dynamics in storage rings
 II.1. Basic equations 
  II.1.1. General EDM equations in MDM spin frame
  II.1.2. General properties of MDM dynamics
   II.1.2.1. MDM motion on a closed orbit
   II.1.2.2. MDM adiabatic invariant, spin field and generalized tune
   II.1.2.3. MDM equations in velocity frame
   II.1.2.4. Correlation theorem of MDM dynamics
   II.1.2.5. Correlation theorem in constant energy orbit configurations 
 II.2. General properties of EDM dynamics
  II.2.1. EDM precession in confront with MDM precession spectrum
   II.2.1.1. Spin equations in terms of the reference spin frame 
   II.2.1.2. EDM in case of a non-resonant MDM dynamics
   II.2.1.3. EDM in a spin-transparent MDM configuration
  II.2.2. EDM precession in confront with electric and magnetic structure of a storage ring
   II.2.2.1. EDM in magnetic configurations
   II.2.2.2. EDM in configurations with transverse electric field
   II.2.2.3. General reduction of EDM correlators for constant energy electric and magnetic orbits
   II.2.2.4. Magic energy case
   II.2.2.5. Reduction of EDM formula for orbits of a modulated energy

III. Proton EDM in small Figure-8 storage rings
 III.1. Proton EDM in constant energy electric and magnetic rings    
  III.1.1. General concept
  III.1.2. Proton and deuteron electric and magnetic racetracks 
  III.1.3. Proton and deuteron EDM in Figure-8 with electric and magnetic configurations
 III.2. Electron EDM in a modulated energy electric storage ring
  III.2.1. Racetrack orbit
  III.2.2. Electron EDM in a modulated energy Figure-8 ring

IV. Polarized source, polarimetry, and final selection
 IV.1. Polarized electron source
 IV.2. Mott polarimetry for electron beam
 IV.3. Systematic uncertainties 
 IV.4. Polarimetry for low energy proton and deuteron beams
 IV.5. Final selection    

V. Beam physics issues and limitations
 V.1. Beam stability 
  V.1.1. Internal and counter-rotating beams Space Charge 
  V.1.2. Synchrotron radiation, CSR, shielding
  V.1.3. External impedances
  V.1.4. Beam micro-wave stability: insability damping by energy spread 
  V.1.5. Feedback for long waves
  V.1.6. Energy loss and compensation
 V.2. Beam scattering
  V.2.1. Emittance growth due to IBS
  V.2.2. Scattering on residual gas 
  V.2.3. Beam-beam scattering
 V.3. Stochastic cooling
 V.4. Beam lifetime
 V.5. Spin diffusion 
  V.5.1. Direct spin scattering
  V.5.2. Spin-orbital diffusion
  V.5.3. Polarization lifetime
 V.6. Tables of parameters

VI. Spin refinement
 VI.1. Compensation for coherent detunes 
  VI.1.1. Vertical misalignment as mimic effect, and compensation 
  VI.1.2. Second order coherent detune, and compensation 
  VI.1.3. Background magnetic field, shield
 VI.2. Compensation for spin decoherency
  VI.2.1. Effects of non-linear transport fields 
  VI.2.2. Space charge and beam current effects 
  VI.2.3. Cancellation of the own and counter-rotating beams space charge effects in Figure-8
  VI.2.4. Cancellation of vertical emittance effect in Figure-8
  VI.2.5. Compensation for decoherency due to the coupling by errors
 VI.3. Stable spin demo

VII. Possibilities of search for Dark Matter and Dark Energy

VIII. Further new ideas
 VIII.1. EDM searches with bunched beams
 VIII.2. Spin Echo methods of spin refinement
 VIII.3. Spin Echo method for electron beams
 VIII.4. Spin Echo method for hadron beams

IX. Conclusion

\end{verbatim}

\newpage 

\bibliography{LowEnergyEDM_arXiv_paper2}% Produces the bibliography via BibTeX.

\begin{thebibliography}{46}%
\makeatletter
\providecommand \@ifxundefined [1]{%
 \@ifx{#1\undefined}
}%
\providecommand \@ifnum [1]{%
 \ifnum #1\expandafter \@firstoftwo
 \else \expandafter \@secondoftwo
 \fi
}%
\providecommand \@ifx [1]{%
 \ifx #1\expandafter \@firstoftwo
 \else \expandafter \@secondoftwo
 \fi
}%
\providecommand \natexlab [1]{#1}%
\providecommand \enquote  [1]{``#1''}%
\providecommand \bibnamefont  [1]{#1}%
\providecommand \bibfnamefont [1]{#1}%
\providecommand \citenamefont [1]{#1}%
\providecommand \href@noop [0]{\@secondoftwo}%
\providecommand \href [0]{\begingroup \@sanitize@url \@href}%
\providecommand \@href[1]{\@@startlink{#1}\@@href}%
\providecommand \@@href[1]{\endgroup#1\@@endlink}%
\providecommand \@sanitize@url [0]{\catcode `\\12\catcode `\$12\catcode
  `\&12\catcode `\#12\catcode `\^12\catcode `\_12\catcode `\%12\relax}%
\providecommand \@@startlink[1]{}%
\providecommand \@@endlink[0]{}%
\providecommand \url  [0]{\begingroup\@sanitize@url \@url }%
\providecommand \@url [1]{\endgroup\@href {#1}{\urlprefix }}%
\providecommand \urlprefix  [0]{URL }%
\providecommand \Eprint [0]{\href }%
\providecommand \doibase [0]{https://doi.org/}%
\providecommand \selectlanguage [0]{\@gobble}%
\providecommand \bibinfo  [0]{\@secondoftwo}%
\providecommand \bibfield  [0]{\@secondoftwo}%
\providecommand \translation [1]{[#1]}%
\providecommand \BibitemOpen [0]{}%
\providecommand \bibitemStop [0]{}%
\providecommand \bibitemNoStop [0]{.\EOS\space}%
\providecommand \EOS [0]{\spacefactor3000\relax}%
\providecommand \BibitemShut  [1]{\csname bibitem#1\endcsname}%
\let\auto@bib@innerbib\@empty
%</preamble>
\bibitem [{\citenamefont {Chupp}\ \emph {et~al.}(2019)\citenamefont {Chupp},
  \citenamefont {Fierlinger}, \citenamefont {Ramsey-Musolf},\ and\
  \citenamefont {Singh}}]{RevModPhys.91.015001}%
  \BibitemOpen
  \bibfield  {author} {\bibinfo {author} {\bibfnamefont {T.~E.}\ \bibnamefont
  {Chupp}}, \bibinfo {author} {\bibfnamefont {P.}~\bibnamefont {Fierlinger}},
  \bibinfo {author} {\bibfnamefont {M.~J.}\ \bibnamefont {Ramsey-Musolf}},\
  and\ \bibinfo {author} {\bibfnamefont {J.~T.}\ \bibnamefont {Singh}},\ }\href
  {https://doi.org/10.1103/RevModPhys.91.015001} {\bibfield  {journal}
  {\bibinfo  {journal} {Rev. Mod. Phys.}\ }\textbf {\bibinfo {volume} {91}},\
  \bibinfo {pages} {015001} (\bibinfo {year} {2019})}\BibitemShut {NoStop}%
\bibitem [{\citenamefont {Fortson}\ \emph {et~al.}(2003)\citenamefont
  {Fortson}, \citenamefont {Sandars},\ and\ \citenamefont
  {Barr}}]{Fortson:2003}%
  \BibitemOpen
  \bibfield  {author} {\bibinfo {author} {\bibfnamefont {N.}~\bibnamefont
  {Fortson}}, \bibinfo {author} {\bibfnamefont {P.}~\bibnamefont {Sandars}},\
  and\ \bibinfo {author} {\bibfnamefont {S.}~\bibnamefont {Barr}},\ }\href
  {https://doi.org/10.1063/1.1595052} {\bibfield  {journal} {\bibinfo
  {journal} {Physics Today}\ }\textbf {\bibinfo {volume} {56}},\ \bibinfo
  {pages} {33} (\bibinfo {year} {2003})}\BibitemShut {NoStop}%
\bibitem [{\citenamefont {Yamanaka}(2017)}]{doi:10.1142/S0218301317300028}%
  \BibitemOpen
  \bibfield  {author} {\bibinfo {author} {\bibfnamefont {N.}~\bibnamefont
  {Yamanaka}},\ }\href {https://doi.org/10.1142/S0218301317300028} {\bibfield
  {journal} {\bibinfo  {journal} {International Journal of Modern Physics E}\
  }\textbf {\bibinfo {volume} {26}},\ \bibinfo {pages} {1730002} (\bibinfo
  {year} {2017})}\BibitemShut {NoStop}%
\bibitem [{\citenamefont {Roussy}\ \emph {et~al.}(2022)\citenamefont {Roussy},
  \citenamefont {Caldwell}, \citenamefont {Wright}, \citenamefont {Cairncross},
  \citenamefont {Shagam}, \citenamefont {Ng}, \citenamefont {Schlossberger},
  \citenamefont {Park}, \citenamefont {Wang}, \citenamefont {Ye},\ and\
  \citenamefont {Cornell}}]{roussy2022new}%
  \BibitemOpen
  \bibfield  {author} {\bibinfo {author} {\bibfnamefont {T.~S.}\ \bibnamefont
  {Roussy}}, \bibinfo {author} {\bibfnamefont {L.}~\bibnamefont {Caldwell}},
  \bibinfo {author} {\bibfnamefont {T.}~\bibnamefont {Wright}}, \bibinfo
  {author} {\bibfnamefont {W.~B.}\ \bibnamefont {Cairncross}}, \bibinfo
  {author} {\bibfnamefont {Y.}~\bibnamefont {Shagam}}, \bibinfo {author}
  {\bibfnamefont {K.~B.}\ \bibnamefont {Ng}}, \bibinfo {author} {\bibfnamefont
  {N.}~\bibnamefont {Schlossberger}}, \bibinfo {author} {\bibfnamefont {S.~Y.}\
  \bibnamefont {Park}}, \bibinfo {author} {\bibfnamefont {A.}~\bibnamefont
  {Wang}}, \bibinfo {author} {\bibfnamefont {J.}~\bibnamefont {Ye}},\ and\
  \bibinfo {author} {\bibfnamefont {E.~A.}\ \bibnamefont {Cornell}},\
  }\href@noop {} {\bibinfo {title} {A new bound on the electron's electric
  dipole moment}} (\bibinfo {year} {2022}),\ \Eprint
  {https://arxiv.org/abs/2212.11841} {arXiv:2212.11841 [physics.atom-ph]}
  \BibitemShut {NoStop}%
\bibitem [{\citenamefont {Graner}\ \emph {et~al.}(2016)\citenamefont {Graner},
  \citenamefont {Chen}, \citenamefont {Lindahl},\ and\ \citenamefont
  {Heckel}}]{PhysRevLett.116.161601}%
  \BibitemOpen
  \bibfield  {author} {\bibinfo {author} {\bibfnamefont {B.}~\bibnamefont
  {Graner}}, \bibinfo {author} {\bibfnamefont {Y.}~\bibnamefont {Chen}},
  \bibinfo {author} {\bibfnamefont {E.~G.}\ \bibnamefont {Lindahl}},\ and\
  \bibinfo {author} {\bibfnamefont {B.~R.}\ \bibnamefont {Heckel}},\ }\href
  {https://doi.org/10.1103/PhysRevLett.116.161601} {\bibfield  {journal}
  {\bibinfo  {journal} {Phys. Rev. Lett.}\ }\textbf {\bibinfo {volume} {116}},\
  \bibinfo {pages} {161601} (\bibinfo {year} {2016})}\BibitemShut {NoStop}%
\bibitem [{\citenamefont {Abel}\ \emph {et~al.}(2020)\citenamefont {Abel},
  \citenamefont {Afach}, \citenamefont {Ayres}, \citenamefont {Baker},
  \citenamefont {Ban}, \citenamefont {Bison}, \citenamefont {Bodek},
  \citenamefont {Bondar}, \citenamefont {Burghoff}, \citenamefont {Chanel},
  \citenamefont {Chowdhuri}, \citenamefont {Chiu}, \citenamefont {Clement},
  \citenamefont {Crawford}, \citenamefont {Daum}, \citenamefont {Emmenegger},
  \citenamefont {Ferraris-Bouchez}, \citenamefont {Fertl}, \citenamefont
  {Flaux}, \citenamefont {Franke}, \citenamefont {Fratangelo}, \citenamefont
  {Geltenbort}, \citenamefont {Green}, \citenamefont {Griffith}, \citenamefont
  {van~der Grinten}, \citenamefont {Gruji\ifmmode~\acute{c}\else \'{c}\fi{}},
  \citenamefont {Harris}, \citenamefont {Hayen}, \citenamefont {Heil},
  \citenamefont {Henneck}, \citenamefont {H\'elaine}, \citenamefont {Hild},
  \citenamefont {Hodge}, \citenamefont {Horras}, \citenamefont {Iaydjiev},
  \citenamefont {Ivanov}, \citenamefont {Kasprzak}, \citenamefont {Kermaidic},
  \citenamefont {Kirch}, \citenamefont {Knecht}, \citenamefont {Knowles},
  \citenamefont {Koch}, \citenamefont {Koss}, \citenamefont {Komposch},
  \citenamefont {Kozela}, \citenamefont {Kraft}, \citenamefont {Krempel},
  \citenamefont {Ku\ifmmode~\acute{z}\else \'{z}\fi{}niak}, \citenamefont
  {Lauss}, \citenamefont {Lefort}, \citenamefont {Lemi\`ere}, \citenamefont
  {Leredde}, \citenamefont {Mohanmurthy}, \citenamefont {Mtchedlishvili},
  \citenamefont {Musgrave}, \citenamefont {Naviliat-Cuncic}, \citenamefont
  {Pais}, \citenamefont {Piegsa}, \citenamefont {Pierre}, \citenamefont
  {Pignol}, \citenamefont {Plonka-Spehr}, \citenamefont {Prashanth},
  \citenamefont {Qu\'em\'ener}, \citenamefont {Rawlik}, \citenamefont
  {Rebreyend}, \citenamefont {Rien\"acker}, \citenamefont {Ries}, \citenamefont
  {Roccia}, \citenamefont {Rogel}, \citenamefont {Rozpedzik}, \citenamefont
  {Schnabel}, \citenamefont {Schmidt-Wellenburg}, \citenamefont {Severijns},
  \citenamefont {Shiers}, \citenamefont {Tavakoli~Dinani}, \citenamefont
  {Thorne}, \citenamefont {Virot}, \citenamefont {Voigt}, \citenamefont {Weis},
  \citenamefont {Wursten}, \citenamefont {Wyszynski}, \citenamefont {Zejma},
  \citenamefont {Zenner},\ and\ \citenamefont
  {Zsigmond}}]{PhysRevLett.124.081803}%
  \BibitemOpen
  \bibfield  {author} {\bibinfo {author} {\bibfnamefont {C.}~\bibnamefont
  {Abel}}, \bibinfo {author} {\bibfnamefont {S.}~\bibnamefont {Afach}},
  \bibinfo {author} {\bibfnamefont {N.~J.}\ \bibnamefont {Ayres}}, \bibinfo
  {author} {\bibfnamefont {C.~A.}\ \bibnamefont {Baker}}, \bibinfo {author}
  {\bibfnamefont {G.}~\bibnamefont {Ban}}, \bibinfo {author} {\bibfnamefont
  {G.}~\bibnamefont {Bison}}, \bibinfo {author} {\bibfnamefont
  {K.}~\bibnamefont {Bodek}}, \bibinfo {author} {\bibfnamefont
  {V.}~\bibnamefont {Bondar}}, \bibinfo {author} {\bibfnamefont
  {M.}~\bibnamefont {Burghoff}}, \bibinfo {author} {\bibfnamefont
  {E.}~\bibnamefont {Chanel}}, \bibinfo {author} {\bibfnamefont
  {Z.}~\bibnamefont {Chowdhuri}}, \bibinfo {author} {\bibfnamefont {P.-J.}\
  \bibnamefont {Chiu}}, \bibinfo {author} {\bibfnamefont {B.}~\bibnamefont
  {Clement}}, \bibinfo {author} {\bibfnamefont {C.~B.}\ \bibnamefont
  {Crawford}}, \bibinfo {author} {\bibfnamefont {M.}~\bibnamefont {Daum}},
  \bibinfo {author} {\bibfnamefont {S.}~\bibnamefont {Emmenegger}}, \bibinfo
  {author} {\bibfnamefont {L.}~\bibnamefont {Ferraris-Bouchez}}, \bibinfo
  {author} {\bibfnamefont {M.}~\bibnamefont {Fertl}}, \bibinfo {author}
  {\bibfnamefont {P.}~\bibnamefont {Flaux}}, \bibinfo {author} {\bibfnamefont
  {B.}~\bibnamefont {Franke}}, \bibinfo {author} {\bibfnamefont
  {A.}~\bibnamefont {Fratangelo}}, \bibinfo {author} {\bibfnamefont
  {P.}~\bibnamefont {Geltenbort}}, \bibinfo {author} {\bibfnamefont
  {K.}~\bibnamefont {Green}}, \bibinfo {author} {\bibfnamefont {W.~C.}\
  \bibnamefont {Griffith}}, \bibinfo {author} {\bibfnamefont {M.}~\bibnamefont
  {van~der Grinten}}, \bibinfo {author} {\bibfnamefont {Z.~D.}\ \bibnamefont
  {Gruji\ifmmode~\acute{c}\else \'{c}\fi{}}}, \bibinfo {author} {\bibfnamefont
  {P.~G.}\ \bibnamefont {Harris}}, \bibinfo {author} {\bibfnamefont
  {L.}~\bibnamefont {Hayen}}, \bibinfo {author} {\bibfnamefont
  {W.}~\bibnamefont {Heil}}, \bibinfo {author} {\bibfnamefont {R.}~\bibnamefont
  {Henneck}}, \bibinfo {author} {\bibfnamefont {V.}~\bibnamefont {H\'elaine}},
  \bibinfo {author} {\bibfnamefont {N.}~\bibnamefont {Hild}}, \bibinfo {author}
  {\bibfnamefont {Z.}~\bibnamefont {Hodge}}, \bibinfo {author} {\bibfnamefont
  {M.}~\bibnamefont {Horras}}, \bibinfo {author} {\bibfnamefont
  {P.}~\bibnamefont {Iaydjiev}}, \bibinfo {author} {\bibfnamefont {S.~N.}\
  \bibnamefont {Ivanov}}, \bibinfo {author} {\bibfnamefont {M.}~\bibnamefont
  {Kasprzak}}, \bibinfo {author} {\bibfnamefont {Y.}~\bibnamefont {Kermaidic}},
  \bibinfo {author} {\bibfnamefont {K.}~\bibnamefont {Kirch}}, \bibinfo
  {author} {\bibfnamefont {A.}~\bibnamefont {Knecht}}, \bibinfo {author}
  {\bibfnamefont {P.}~\bibnamefont {Knowles}}, \bibinfo {author} {\bibfnamefont
  {H.-C.}\ \bibnamefont {Koch}}, \bibinfo {author} {\bibfnamefont {P.~A.}\
  \bibnamefont {Koss}}, \bibinfo {author} {\bibfnamefont {S.}~\bibnamefont
  {Komposch}}, \bibinfo {author} {\bibfnamefont {A.}~\bibnamefont {Kozela}},
  \bibinfo {author} {\bibfnamefont {A.}~\bibnamefont {Kraft}}, \bibinfo
  {author} {\bibfnamefont {J.}~\bibnamefont {Krempel}}, \bibinfo {author}
  {\bibfnamefont {M.}~\bibnamefont {Ku\ifmmode~\acute{z}\else \'{z}\fi{}niak}},
  \bibinfo {author} {\bibfnamefont {B.}~\bibnamefont {Lauss}}, \bibinfo
  {author} {\bibfnamefont {T.}~\bibnamefont {Lefort}}, \bibinfo {author}
  {\bibfnamefont {Y.}~\bibnamefont {Lemi\`ere}}, \bibinfo {author}
  {\bibfnamefont {A.}~\bibnamefont {Leredde}}, \bibinfo {author} {\bibfnamefont
  {P.}~\bibnamefont {Mohanmurthy}}, \bibinfo {author} {\bibfnamefont
  {A.}~\bibnamefont {Mtchedlishvili}}, \bibinfo {author} {\bibfnamefont
  {M.}~\bibnamefont {Musgrave}}, \bibinfo {author} {\bibfnamefont
  {O.}~\bibnamefont {Naviliat-Cuncic}}, \bibinfo {author} {\bibfnamefont
  {D.}~\bibnamefont {Pais}}, \bibinfo {author} {\bibfnamefont {F.~M.}\
  \bibnamefont {Piegsa}}, \bibinfo {author} {\bibfnamefont {E.}~\bibnamefont
  {Pierre}}, \bibinfo {author} {\bibfnamefont {G.}~\bibnamefont {Pignol}},
  \bibinfo {author} {\bibfnamefont {C.}~\bibnamefont {Plonka-Spehr}}, \bibinfo
  {author} {\bibfnamefont {P.~N.}\ \bibnamefont {Prashanth}}, \bibinfo {author}
  {\bibfnamefont {G.}~\bibnamefont {Qu\'em\'ener}}, \bibinfo {author}
  {\bibfnamefont {M.}~\bibnamefont {Rawlik}}, \bibinfo {author} {\bibfnamefont
  {D.}~\bibnamefont {Rebreyend}}, \bibinfo {author} {\bibfnamefont
  {I.}~\bibnamefont {Rien\"acker}}, \bibinfo {author} {\bibfnamefont
  {D.}~\bibnamefont {Ries}}, \bibinfo {author} {\bibfnamefont {S.}~\bibnamefont
  {Roccia}}, \bibinfo {author} {\bibfnamefont {G.}~\bibnamefont {Rogel}},
  \bibinfo {author} {\bibfnamefont {D.}~\bibnamefont {Rozpedzik}}, \bibinfo
  {author} {\bibfnamefont {A.}~\bibnamefont {Schnabel}}, \bibinfo {author}
  {\bibfnamefont {P.}~\bibnamefont {Schmidt-Wellenburg}}, \bibinfo {author}
  {\bibfnamefont {N.}~\bibnamefont {Severijns}}, \bibinfo {author}
  {\bibfnamefont {D.}~\bibnamefont {Shiers}}, \bibinfo {author} {\bibfnamefont
  {R.}~\bibnamefont {Tavakoli~Dinani}}, \bibinfo {author} {\bibfnamefont
  {J.~A.}\ \bibnamefont {Thorne}}, \bibinfo {author} {\bibfnamefont
  {R.}~\bibnamefont {Virot}}, \bibinfo {author} {\bibfnamefont
  {J.}~\bibnamefont {Voigt}}, \bibinfo {author} {\bibfnamefont
  {A.}~\bibnamefont {Weis}}, \bibinfo {author} {\bibfnamefont {E.}~\bibnamefont
  {Wursten}}, \bibinfo {author} {\bibfnamefont {G.}~\bibnamefont {Wyszynski}},
  \bibinfo {author} {\bibfnamefont {J.}~\bibnamefont {Zejma}}, \bibinfo
  {author} {\bibfnamefont {J.}~\bibnamefont {Zenner}},\ and\ \bibinfo {author}
  {\bibfnamefont {G.}~\bibnamefont {Zsigmond}},\ }\href
  {https://doi.org/10.1103/PhysRevLett.124.081803} {\bibfield  {journal}
  {\bibinfo  {journal} {Phys. Rev. Lett.}\ }\textbf {\bibinfo {volume} {124}},\
  \bibinfo {pages} {081803} (\bibinfo {year} {2020})}\BibitemShut {NoStop}%
\bibitem [{\citenamefont {Bennett}\ \emph {et~al.}(2009)\citenamefont
  {Bennett}, \citenamefont {Bousquet}, \citenamefont {Brown}, \citenamefont
  {Bunce}, \citenamefont {Carey}, \citenamefont {Cushman}, \citenamefont
  {Danby}, \citenamefont {Debevec}, \citenamefont {Deile}, \citenamefont
  {Deng}, \citenamefont {Deninger}, \citenamefont {Dhawan}, \citenamefont
  {Druzhinin}, \citenamefont {Duong}, \citenamefont {Efstathiadis},
  \citenamefont {Farley}, \citenamefont {Fedotovich}, \citenamefont {Giron},
  \citenamefont {Gray}, \citenamefont {Grigoriev}, \citenamefont
  {Grosse-Perdekamp}, \citenamefont {Grossmann}, \citenamefont {Hare},
  \citenamefont {Hertzog}, \citenamefont {Huang}, \citenamefont {Hughes},
  \citenamefont {Iwasaki}, \citenamefont {Jungmann}, \citenamefont {Kawall},
  \citenamefont {Kawamura}, \citenamefont {Khazin}, \citenamefont {Kindem},
  \citenamefont {Krienen}, \citenamefont {Kronkvist}, \citenamefont {Lam},
  \citenamefont {Larsen}, \citenamefont {Lee}, \citenamefont {Logashenko},
  \citenamefont {McNabb}, \citenamefont {Meng}, \citenamefont {Mi},
  \citenamefont {Miller}, \citenamefont {Mizumachi}, \citenamefont {Morse},
  \citenamefont {Nikas}, \citenamefont {Onderwater}, \citenamefont {Orlov},
  \citenamefont {\"Ozben}, \citenamefont {Paley}, \citenamefont {Peng},
  \citenamefont {Polly}, \citenamefont {Pretz}, \citenamefont {Prigl},
  \citenamefont {zu~Putlitz}, \citenamefont {Qian}, \citenamefont {Redin},
  \citenamefont {Rind}, \citenamefont {Roberts}, \citenamefont {Ryskulov},
  \citenamefont {Sedykh}, \citenamefont {Semertzidis}, \citenamefont {Shagin},
  \citenamefont {Shatunov}, \citenamefont {Sichtermann}, \citenamefont
  {Solodov}, \citenamefont {Sossong}, \citenamefont {Steinmetz}, \citenamefont
  {Sulak}, \citenamefont {Timmermans}, \citenamefont {Trofimov}, \citenamefont
  {Urner}, \citenamefont {von Walter}, \citenamefont {Warburton}, \citenamefont
  {Winn}, \citenamefont {Yamamoto},\ and\ \citenamefont
  {Zimmerman}}]{PhysRevD.80.052008}%
  \BibitemOpen
  \bibfield  {author} {\bibinfo {author} {\bibfnamefont {G.~W.}\ \bibnamefont
  {Bennett}}, \bibinfo {author} {\bibfnamefont {B.}~\bibnamefont {Bousquet}},
  \bibinfo {author} {\bibfnamefont {H.~N.}\ \bibnamefont {Brown}}, \bibinfo
  {author} {\bibfnamefont {G.}~\bibnamefont {Bunce}}, \bibinfo {author}
  {\bibfnamefont {R.~M.}\ \bibnamefont {Carey}}, \bibinfo {author}
  {\bibfnamefont {P.}~\bibnamefont {Cushman}}, \bibinfo {author} {\bibfnamefont
  {G.~T.}\ \bibnamefont {Danby}}, \bibinfo {author} {\bibfnamefont {P.~T.}\
  \bibnamefont {Debevec}}, \bibinfo {author} {\bibfnamefont {M.}~\bibnamefont
  {Deile}}, \bibinfo {author} {\bibfnamefont {H.}~\bibnamefont {Deng}},
  \bibinfo {author} {\bibfnamefont {W.}~\bibnamefont {Deninger}}, \bibinfo
  {author} {\bibfnamefont {S.~K.}\ \bibnamefont {Dhawan}}, \bibinfo {author}
  {\bibfnamefont {V.~P.}\ \bibnamefont {Druzhinin}}, \bibinfo {author}
  {\bibfnamefont {L.}~\bibnamefont {Duong}}, \bibinfo {author} {\bibfnamefont
  {E.}~\bibnamefont {Efstathiadis}}, \bibinfo {author} {\bibfnamefont
  {F.~J.~M.}\ \bibnamefont {Farley}}, \bibinfo {author} {\bibfnamefont {G.~V.}\
  \bibnamefont {Fedotovich}}, \bibinfo {author} {\bibfnamefont
  {S.}~\bibnamefont {Giron}}, \bibinfo {author} {\bibfnamefont {F.~E.}\
  \bibnamefont {Gray}}, \bibinfo {author} {\bibfnamefont {D.}~\bibnamefont
  {Grigoriev}}, \bibinfo {author} {\bibfnamefont {M.}~\bibnamefont
  {Grosse-Perdekamp}}, \bibinfo {author} {\bibfnamefont {A.}~\bibnamefont
  {Grossmann}}, \bibinfo {author} {\bibfnamefont {M.~F.}\ \bibnamefont {Hare}},
  \bibinfo {author} {\bibfnamefont {D.~W.}\ \bibnamefont {Hertzog}}, \bibinfo
  {author} {\bibfnamefont {X.}~\bibnamefont {Huang}}, \bibinfo {author}
  {\bibfnamefont {V.~W.}\ \bibnamefont {Hughes}}, \bibinfo {author}
  {\bibfnamefont {M.}~\bibnamefont {Iwasaki}}, \bibinfo {author} {\bibfnamefont
  {K.}~\bibnamefont {Jungmann}}, \bibinfo {author} {\bibfnamefont
  {D.}~\bibnamefont {Kawall}}, \bibinfo {author} {\bibfnamefont
  {M.}~\bibnamefont {Kawamura}}, \bibinfo {author} {\bibfnamefont {B.~I.}\
  \bibnamefont {Khazin}}, \bibinfo {author} {\bibfnamefont {J.}~\bibnamefont
  {Kindem}}, \bibinfo {author} {\bibfnamefont {F.}~\bibnamefont {Krienen}},
  \bibinfo {author} {\bibfnamefont {I.}~\bibnamefont {Kronkvist}}, \bibinfo
  {author} {\bibfnamefont {A.}~\bibnamefont {Lam}}, \bibinfo {author}
  {\bibfnamefont {R.}~\bibnamefont {Larsen}}, \bibinfo {author} {\bibfnamefont
  {Y.~Y.}\ \bibnamefont {Lee}}, \bibinfo {author} {\bibfnamefont
  {I.}~\bibnamefont {Logashenko}}, \bibinfo {author} {\bibfnamefont
  {R.}~\bibnamefont {McNabb}}, \bibinfo {author} {\bibfnamefont
  {W.}~\bibnamefont {Meng}}, \bibinfo {author} {\bibfnamefont {J.}~\bibnamefont
  {Mi}}, \bibinfo {author} {\bibfnamefont {J.~P.}\ \bibnamefont {Miller}},
  \bibinfo {author} {\bibfnamefont {Y.}~\bibnamefont {Mizumachi}}, \bibinfo
  {author} {\bibfnamefont {W.~M.}\ \bibnamefont {Morse}}, \bibinfo {author}
  {\bibfnamefont {D.}~\bibnamefont {Nikas}}, \bibinfo {author} {\bibfnamefont
  {C.~J.~G.}\ \bibnamefont {Onderwater}}, \bibinfo {author} {\bibfnamefont
  {Y.}~\bibnamefont {Orlov}}, \bibinfo {author} {\bibfnamefont {C.~S.}\
  \bibnamefont {\"Ozben}}, \bibinfo {author} {\bibfnamefont {J.~M.}\
  \bibnamefont {Paley}}, \bibinfo {author} {\bibfnamefont {Q.}~\bibnamefont
  {Peng}}, \bibinfo {author} {\bibfnamefont {C.~C.}\ \bibnamefont {Polly}},
  \bibinfo {author} {\bibfnamefont {J.}~\bibnamefont {Pretz}}, \bibinfo
  {author} {\bibfnamefont {R.}~\bibnamefont {Prigl}}, \bibinfo {author}
  {\bibfnamefont {G.}~\bibnamefont {zu~Putlitz}}, \bibinfo {author}
  {\bibfnamefont {T.}~\bibnamefont {Qian}}, \bibinfo {author} {\bibfnamefont
  {S.~I.}\ \bibnamefont {Redin}}, \bibinfo {author} {\bibfnamefont
  {O.}~\bibnamefont {Rind}}, \bibinfo {author} {\bibfnamefont {B.~L.}\
  \bibnamefont {Roberts}}, \bibinfo {author} {\bibfnamefont {N.}~\bibnamefont
  {Ryskulov}}, \bibinfo {author} {\bibfnamefont {S.}~\bibnamefont {Sedykh}},
  \bibinfo {author} {\bibfnamefont {Y.~K.}\ \bibnamefont {Semertzidis}},
  \bibinfo {author} {\bibfnamefont {P.}~\bibnamefont {Shagin}}, \bibinfo
  {author} {\bibfnamefont {Y.~M.}\ \bibnamefont {Shatunov}}, \bibinfo {author}
  {\bibfnamefont {E.~P.}\ \bibnamefont {Sichtermann}}, \bibinfo {author}
  {\bibfnamefont {E.}~\bibnamefont {Solodov}}, \bibinfo {author} {\bibfnamefont
  {M.}~\bibnamefont {Sossong}}, \bibinfo {author} {\bibfnamefont
  {A.}~\bibnamefont {Steinmetz}}, \bibinfo {author} {\bibfnamefont {L.~R.}\
  \bibnamefont {Sulak}}, \bibinfo {author} {\bibfnamefont {C.}~\bibnamefont
  {Timmermans}}, \bibinfo {author} {\bibfnamefont {A.}~\bibnamefont
  {Trofimov}}, \bibinfo {author} {\bibfnamefont {D.}~\bibnamefont {Urner}},
  \bibinfo {author} {\bibfnamefont {P.}~\bibnamefont {von Walter}}, \bibinfo
  {author} {\bibfnamefont {D.}~\bibnamefont {Warburton}}, \bibinfo {author}
  {\bibfnamefont {D.}~\bibnamefont {Winn}}, \bibinfo {author} {\bibfnamefont
  {A.}~\bibnamefont {Yamamoto}},\ and\ \bibinfo {author} {\bibfnamefont
  {D.}~\bibnamefont {Zimmerman}} (\bibinfo {collaboration} {Muon ($g-2$)
  Collaboration}),\ }\href {https://doi.org/10.1103/PhysRevD.80.052008}
  {\bibfield  {journal} {\bibinfo  {journal} {Phys. Rev. D}\ }\textbf {\bibinfo
  {volume} {80}},\ \bibinfo {pages} {052008} (\bibinfo {year}
  {2009})}\BibitemShut {NoStop}%
\bibitem [{\citenamefont {Fukuyama}\ and\ \citenamefont
  {Silenko}(2013)}]{Fukuyama2013}%
  \BibitemOpen
  \bibfield  {author} {\bibinfo {author} {\bibfnamefont {T.}~\bibnamefont
  {Fukuyama}}\ and\ \bibinfo {author} {\bibfnamefont {A.}~\bibnamefont
  {Silenko}},\ }\href {https://doi.org/10.1142/S0217751X13501479} {\bibfield
  {journal} {\bibinfo  {journal} {International Journal of Modern Physics A}\
  }\textbf {\bibinfo {volume} {28}},\ \bibinfo {pages} {1350147} (\bibinfo
  {year} {2013})}\BibitemShut {NoStop}%
\bibitem [{\citenamefont {Thomas}(1927)}]{doi:10.1080/14786440108564170}%
  \BibitemOpen
  \bibfield  {author} {\bibinfo {author} {\bibfnamefont {L.~H.}\ \bibnamefont
  {Thomas}},\ }\href {https://doi.org/10.1080/14786440108564170} {\bibfield
  {journal} {\bibinfo  {journal} {The London, Edinburgh, and Dublin
  Philosophical Magazine and Journal of Science}\ }\textbf {\bibinfo {volume}
  {3}},\ \bibinfo {pages} {1} (\bibinfo {year} {1927})}\BibitemShut {NoStop}%
\bibitem [{\citenamefont {Abusaif}\ \emph {et~al.}(2019)\citenamefont
  {Abusaif}, \citenamefont {Aggarwal}, \citenamefont {Aksentev}, \citenamefont
  {Alberdi-Esuain}, \citenamefont {Atanasov}, \citenamefont {Barion},
  \citenamefont {Basile}, \citenamefont {Berz}, \citenamefont {Bey\ss},
  \citenamefont {B\"ohme}, \citenamefont {B\"oker}, \citenamefont {Borburgh},
  \citenamefont {Carli}, \citenamefont {Ciepa\l{}}, \citenamefont {Ciullo},
  \citenamefont {Contalbrigo}, \citenamefont {Conto}, \citenamefont {Dymov},
  \citenamefont {Felden}, \citenamefont {Gagoshidze}, \citenamefont {Gaisser},
  \citenamefont {Gebel}, \citenamefont {Giese}, \citenamefont {Grigoryev},
  \citenamefont {Grzonka}, \citenamefont {Tahar}, \citenamefont {Hahnraths},
  \citenamefont {Heberling}, \citenamefont {Hejny}, \citenamefont {Hetzel},
  \citenamefont {H\"olscher}, \citenamefont {Javakhishvili}, \citenamefont
  {Jorat}, \citenamefont {Kacharava}, \citenamefont {Kamerdzhiev},
  \citenamefont {Karanth}, \citenamefont {K\"aseberg}, \citenamefont
  {Keshelashvili}, \citenamefont {Koop}, \citenamefont {Kulikov}, \citenamefont
  {Laihem}, \citenamefont {Lamont}, \citenamefont {Lehrach}, \citenamefont
  {Lenisa}, \citenamefont {Lomidze}, \citenamefont {Lorentz}, \citenamefont
  {Macharashvili}, \citenamefont {Magiera}, \citenamefont {Makino},
  \citenamefont {Martin}, \citenamefont {Mchedlishvili}, \citenamefont
  {Mei{\ss}ner}, \citenamefont {Metreveli}, \citenamefont {Michaud},
  \citenamefont {M\"uller}, \citenamefont {Nass}, \citenamefont {Natour},
  \citenamefont {Nikolaev}, \citenamefont {Nogga}, \citenamefont {Pesce},
  \citenamefont {Poncza}, \citenamefont {Prasuhn}, \citenamefont {Pretz},
  \citenamefont {Rathmann}, \citenamefont {Ritman}, \citenamefont {Rosenthal},
  \citenamefont {Saleev}, \citenamefont {Schott}, \citenamefont {Sefzick},
  \citenamefont {Senichev}, \citenamefont {Shergelashvili}, \citenamefont
  {Shmakova}, \citenamefont {Siddique}, \citenamefont {Silenko}, \citenamefont
  {Simon}, \citenamefont {Slim}, \citenamefont {Soltner}, \citenamefont
  {Stahl}, \citenamefont {Stassen}, \citenamefont {Stephenson}, \citenamefont
  {Straatmann}, \citenamefont {Str\"oher}, \citenamefont {Tabidze},
  \citenamefont {Tagliente}, \citenamefont {Talman}, \citenamefont {Uzikov},
  \citenamefont {Valdau}, \citenamefont {Valetov}, \citenamefont {Wagner},
  \citenamefont {Weidemann}, \citenamefont {Wirzba}, \citenamefont
  {Wro\'{n}ska}, \citenamefont {W\"ustner}, \citenamefont {Zupranski},\ and\
  \citenamefont {\.Zurek}}]{abusaif2019storage}%
  \BibitemOpen
  \bibfield  {author} {\bibinfo {author} {\bibfnamefont {F.}~\bibnamefont
  {Abusaif}}, \bibinfo {author} {\bibfnamefont {A.}~\bibnamefont {Aggarwal}},
  \bibinfo {author} {\bibfnamefont {A.}~\bibnamefont {Aksentev}}, \bibinfo
  {author} {\bibfnamefont {B.}~\bibnamefont {Alberdi-Esuain}}, \bibinfo
  {author} {\bibfnamefont {A.}~\bibnamefont {Atanasov}}, \bibinfo {author}
  {\bibfnamefont {L.}~\bibnamefont {Barion}}, \bibinfo {author} {\bibfnamefont
  {S.}~\bibnamefont {Basile}}, \bibinfo {author} {\bibfnamefont
  {M.}~\bibnamefont {Berz}}, \bibinfo {author} {\bibfnamefont {M.}~\bibnamefont
  {Bey\ss}}, \bibinfo {author} {\bibfnamefont {C.}~\bibnamefont {B\"ohme}},
  \bibinfo {author} {\bibfnamefont {J.}~\bibnamefont {B\"oker}}, \bibinfo
  {author} {\bibfnamefont {J.}~\bibnamefont {Borburgh}}, \bibinfo {author}
  {\bibfnamefont {C.}~\bibnamefont {Carli}}, \bibinfo {author} {\bibfnamefont
  {I.}~\bibnamefont {Ciepa\l{}}}, \bibinfo {author} {\bibfnamefont
  {G.}~\bibnamefont {Ciullo}}, \bibinfo {author} {\bibfnamefont
  {M.}~\bibnamefont {Contalbrigo}}, \bibinfo {author} {\bibfnamefont
  {J.~M.~D.}\ \bibnamefont {Conto}}, \bibinfo {author} {\bibfnamefont
  {S.}~\bibnamefont {Dymov}}, \bibinfo {author} {\bibfnamefont
  {O.}~\bibnamefont {Felden}}, \bibinfo {author} {\bibfnamefont
  {M.}~\bibnamefont {Gagoshidze}}, \bibinfo {author} {\bibfnamefont
  {M.}~\bibnamefont {Gaisser}}, \bibinfo {author} {\bibfnamefont
  {R.}~\bibnamefont {Gebel}}, \bibinfo {author} {\bibfnamefont
  {N.}~\bibnamefont {Giese}}, \bibinfo {author} {\bibfnamefont
  {K.}~\bibnamefont {Grigoryev}}, \bibinfo {author} {\bibfnamefont
  {D.}~\bibnamefont {Grzonka}}, \bibinfo {author} {\bibfnamefont {M.~H.}\
  \bibnamefont {Tahar}}, \bibinfo {author} {\bibfnamefont {T.}~\bibnamefont
  {Hahnraths}}, \bibinfo {author} {\bibfnamefont {D.}~\bibnamefont
  {Heberling}}, \bibinfo {author} {\bibfnamefont {V.}~\bibnamefont {Hejny}},
  \bibinfo {author} {\bibfnamefont {J.}~\bibnamefont {Hetzel}}, \bibinfo
  {author} {\bibfnamefont {D.}~\bibnamefont {H\"olscher}}, \bibinfo {author}
  {\bibfnamefont {O.}~\bibnamefont {Javakhishvili}}, \bibinfo {author}
  {\bibfnamefont {L.}~\bibnamefont {Jorat}}, \bibinfo {author} {\bibfnamefont
  {A.}~\bibnamefont {Kacharava}}, \bibinfo {author} {\bibfnamefont
  {V.}~\bibnamefont {Kamerdzhiev}}, \bibinfo {author} {\bibfnamefont
  {S.}~\bibnamefont {Karanth}}, \bibinfo {author} {\bibfnamefont
  {C.}~\bibnamefont {K\"aseberg}}, \bibinfo {author} {\bibfnamefont
  {I.}~\bibnamefont {Keshelashvili}}, \bibinfo {author} {\bibfnamefont
  {I.}~\bibnamefont {Koop}}, \bibinfo {author} {\bibfnamefont {A.}~\bibnamefont
  {Kulikov}}, \bibinfo {author} {\bibfnamefont {K.}~\bibnamefont {Laihem}},
  \bibinfo {author} {\bibfnamefont {M.}~\bibnamefont {Lamont}}, \bibinfo
  {author} {\bibfnamefont {A.}~\bibnamefont {Lehrach}}, \bibinfo {author}
  {\bibfnamefont {P.}~\bibnamefont {Lenisa}}, \bibinfo {author} {\bibfnamefont
  {N.}~\bibnamefont {Lomidze}}, \bibinfo {author} {\bibfnamefont
  {B.}~\bibnamefont {Lorentz}}, \bibinfo {author} {\bibfnamefont
  {G.}~\bibnamefont {Macharashvili}}, \bibinfo {author} {\bibfnamefont
  {A.}~\bibnamefont {Magiera}}, \bibinfo {author} {\bibfnamefont
  {K.}~\bibnamefont {Makino}}, \bibinfo {author} {\bibfnamefont
  {S.}~\bibnamefont {Martin}}, \bibinfo {author} {\bibfnamefont
  {D.}~\bibnamefont {Mchedlishvili}}, \bibinfo {author} {\bibfnamefont {U.~G.}\
  \bibnamefont {Mei{\ss}ner}}, \bibinfo {author} {\bibfnamefont
  {Z.}~\bibnamefont {Metreveli}}, \bibinfo {author} {\bibfnamefont
  {J.}~\bibnamefont {Michaud}}, \bibinfo {author} {\bibfnamefont
  {F.}~\bibnamefont {M\"uller}}, \bibinfo {author} {\bibfnamefont
  {A.}~\bibnamefont {Nass}}, \bibinfo {author} {\bibfnamefont {G.}~\bibnamefont
  {Natour}}, \bibinfo {author} {\bibfnamefont {N.}~\bibnamefont {Nikolaev}},
  \bibinfo {author} {\bibfnamefont {A.}~\bibnamefont {Nogga}}, \bibinfo
  {author} {\bibfnamefont {A.}~\bibnamefont {Pesce}}, \bibinfo {author}
  {\bibfnamefont {V.}~\bibnamefont {Poncza}}, \bibinfo {author} {\bibfnamefont
  {D.}~\bibnamefont {Prasuhn}}, \bibinfo {author} {\bibfnamefont
  {J.}~\bibnamefont {Pretz}}, \bibinfo {author} {\bibfnamefont
  {F.}~\bibnamefont {Rathmann}}, \bibinfo {author} {\bibfnamefont
  {J.}~\bibnamefont {Ritman}}, \bibinfo {author} {\bibfnamefont
  {M.}~\bibnamefont {Rosenthal}}, \bibinfo {author} {\bibfnamefont
  {A.}~\bibnamefont {Saleev}}, \bibinfo {author} {\bibfnamefont
  {M.}~\bibnamefont {Schott}}, \bibinfo {author} {\bibfnamefont
  {T.}~\bibnamefont {Sefzick}}, \bibinfo {author} {\bibfnamefont
  {Y.}~\bibnamefont {Senichev}}, \bibinfo {author} {\bibfnamefont
  {D.}~\bibnamefont {Shergelashvili}}, \bibinfo {author} {\bibfnamefont
  {V.}~\bibnamefont {Shmakova}}, \bibinfo {author} {\bibfnamefont
  {S.}~\bibnamefont {Siddique}}, \bibinfo {author} {\bibfnamefont
  {A.}~\bibnamefont {Silenko}}, \bibinfo {author} {\bibfnamefont
  {M.}~\bibnamefont {Simon}}, \bibinfo {author} {\bibfnamefont
  {J.}~\bibnamefont {Slim}}, \bibinfo {author} {\bibfnamefont {H.}~\bibnamefont
  {Soltner}}, \bibinfo {author} {\bibfnamefont {A.}~\bibnamefont {Stahl}},
  \bibinfo {author} {\bibfnamefont {R.}~\bibnamefont {Stassen}}, \bibinfo
  {author} {\bibfnamefont {E.}~\bibnamefont {Stephenson}}, \bibinfo {author}
  {\bibfnamefont {H.}~\bibnamefont {Straatmann}}, \bibinfo {author}
  {\bibfnamefont {H.}~\bibnamefont {Str\"oher}}, \bibinfo {author}
  {\bibfnamefont {M.}~\bibnamefont {Tabidze}}, \bibinfo {author} {\bibfnamefont
  {G.}~\bibnamefont {Tagliente}}, \bibinfo {author} {\bibfnamefont
  {R.}~\bibnamefont {Talman}}, \bibinfo {author} {\bibfnamefont
  {Y.}~\bibnamefont {Uzikov}}, \bibinfo {author} {\bibfnamefont
  {Y.}~\bibnamefont {Valdau}}, \bibinfo {author} {\bibfnamefont
  {E.}~\bibnamefont {Valetov}}, \bibinfo {author} {\bibfnamefont
  {T.}~\bibnamefont {Wagner}}, \bibinfo {author} {\bibfnamefont
  {C.}~\bibnamefont {Weidemann}}, \bibinfo {author} {\bibfnamefont
  {A.}~\bibnamefont {Wirzba}}, \bibinfo {author} {\bibfnamefont
  {A.}~\bibnamefont {Wro\'{n}ska}}, \bibinfo {author} {\bibfnamefont
  {P.}~\bibnamefont {W\"ustner}}, \bibinfo {author} {\bibfnamefont
  {P.}~\bibnamefont {Zupranski}},\ and\ \bibinfo {author} {\bibfnamefont
  {M.}~\bibnamefont {\.Zurek}},\ }\href@noop {} {\bibinfo {title} {{Storage
  Ring to Search for Electric Dipole Moments of Charged Particles --
  Feasibility Study}}} (\bibinfo {year} {2019}),\ \Eprint
  {https://arxiv.org/abs/1912.07881} {arXiv:1912.07881 [hep-ex]} \BibitemShut
  {NoStop}%
\bibitem [{\citenamefont {Anastassopoulos}\ \emph {et~al.}(2011)\citenamefont
  {Anastassopoulos} \emph {et~al.}}]{1865072}%
  \BibitemOpen
  \bibfield  {author} {\bibinfo {author} {\bibfnamefont {V.}~\bibnamefont
  {Anastassopoulos}} \emph {et~al.} (\bibinfo {collaboration} {Storage Ring EDM
  Collaboration}),\ }\href {http://www.bnl.gov/edm} {\bibinfo {title} {A
  proposal to measure the proton electric dipole moment with $10^{-29}$ $e
  \cdot \text{cm}$ sensitivity}} (\bibinfo {year} {2011})\BibitemShut {NoStop}%
\bibitem [{\citenamefont {Anastassopoulos}\ \emph {et~al.}(2016)\citenamefont
  {Anastassopoulos}, \citenamefont {Andrianov}, \citenamefont {Baartman},
  \citenamefont {Baessler}, \citenamefont {Bai}, \citenamefont {Benante},
  \citenamefont {Berz}, \citenamefont {Blaskiewicz}, \citenamefont {Bowcock},
  \citenamefont {Brown}, \citenamefont {Casey}, \citenamefont {Conte},
  \citenamefont {Crnkovic}, \citenamefont {D'Imperio}, \citenamefont
  {Fanourakis}, \citenamefont {Fedotov}, \citenamefont {Fierlinger},
  \citenamefont {Fischer}, \citenamefont {Gaisser}, \citenamefont {Giomataris},
  \citenamefont {Grosse-Perdekamp}, \citenamefont {Guidoboni}, \citenamefont
  {Hac\ifmmode \imath \else \i \fi{}\"omero\ifmmode~\breve{g}\else
  \u{g}\fi{}lu}, \citenamefont {Hoffstaetter}, \citenamefont {Huang},
  \citenamefont {Incagli}, \citenamefont {Ivanov}, \citenamefont {Kawall},
  \citenamefont {Kim}, \citenamefont {King}, \citenamefont {Koop},
  \citenamefont {Lazarus}, \citenamefont {Lebedev}, \citenamefont {Lee},
  \citenamefont {Lee}, \citenamefont {Lee}, \citenamefont {Lehrach},
  \citenamefont {Lenisa}, \citenamefont {Levi~Sandri}, \citenamefont {Luccio},
  \citenamefont {Lyapin}, \citenamefont {MacKay}, \citenamefont {Maier},
  \citenamefont {Makino}, \citenamefont {Malitsky}, \citenamefont {Marciano},
  \citenamefont {Meng}, \citenamefont {Meot}, \citenamefont {Metodiev},
  \citenamefont {Miceli}, \citenamefont {Moricciani}, \citenamefont {Morse},
  \citenamefont {Nagaitsev}, \citenamefont {Nayak}, \citenamefont {Orlov},
  \citenamefont {Ozben}, \citenamefont {Park}, \citenamefont {Pesce},
  \citenamefont {Petrakou}, \citenamefont {Pile}, \citenamefont {Podobedov},
  \citenamefont {Polychronakos}, \citenamefont {Pretz}, \citenamefont
  {Ptitsyn}, \citenamefont {Ramberg}, \citenamefont {Raparia}, \citenamefont
  {Rathmann}, \citenamefont {Rescia}, \citenamefont {Roser}, \citenamefont
  {Kamal~Sayed}, \citenamefont {Semertzidis}, \citenamefont {Senichev},
  \citenamefont {Sidorin}, \citenamefont {Silenko}, \citenamefont {Simos},
  \citenamefont {Stahl}, \citenamefont {Stephenson}, \citenamefont {Str\"oher},
  \citenamefont {Syphers}, \citenamefont {Talman}, \citenamefont {Talman},
  \citenamefont {Tishchenko}, \citenamefont {Touramanis}, \citenamefont
  {Tsoupas}, \citenamefont {Venanzoni}, \citenamefont {Vetter}, \citenamefont
  {Vlassis}, \citenamefont {Won}, \citenamefont {Zavattini}, \citenamefont
  {Zelenski},\ and\ \citenamefont {Zioutas}}]{doi:10.1063/1.4967465}%
  \BibitemOpen
  \bibfield  {author} {\bibinfo {author} {\bibfnamefont {V.}~\bibnamefont
  {Anastassopoulos}}, \bibinfo {author} {\bibfnamefont {S.}~\bibnamefont
  {Andrianov}}, \bibinfo {author} {\bibfnamefont {R.}~\bibnamefont {Baartman}},
  \bibinfo {author} {\bibfnamefont {S.}~\bibnamefont {Baessler}}, \bibinfo
  {author} {\bibfnamefont {M.}~\bibnamefont {Bai}}, \bibinfo {author}
  {\bibfnamefont {J.}~\bibnamefont {Benante}}, \bibinfo {author} {\bibfnamefont
  {M.}~\bibnamefont {Berz}}, \bibinfo {author} {\bibfnamefont {M.}~\bibnamefont
  {Blaskiewicz}}, \bibinfo {author} {\bibfnamefont {T.}~\bibnamefont
  {Bowcock}}, \bibinfo {author} {\bibfnamefont {K.}~\bibnamefont {Brown}},
  \bibinfo {author} {\bibfnamefont {B.}~\bibnamefont {Casey}}, \bibinfo
  {author} {\bibfnamefont {M.}~\bibnamefont {Conte}}, \bibinfo {author}
  {\bibfnamefont {J.~D.}\ \bibnamefont {Crnkovic}}, \bibinfo {author}
  {\bibfnamefont {N.}~\bibnamefont {D'Imperio}}, \bibinfo {author}
  {\bibfnamefont {G.}~\bibnamefont {Fanourakis}}, \bibinfo {author}
  {\bibfnamefont {A.}~\bibnamefont {Fedotov}}, \bibinfo {author} {\bibfnamefont
  {P.}~\bibnamefont {Fierlinger}}, \bibinfo {author} {\bibfnamefont
  {W.}~\bibnamefont {Fischer}}, \bibinfo {author} {\bibfnamefont {M.~O.}\
  \bibnamefont {Gaisser}}, \bibinfo {author} {\bibfnamefont {Y.}~\bibnamefont
  {Giomataris}}, \bibinfo {author} {\bibfnamefont {M.}~\bibnamefont
  {Grosse-Perdekamp}}, \bibinfo {author} {\bibfnamefont {G.}~\bibnamefont
  {Guidoboni}}, \bibinfo {author} {\bibfnamefont {S.}~\bibnamefont {Hac\ifmmode
  \imath \else \i \fi{}\"omero\ifmmode~\breve{g}\else \u{g}\fi{}lu}}, \bibinfo
  {author} {\bibfnamefont {G.}~\bibnamefont {Hoffstaetter}}, \bibinfo {author}
  {\bibfnamefont {H.}~\bibnamefont {Huang}}, \bibinfo {author} {\bibfnamefont
  {M.}~\bibnamefont {Incagli}}, \bibinfo {author} {\bibfnamefont
  {A.}~\bibnamefont {Ivanov}}, \bibinfo {author} {\bibfnamefont
  {D.}~\bibnamefont {Kawall}}, \bibinfo {author} {\bibfnamefont {Y.~I.}\
  \bibnamefont {Kim}}, \bibinfo {author} {\bibfnamefont {B.}~\bibnamefont
  {King}}, \bibinfo {author} {\bibfnamefont {I.~A.}\ \bibnamefont {Koop}},
  \bibinfo {author} {\bibfnamefont {D.~M.}\ \bibnamefont {Lazarus}}, \bibinfo
  {author} {\bibfnamefont {V.}~\bibnamefont {Lebedev}}, \bibinfo {author}
  {\bibfnamefont {M.~J.}\ \bibnamefont {Lee}}, \bibinfo {author} {\bibfnamefont
  {S.}~\bibnamefont {Lee}}, \bibinfo {author} {\bibfnamefont {Y.~H.}\
  \bibnamefont {Lee}}, \bibinfo {author} {\bibfnamefont {A.}~\bibnamefont
  {Lehrach}}, \bibinfo {author} {\bibfnamefont {P.}~\bibnamefont {Lenisa}},
  \bibinfo {author} {\bibfnamefont {P.}~\bibnamefont {Levi~Sandri}}, \bibinfo
  {author} {\bibfnamefont {A.~U.}\ \bibnamefont {Luccio}}, \bibinfo {author}
  {\bibfnamefont {A.}~\bibnamefont {Lyapin}}, \bibinfo {author} {\bibfnamefont
  {W.}~\bibnamefont {MacKay}}, \bibinfo {author} {\bibfnamefont
  {R.}~\bibnamefont {Maier}}, \bibinfo {author} {\bibfnamefont
  {K.}~\bibnamefont {Makino}}, \bibinfo {author} {\bibfnamefont
  {N.}~\bibnamefont {Malitsky}}, \bibinfo {author} {\bibfnamefont {W.~J.}\
  \bibnamefont {Marciano}}, \bibinfo {author} {\bibfnamefont {W.}~\bibnamefont
  {Meng}}, \bibinfo {author} {\bibfnamefont {F.}~\bibnamefont {Meot}}, \bibinfo
  {author} {\bibfnamefont {E.~M.}\ \bibnamefont {Metodiev}}, \bibinfo {author}
  {\bibfnamefont {L.}~\bibnamefont {Miceli}}, \bibinfo {author} {\bibfnamefont
  {D.}~\bibnamefont {Moricciani}}, \bibinfo {author} {\bibfnamefont {W.~M.}\
  \bibnamefont {Morse}}, \bibinfo {author} {\bibfnamefont {S.}~\bibnamefont
  {Nagaitsev}}, \bibinfo {author} {\bibfnamefont {S.~K.}\ \bibnamefont
  {Nayak}}, \bibinfo {author} {\bibfnamefont {Y.~F.}\ \bibnamefont {Orlov}},
  \bibinfo {author} {\bibfnamefont {C.~S.}\ \bibnamefont {Ozben}}, \bibinfo
  {author} {\bibfnamefont {S.~T.}\ \bibnamefont {Park}}, \bibinfo {author}
  {\bibfnamefont {A.}~\bibnamefont {Pesce}}, \bibinfo {author} {\bibfnamefont
  {E.}~\bibnamefont {Petrakou}}, \bibinfo {author} {\bibfnamefont
  {P.}~\bibnamefont {Pile}}, \bibinfo {author} {\bibfnamefont {B.}~\bibnamefont
  {Podobedov}}, \bibinfo {author} {\bibfnamefont {V.}~\bibnamefont
  {Polychronakos}}, \bibinfo {author} {\bibfnamefont {J.}~\bibnamefont
  {Pretz}}, \bibinfo {author} {\bibfnamefont {V.}~\bibnamefont {Ptitsyn}},
  \bibinfo {author} {\bibfnamefont {E.}~\bibnamefont {Ramberg}}, \bibinfo
  {author} {\bibfnamefont {D.}~\bibnamefont {Raparia}}, \bibinfo {author}
  {\bibfnamefont {F.}~\bibnamefont {Rathmann}}, \bibinfo {author}
  {\bibfnamefont {S.}~\bibnamefont {Rescia}}, \bibinfo {author} {\bibfnamefont
  {T.}~\bibnamefont {Roser}}, \bibinfo {author} {\bibfnamefont
  {H.}~\bibnamefont {Kamal~Sayed}}, \bibinfo {author} {\bibfnamefont {Y.~K.}\
  \bibnamefont {Semertzidis}}, \bibinfo {author} {\bibfnamefont
  {Y.}~\bibnamefont {Senichev}}, \bibinfo {author} {\bibfnamefont
  {A.}~\bibnamefont {Sidorin}}, \bibinfo {author} {\bibfnamefont
  {A.}~\bibnamefont {Silenko}}, \bibinfo {author} {\bibfnamefont
  {N.}~\bibnamefont {Simos}}, \bibinfo {author} {\bibfnamefont
  {A.}~\bibnamefont {Stahl}}, \bibinfo {author} {\bibfnamefont {E.~J.}\
  \bibnamefont {Stephenson}}, \bibinfo {author} {\bibfnamefont
  {H.}~\bibnamefont {Str\"oher}}, \bibinfo {author} {\bibfnamefont {M.~J.}\
  \bibnamefont {Syphers}}, \bibinfo {author} {\bibfnamefont {J.}~\bibnamefont
  {Talman}}, \bibinfo {author} {\bibfnamefont {R.~M.}\ \bibnamefont {Talman}},
  \bibinfo {author} {\bibfnamefont {V.}~\bibnamefont {Tishchenko}}, \bibinfo
  {author} {\bibfnamefont {C.}~\bibnamefont {Touramanis}}, \bibinfo {author}
  {\bibfnamefont {N.}~\bibnamefont {Tsoupas}}, \bibinfo {author} {\bibfnamefont
  {G.}~\bibnamefont {Venanzoni}}, \bibinfo {author} {\bibfnamefont
  {K.}~\bibnamefont {Vetter}}, \bibinfo {author} {\bibfnamefont
  {S.}~\bibnamefont {Vlassis}}, \bibinfo {author} {\bibfnamefont
  {E.}~\bibnamefont {Won}}, \bibinfo {author} {\bibfnamefont {G.}~\bibnamefont
  {Zavattini}}, \bibinfo {author} {\bibfnamefont {A.}~\bibnamefont
  {Zelenski}},\ and\ \bibinfo {author} {\bibfnamefont {K.}~\bibnamefont
  {Zioutas}},\ }\href {https://doi.org/10.1063/1.4967465} {\bibfield  {journal}
  {\bibinfo  {journal} {Review of Scientific Instruments}\ }\textbf {\bibinfo
  {volume} {87}},\ \bibinfo {pages} {115116} (\bibinfo {year}
  {2016})}\BibitemShut {NoStop}%
\bibitem [{\citenamefont {Hac\ifmmode \imath \else \i
  \fi{}\"omero\ifmmode~\breve{g}\else \u{g}\fi{}lu}\ and\ \citenamefont
  {Semertzidis}(2019)}]{PhysRevAccelBeams.22.034001}%
  \BibitemOpen
  \bibfield  {author} {\bibinfo {author} {\bibfnamefont {S.}~\bibnamefont
  {Hac\ifmmode \imath \else \i \fi{}\"omero\ifmmode~\breve{g}\else
  \u{g}\fi{}lu}}\ and\ \bibinfo {author} {\bibfnamefont {Y.~K.}\ \bibnamefont
  {Semertzidis}},\ }\href {https://doi.org/10.1103/PhysRevAccelBeams.22.034001}
  {\bibfield  {journal} {\bibinfo  {journal} {Phys. Rev. Accel. Beams}\
  }\textbf {\bibinfo {volume} {22}},\ \bibinfo {pages} {034001} (\bibinfo
  {year} {2019})}\BibitemShut {NoStop}%
\bibitem [{\citenamefont {Omarov}\ \emph {et~al.}(2022)\citenamefont {Omarov},
  \citenamefont {Davoudiasl}, \citenamefont {Hac\ifmmode \imath \else \i
  \fi{}\"omero\ifmmode~\breve{g}\else \u{g}\fi{}lu}, \citenamefont {Lebedev},
  \citenamefont {Morse}, \citenamefont {Semertzidis}, \citenamefont {Silenko},
  \citenamefont {Stephenson},\ and\ \citenamefont
  {Suleiman}}]{PhysRevD.105.032001}%
  \BibitemOpen
  \bibfield  {author} {\bibinfo {author} {\bibfnamefont {Z.}~\bibnamefont
  {Omarov}}, \bibinfo {author} {\bibfnamefont {H.}~\bibnamefont {Davoudiasl}},
  \bibinfo {author} {\bibfnamefont {S.}~\bibnamefont {Hac\ifmmode \imath \else
  \i \fi{}\"omero\ifmmode~\breve{g}\else \u{g}\fi{}lu}}, \bibinfo {author}
  {\bibfnamefont {V.}~\bibnamefont {Lebedev}}, \bibinfo {author} {\bibfnamefont
  {W.~M.}\ \bibnamefont {Morse}}, \bibinfo {author} {\bibfnamefont {Y.~K.}\
  \bibnamefont {Semertzidis}}, \bibinfo {author} {\bibfnamefont {A.~J.}\
  \bibnamefont {Silenko}}, \bibinfo {author} {\bibfnamefont {E.~J.}\
  \bibnamefont {Stephenson}},\ and\ \bibinfo {author} {\bibfnamefont
  {R.}~\bibnamefont {Suleiman}},\ }\href
  {https://doi.org/10.1103/PhysRevD.105.032001} {\bibfield  {journal} {\bibinfo
   {journal} {Phys. Rev. D}\ }\textbf {\bibinfo {volume} {105}},\ \bibinfo
  {pages} {032001} (\bibinfo {year} {2022})}\BibitemShut {NoStop}%
\bibitem [{\citenamefont {Guidoboni}\ \emph {et~al.}(2016)\citenamefont
  {Guidoboni}, \citenamefont {Stephenson}, \citenamefont {Andrianov},
  \citenamefont {Augustyniak}, \citenamefont {Bagdasarian}, \citenamefont
  {Bai}, \citenamefont {Baylac}, \citenamefont {Bernreuther}, \citenamefont
  {Bertelli}, \citenamefont {Berz}, \citenamefont {B\"oker}, \citenamefont
  {B\"ohme}, \citenamefont {Bsaisou}, \citenamefont {Chekmenev}, \citenamefont
  {Chiladze}, \citenamefont {Ciullo}, \citenamefont {Contalbrigo},
  \citenamefont {de~Conto}, \citenamefont {Dymov}, \citenamefont {Engels},
  \citenamefont {Esser}, \citenamefont {Eversmann}, \citenamefont {Felden},
  \citenamefont {Gaisser}, \citenamefont {Gebel}, \citenamefont {Gl\"uckler},
  \citenamefont {Goldenbaum}, \citenamefont {Grigoryev}, \citenamefont
  {Grzonka}, \citenamefont {Hahnraths}, \citenamefont {Heberling},
  \citenamefont {Hejny}, \citenamefont {Hempelmann}, \citenamefont {Hetzel},
  \citenamefont {Hinder}, \citenamefont {Hipple}, \citenamefont {H\"olscher},
  \citenamefont {Ivanov}, \citenamefont {Kacharava}, \citenamefont
  {Kamerdzhiev}, \citenamefont {Kamys}, \citenamefont {Keshelashvili},
  \citenamefont {Khoukaz}, \citenamefont {Koop}, \citenamefont {Krause},
  \citenamefont {Krewald}, \citenamefont {Kulikov}, \citenamefont {Lehrach},
  \citenamefont {Lenisa}, \citenamefont {Lomidze}, \citenamefont {Lorentz},
  \citenamefont {Maanen}, \citenamefont {Macharashvili}, \citenamefont
  {Magiera}, \citenamefont {Maier}, \citenamefont {Makino}, \citenamefont
  {Maria\ifmmode~\acute{n}\else \'{n}\fi{}ski}, \citenamefont {Mchedlishvili},
  \citenamefont {Mei\ss{}ner}, \citenamefont {Mey}, \citenamefont {Morse},
  \citenamefont {M\"uller}, \citenamefont {Nass}, \citenamefont {Natour},
  \citenamefont {Nikolaev}, \citenamefont {Nioradze}, \citenamefont
  {Nowakowski}, \citenamefont {Orlov}, \citenamefont {Pesce}, \citenamefont
  {Prasuhn}, \citenamefont {Pretz}, \citenamefont {Rathmann}, \citenamefont
  {Ritman}, \citenamefont {Rosenthal}, \citenamefont {Rudy}, \citenamefont
  {Saleev}, \citenamefont {Sefzick}, \citenamefont {Semertzidis}, \citenamefont
  {Senichev}, \citenamefont {Shmakova}, \citenamefont {Silenko}, \citenamefont
  {Simon}, \citenamefont {Slim}, \citenamefont {Soltner}, \citenamefont
  {Stahl}, \citenamefont {Stassen}, \citenamefont {Statera}, \citenamefont
  {Stockhorst}, \citenamefont {Straatmann}, \citenamefont {Str\"oher},
  \citenamefont {Tabidze}, \citenamefont {Talman}, \citenamefont
  {Th\"orngren~Engblom}, \citenamefont {Trinkel}, \citenamefont
  {Trzci\ifmmode~\acute{n}\else \'{n}\fi{}ski}, \citenamefont {Uzikov},
  \citenamefont {Valdau}, \citenamefont {Valetov}, \citenamefont {Vassiliev},
  \citenamefont {Weidemann}, \citenamefont {Wilkin}, \citenamefont
  {Wro\ifmmode~\acute{n}\else \'{n}\fi{}ska}, \citenamefont {W\"ustner},
  \citenamefont {Zakrzewska}, \citenamefont {Zupra\ifmmode~\acute{n}\else
  \'{n}\fi{}ski},\ and\ \citenamefont {Zyuzin}}]{PhysRevLett.117.054801}%
  \BibitemOpen
  \bibfield  {author} {\bibinfo {author} {\bibfnamefont {G.}~\bibnamefont
  {Guidoboni}}, \bibinfo {author} {\bibfnamefont {E.}~\bibnamefont
  {Stephenson}}, \bibinfo {author} {\bibfnamefont {S.}~\bibnamefont
  {Andrianov}}, \bibinfo {author} {\bibfnamefont {W.}~\bibnamefont
  {Augustyniak}}, \bibinfo {author} {\bibfnamefont {Z.}~\bibnamefont
  {Bagdasarian}}, \bibinfo {author} {\bibfnamefont {M.}~\bibnamefont {Bai}},
  \bibinfo {author} {\bibfnamefont {M.}~\bibnamefont {Baylac}}, \bibinfo
  {author} {\bibfnamefont {W.}~\bibnamefont {Bernreuther}}, \bibinfo {author}
  {\bibfnamefont {S.}~\bibnamefont {Bertelli}}, \bibinfo {author}
  {\bibfnamefont {M.}~\bibnamefont {Berz}}, \bibinfo {author} {\bibfnamefont
  {J.}~\bibnamefont {B\"oker}}, \bibinfo {author} {\bibfnamefont
  {C.}~\bibnamefont {B\"ohme}}, \bibinfo {author} {\bibfnamefont
  {J.}~\bibnamefont {Bsaisou}}, \bibinfo {author} {\bibfnamefont
  {S.}~\bibnamefont {Chekmenev}}, \bibinfo {author} {\bibfnamefont
  {D.}~\bibnamefont {Chiladze}}, \bibinfo {author} {\bibfnamefont
  {G.}~\bibnamefont {Ciullo}}, \bibinfo {author} {\bibfnamefont
  {M.}~\bibnamefont {Contalbrigo}}, \bibinfo {author} {\bibfnamefont {J.-M.}\
  \bibnamefont {de~Conto}}, \bibinfo {author} {\bibfnamefont {S.}~\bibnamefont
  {Dymov}}, \bibinfo {author} {\bibfnamefont {R.}~\bibnamefont {Engels}},
  \bibinfo {author} {\bibfnamefont {F.~M.}\ \bibnamefont {Esser}}, \bibinfo
  {author} {\bibfnamefont {D.}~\bibnamefont {Eversmann}}, \bibinfo {author}
  {\bibfnamefont {O.}~\bibnamefont {Felden}}, \bibinfo {author} {\bibfnamefont
  {M.}~\bibnamefont {Gaisser}}, \bibinfo {author} {\bibfnamefont
  {R.}~\bibnamefont {Gebel}}, \bibinfo {author} {\bibfnamefont
  {H.}~\bibnamefont {Gl\"uckler}}, \bibinfo {author} {\bibfnamefont
  {F.}~\bibnamefont {Goldenbaum}}, \bibinfo {author} {\bibfnamefont
  {K.}~\bibnamefont {Grigoryev}}, \bibinfo {author} {\bibfnamefont
  {D.}~\bibnamefont {Grzonka}}, \bibinfo {author} {\bibfnamefont
  {T.}~\bibnamefont {Hahnraths}}, \bibinfo {author} {\bibfnamefont
  {D.}~\bibnamefont {Heberling}}, \bibinfo {author} {\bibfnamefont
  {V.}~\bibnamefont {Hejny}}, \bibinfo {author} {\bibfnamefont
  {N.}~\bibnamefont {Hempelmann}}, \bibinfo {author} {\bibfnamefont
  {J.}~\bibnamefont {Hetzel}}, \bibinfo {author} {\bibfnamefont
  {F.}~\bibnamefont {Hinder}}, \bibinfo {author} {\bibfnamefont
  {R.}~\bibnamefont {Hipple}}, \bibinfo {author} {\bibfnamefont
  {D.}~\bibnamefont {H\"olscher}}, \bibinfo {author} {\bibfnamefont
  {A.}~\bibnamefont {Ivanov}}, \bibinfo {author} {\bibfnamefont
  {A.}~\bibnamefont {Kacharava}}, \bibinfo {author} {\bibfnamefont
  {V.}~\bibnamefont {Kamerdzhiev}}, \bibinfo {author} {\bibfnamefont
  {B.}~\bibnamefont {Kamys}}, \bibinfo {author} {\bibfnamefont
  {I.}~\bibnamefont {Keshelashvili}}, \bibinfo {author} {\bibfnamefont
  {A.}~\bibnamefont {Khoukaz}}, \bibinfo {author} {\bibfnamefont
  {I.}~\bibnamefont {Koop}}, \bibinfo {author} {\bibfnamefont {H.-J.}\
  \bibnamefont {Krause}}, \bibinfo {author} {\bibfnamefont {S.}~\bibnamefont
  {Krewald}}, \bibinfo {author} {\bibfnamefont {A.}~\bibnamefont {Kulikov}},
  \bibinfo {author} {\bibfnamefont {A.}~\bibnamefont {Lehrach}}, \bibinfo
  {author} {\bibfnamefont {P.}~\bibnamefont {Lenisa}}, \bibinfo {author}
  {\bibfnamefont {N.}~\bibnamefont {Lomidze}}, \bibinfo {author} {\bibfnamefont
  {B.}~\bibnamefont {Lorentz}}, \bibinfo {author} {\bibfnamefont
  {P.}~\bibnamefont {Maanen}}, \bibinfo {author} {\bibfnamefont
  {G.}~\bibnamefont {Macharashvili}}, \bibinfo {author} {\bibfnamefont
  {A.}~\bibnamefont {Magiera}}, \bibinfo {author} {\bibfnamefont
  {R.}~\bibnamefont {Maier}}, \bibinfo {author} {\bibfnamefont
  {K.}~\bibnamefont {Makino}}, \bibinfo {author} {\bibfnamefont
  {B.}~\bibnamefont {Maria\ifmmode~\acute{n}\else \'{n}\fi{}ski}}, \bibinfo
  {author} {\bibfnamefont {D.}~\bibnamefont {Mchedlishvili}}, \bibinfo {author}
  {\bibfnamefont {U.-G.}\ \bibnamefont {Mei\ss{}ner}}, \bibinfo {author}
  {\bibfnamefont {S.}~\bibnamefont {Mey}}, \bibinfo {author} {\bibfnamefont
  {W.}~\bibnamefont {Morse}}, \bibinfo {author} {\bibfnamefont
  {F.}~\bibnamefont {M\"uller}}, \bibinfo {author} {\bibfnamefont
  {A.}~\bibnamefont {Nass}}, \bibinfo {author} {\bibfnamefont {G.}~\bibnamefont
  {Natour}}, \bibinfo {author} {\bibfnamefont {N.}~\bibnamefont {Nikolaev}},
  \bibinfo {author} {\bibfnamefont {M.}~\bibnamefont {Nioradze}}, \bibinfo
  {author} {\bibfnamefont {K.}~\bibnamefont {Nowakowski}}, \bibinfo {author}
  {\bibfnamefont {Y.}~\bibnamefont {Orlov}}, \bibinfo {author} {\bibfnamefont
  {A.}~\bibnamefont {Pesce}}, \bibinfo {author} {\bibfnamefont
  {D.}~\bibnamefont {Prasuhn}}, \bibinfo {author} {\bibfnamefont
  {J.}~\bibnamefont {Pretz}}, \bibinfo {author} {\bibfnamefont
  {F.}~\bibnamefont {Rathmann}}, \bibinfo {author} {\bibfnamefont
  {J.}~\bibnamefont {Ritman}}, \bibinfo {author} {\bibfnamefont
  {M.}~\bibnamefont {Rosenthal}}, \bibinfo {author} {\bibfnamefont
  {Z.}~\bibnamefont {Rudy}}, \bibinfo {author} {\bibfnamefont {A.}~\bibnamefont
  {Saleev}}, \bibinfo {author} {\bibfnamefont {T.}~\bibnamefont {Sefzick}},
  \bibinfo {author} {\bibfnamefont {Y.}~\bibnamefont {Semertzidis}}, \bibinfo
  {author} {\bibfnamefont {Y.}~\bibnamefont {Senichev}}, \bibinfo {author}
  {\bibfnamefont {V.}~\bibnamefont {Shmakova}}, \bibinfo {author}
  {\bibfnamefont {A.}~\bibnamefont {Silenko}}, \bibinfo {author} {\bibfnamefont
  {M.}~\bibnamefont {Simon}}, \bibinfo {author} {\bibfnamefont
  {J.}~\bibnamefont {Slim}}, \bibinfo {author} {\bibfnamefont {H.}~\bibnamefont
  {Soltner}}, \bibinfo {author} {\bibfnamefont {A.}~\bibnamefont {Stahl}},
  \bibinfo {author} {\bibfnamefont {R.}~\bibnamefont {Stassen}}, \bibinfo
  {author} {\bibfnamefont {M.}~\bibnamefont {Statera}}, \bibinfo {author}
  {\bibfnamefont {H.}~\bibnamefont {Stockhorst}}, \bibinfo {author}
  {\bibfnamefont {H.}~\bibnamefont {Straatmann}}, \bibinfo {author}
  {\bibfnamefont {H.}~\bibnamefont {Str\"oher}}, \bibinfo {author}
  {\bibfnamefont {M.}~\bibnamefont {Tabidze}}, \bibinfo {author} {\bibfnamefont
  {R.}~\bibnamefont {Talman}}, \bibinfo {author} {\bibfnamefont
  {P.}~\bibnamefont {Th\"orngren~Engblom}}, \bibinfo {author} {\bibfnamefont
  {F.}~\bibnamefont {Trinkel}}, \bibinfo {author} {\bibfnamefont
  {A.}~\bibnamefont {Trzci\ifmmode~\acute{n}\else \'{n}\fi{}ski}}, \bibinfo
  {author} {\bibfnamefont {Y.}~\bibnamefont {Uzikov}}, \bibinfo {author}
  {\bibfnamefont {Y.}~\bibnamefont {Valdau}}, \bibinfo {author} {\bibfnamefont
  {E.}~\bibnamefont {Valetov}}, \bibinfo {author} {\bibfnamefont
  {A.}~\bibnamefont {Vassiliev}}, \bibinfo {author} {\bibfnamefont
  {C.}~\bibnamefont {Weidemann}}, \bibinfo {author} {\bibfnamefont
  {C.}~\bibnamefont {Wilkin}}, \bibinfo {author} {\bibfnamefont
  {A.}~\bibnamefont {Wro\ifmmode~\acute{n}\else \'{n}\fi{}ska}}, \bibinfo
  {author} {\bibfnamefont {P.}~\bibnamefont {W\"ustner}}, \bibinfo {author}
  {\bibfnamefont {M.}~\bibnamefont {Zakrzewska}}, \bibinfo {author}
  {\bibfnamefont {P.}~\bibnamefont {Zupra\ifmmode~\acute{n}\else
  \'{n}\fi{}ski}},\ and\ \bibinfo {author} {\bibfnamefont {D.}~\bibnamefont
  {Zyuzin}} (\bibinfo {collaboration} {JEDI Collaboration}),\ }\href
  {https://doi.org/10.1103/PhysRevLett.117.054801} {\bibfield  {journal}
  {\bibinfo  {journal} {Phys. Rev. Lett.}\ }\textbf {\bibinfo {volume} {117}},\
  \bibinfo {pages} {054801} (\bibinfo {year} {2016})}\BibitemShut {NoStop}%
\bibitem [{\citenamefont {Filatov}\ \emph
  {et~al.}(2020{\natexlab{a}})\citenamefont {Filatov}, \citenamefont
  {Kondratenko}, \citenamefont {Kondratenko}, \citenamefont {Derbenev},\ and\
  \citenamefont {Morozov}}]{PhysRevLett.124.194801}%
  \BibitemOpen
  \bibfield  {author} {\bibinfo {author} {\bibfnamefont {Y.~N.}\ \bibnamefont
  {Filatov}}, \bibinfo {author} {\bibfnamefont {A.~M.}\ \bibnamefont
  {Kondratenko}}, \bibinfo {author} {\bibfnamefont {M.~A.}\ \bibnamefont
  {Kondratenko}}, \bibinfo {author} {\bibfnamefont {Y.~S.}\ \bibnamefont
  {Derbenev}},\ and\ \bibinfo {author} {\bibfnamefont {V.~S.}\ \bibnamefont
  {Morozov}},\ }\href {https://doi.org/10.1103/PhysRevLett.124.194801}
  {\bibfield  {journal} {\bibinfo  {journal} {Phys. Rev. Lett.}\ }\textbf
  {\bibinfo {volume} {124}},\ \bibinfo {pages} {194801} (\bibinfo {year}
  {2020}{\natexlab{a}})}\BibitemShut {NoStop}%
\bibitem [{\citenamefont {Kondratenko}\ \emph {et~al.}(2019)\citenamefont
  {Kondratenko}, \citenamefont {Kondratenko}, \citenamefont {Filatov},
  \citenamefont {Kovalenko}, \citenamefont {Derbenev},\ and\ \citenamefont
  {Morozov}}]{Kondratenko:2019ubq}%
  \BibitemOpen
  \bibfield  {author} {\bibinfo {author} {\bibfnamefont {A.}~\bibnamefont
  {Kondratenko}}, \bibinfo {author} {\bibfnamefont {M.}~\bibnamefont
  {Kondratenko}}, \bibinfo {author} {\bibfnamefont {Y.}~\bibnamefont
  {Filatov}}, \bibinfo {author} {\bibfnamefont {A.}~\bibnamefont {Kovalenko}},
  \bibinfo {author} {\bibfnamefont {Y.}~\bibnamefont {Derbenev}},\ and\
  \bibinfo {author} {\bibfnamefont {V.}~\bibnamefont {Morozov}},\ }\href
  {https://doi.org/10.1051/epjconf/201920410013} {\bibfield  {journal}
  {\bibinfo  {journal} {EPJ Web Conf.}\ }\textbf {\bibinfo {volume} {204}},\
  \bibinfo {pages} {10013} (\bibinfo {year} {2019})}\BibitemShut {NoStop}%
\bibitem [{\citenamefont {Graham}\ \emph {et~al.}(2021)\citenamefont {Graham},
  \citenamefont {Hac\ifmmode \imath \else \i
  \fi{}\"omero\ifmmode~\breve{g}\else \u{g}\fi{}lu}, \citenamefont {Kaplan},
  \citenamefont {Omarov}, \citenamefont {Rajendran},\ and\ \citenamefont
  {Semertzidis}}]{PhysRevD.103.055010}%
  \BibitemOpen
  \bibfield  {author} {\bibinfo {author} {\bibfnamefont {P.~W.}\ \bibnamefont
  {Graham}}, \bibinfo {author} {\bibfnamefont {S.}~\bibnamefont {Hac\ifmmode
  \imath \else \i \fi{}\"omero\ifmmode~\breve{g}\else \u{g}\fi{}lu}}, \bibinfo
  {author} {\bibfnamefont {D.~E.}\ \bibnamefont {Kaplan}}, \bibinfo {author}
  {\bibfnamefont {Z.}~\bibnamefont {Omarov}}, \bibinfo {author} {\bibfnamefont
  {S.}~\bibnamefont {Rajendran}},\ and\ \bibinfo {author} {\bibfnamefont
  {Y.~K.}\ \bibnamefont {Semertzidis}},\ }\href
  {https://doi.org/10.1103/PhysRevD.103.055010} {\bibfield  {journal} {\bibinfo
   {journal} {Phys. Rev. D}\ }\textbf {\bibinfo {volume} {103}},\ \bibinfo
  {pages} {055010} (\bibinfo {year} {2021})}\BibitemShut {NoStop}%
\bibitem [{\citenamefont {Chang}\ \emph {et~al.}(2019)\citenamefont {Chang},
  \citenamefont {Hac\ifmmode \imath \else \i
  \fi{}\"omero\ifmmode~\breve{g}\else \u{g}\fi{}lu}, \citenamefont {Kim},
  \citenamefont {Lee}, \citenamefont {Park},\ and\ \citenamefont
  {Semertzidis}}]{PhysRevD.99.083002}%
  \BibitemOpen
  \bibfield  {author} {\bibinfo {author} {\bibfnamefont {S.~P.}\ \bibnamefont
  {Chang}}, \bibinfo {author} {\bibfnamefont {S.}~\bibnamefont {Hac\ifmmode
  \imath \else \i \fi{}\"omero\ifmmode~\breve{g}\else \u{g}\fi{}lu}}, \bibinfo
  {author} {\bibfnamefont {O.}~\bibnamefont {Kim}}, \bibinfo {author}
  {\bibfnamefont {S.}~\bibnamefont {Lee}}, \bibinfo {author} {\bibfnamefont
  {S.}~\bibnamefont {Park}},\ and\ \bibinfo {author} {\bibfnamefont {Y.~K.}\
  \bibnamefont {Semertzidis}},\ }\href
  {https://doi.org/10.1103/PhysRevD.99.083002} {\bibfield  {journal} {\bibinfo
  {journal} {Phys. Rev. D}\ }\textbf {\bibinfo {volume} {99}},\ \bibinfo
  {pages} {083002} (\bibinfo {year} {2019})}\BibitemShut {NoStop}%
\bibitem [{\citenamefont {Flanz}\ and\ \citenamefont
  {Sargent}(1985)}]{FLANZ1985325}%
  \BibitemOpen
  \bibfield  {author} {\bibinfo {author} {\bibfnamefont {J.}~\bibnamefont
  {Flanz}}\ and\ \bibinfo {author} {\bibfnamefont {C.}~\bibnamefont
  {Sargent}},\ }\href {https://doi.org/10.1016/0168-9002(85)90585-6} {\bibfield
   {journal} {\bibinfo  {journal} {Nuclear Instruments and Methods in Physics
  Research Section A: Accelerators, Spectrometers, Detectors and Associated
  Equipment}\ }\textbf {\bibinfo {volume} {241}},\ \bibinfo {pages} {325}
  (\bibinfo {year} {1985})}\BibitemShut {NoStop}%
\bibitem [{\citenamefont {Conte}\ and\ \citenamefont
  {Martini}(1985)}]{Conte:1985sm}%
  \BibitemOpen
  \bibfield  {author} {\bibinfo {author} {\bibfnamefont {M.}~\bibnamefont
  {Conte}}\ and\ \bibinfo {author} {\bibfnamefont {M.}~\bibnamefont
  {Martini}},\ }\href {http://cdsweb.cern.ch/record/457572/files/p1.pdf}
  {\bibfield  {journal} {\bibinfo  {journal} {Part. Accel.}\ }\textbf {\bibinfo
  {volume} {17}},\ \bibinfo {pages} {1} (\bibinfo {year} {1985})}\BibitemShut
  {NoStop}%
\bibitem [{\citenamefont {MAD-X}(2022)}]{MAD-X}%
  \BibitemOpen
  \bibfield  {author} {\bibinfo {author} {\bibnamefont {MAD-X}},\ }\href
  {http://mad.web.cern.ch/mad} {\bibinfo {title} {{Methodical Accelerator
  Design}}} (\bibinfo {year} {2022})\BibitemShut {NoStop}%
\bibitem [{\citenamefont {Marriner}(2004)}]{MARRINER200411}%
  \BibitemOpen
  \bibfield  {author} {\bibinfo {author} {\bibfnamefont {J.}~\bibnamefont
  {Marriner}},\ }\href {https://doi.org/10.1016/j.nima.2004.06.025} {\bibfield
  {journal} {\bibinfo  {journal} {Nuclear Instruments and Methods in Physics
  Research Section A: Accelerators, Spectrometers, Detectors and Associated
  Equipment}\ }\textbf {\bibinfo {volume} {532}},\ \bibinfo {pages} {11}
  (\bibinfo {year} {2004})},\ \bibinfo {note} {{International Workshop on Beam
  Cooling and Related Topics}}\BibitemShut {NoStop}%
\bibitem [{\citenamefont {M\"ohl}(1983)}]{Mohl1983hmt}%
  \BibitemOpen
  \bibfield  {author} {\bibinfo {author} {\bibfnamefont {D.}~\bibnamefont
  {M\"ohl}},\ }\href
  {https://inspirehep.net/files/bdfebe85390380a11e2f80e27621d81f} {\bibfield
  {journal} {\bibinfo  {journal} {Conf. Proc. C}\ }\textbf {\bibinfo {volume}
  {831011}},\ \bibinfo {pages} {97} (\bibinfo {year} {1983})}\BibitemShut
  {NoStop}%
\bibitem [{\citenamefont {Lee}(2018)}]{Lee2018}%
  \BibitemOpen
  \bibfield  {author} {\bibinfo {author} {\bibfnamefont {S.~Y.}\ \bibnamefont
  {Lee}},\ }\href {https://doi.org/10.1142/11111} {\emph {\bibinfo {title}
  {Accelerator Physics (Fourth Edition)}}}\ (\bibinfo  {publisher} {World
  Scientific Publishing Company},\ \bibinfo {year} {2018})\ p.\ \bibinfo
  {pages} {248}\BibitemShut {NoStop}%
\bibitem [{\citenamefont {Eldred}\ \emph {et~al.}(2021)\citenamefont {Eldred},
  \citenamefont {Lebedev}, \citenamefont {Seiya},\ and\ \citenamefont
  {Shiltsev}}]{PhysRevAccelBeams.24.044001}%
  \BibitemOpen
  \bibfield  {author} {\bibinfo {author} {\bibfnamefont {J.}~\bibnamefont
  {Eldred}}, \bibinfo {author} {\bibfnamefont {V.}~\bibnamefont {Lebedev}},
  \bibinfo {author} {\bibfnamefont {K.}~\bibnamefont {Seiya}},\ and\ \bibinfo
  {author} {\bibfnamefont {V.}~\bibnamefont {Shiltsev}},\ }\href
  {https://doi.org/10.1103/PhysRevAccelBeams.24.044001} {\bibfield  {journal}
  {\bibinfo  {journal} {Phys. Rev. Accel. Beams}\ }\textbf {\bibinfo {volume}
  {24}},\ \bibinfo {pages} {044001} (\bibinfo {year} {2021})}\BibitemShut
  {NoStop}%
\bibitem [{\citenamefont {Murphy}\ \emph {et~al.}(1997)\citenamefont {Murphy},
  \citenamefont {Krinsky},\ and\ \citenamefont {Gluckstern}}]{Murphy:1996yt}%
  \BibitemOpen
  \bibfield  {author} {\bibinfo {author} {\bibfnamefont {J.~B.}\ \bibnamefont
  {Murphy}}, \bibinfo {author} {\bibfnamefont {S.}~\bibnamefont {Krinsky}},\
  and\ \bibinfo {author} {\bibfnamefont {R.~L.}\ \bibnamefont {Gluckstern}},\
  }\href {https://inspirehep.net/files/1afd5dcc4e791d233fb0ffb5a2f270a4}
  {\bibfield  {journal} {\bibinfo  {journal} {Part. Accel.}\ }\textbf {\bibinfo
  {volume} {57}},\ \bibinfo {pages} {9} (\bibinfo {year} {1997})}\BibitemShut
  {NoStop}%
\bibitem [{\citenamefont {Warnock}\ and\ \citenamefont
  {Morton}(1990)}]{Warnock:1990}%
  \BibitemOpen
  \bibfield  {author} {\bibinfo {author} {\bibfnamefont {R.}~\bibnamefont
  {Warnock}}\ and\ \bibinfo {author} {\bibfnamefont {P.~M.}\ \bibnamefont
  {Morton}},\ }\href
  {http://slac.stanford.edu/pubs/slacpubs/4500/slac-pub-4562.pdf} {\bibfield
  {journal} {\bibinfo  {journal} {Part. Accel.}\ }\textbf {\bibinfo {volume}
  {25}},\ \bibinfo {pages} {113} (\bibinfo {year} {1990})}\BibitemShut
  {NoStop}%
\bibitem [{\citenamefont {Stutzman}\ \emph {et~al.}(2018)\citenamefont
  {Stutzman}, \citenamefont {Adderley}, \citenamefont {Mamun},\ and\
  \citenamefont {Poelker}}]{doi:10.1116/1.5010154}%
  \BibitemOpen
  \bibfield  {author} {\bibinfo {author} {\bibfnamefont {M.~L.}\ \bibnamefont
  {Stutzman}}, \bibinfo {author} {\bibfnamefont {P.~A.}\ \bibnamefont
  {Adderley}}, \bibinfo {author} {\bibfnamefont {M.~A.~A.}\ \bibnamefont
  {Mamun}},\ and\ \bibinfo {author} {\bibfnamefont {M.}~\bibnamefont
  {Poelker}},\ }\href {https://doi.org/10.1116/1.5010154} {\bibfield  {journal}
  {\bibinfo  {journal} {Journal of Vacuum Science \& Technology A}\ }\textbf
  {\bibinfo {volume} {36}},\ \bibinfo {pages} {031603} (\bibinfo {year}
  {2018})}\BibitemShut {NoStop}%
\bibitem [{\citenamefont {Filatov}\ \emph
  {et~al.}(2020{\natexlab{b}})\citenamefont {Filatov} \emph
  {et~al.}}]{Filatov2020}%
  \BibitemOpen
  \bibfield  {author} {\bibinfo {author} {\bibfnamefont {Y.~N.}\ \bibnamefont
  {Filatov}} \emph {et~al.},\ }\href
  {https://doi.org/10.1140/epjc/s10052-020-8344-5} {\bibfield  {journal}
  {\bibinfo  {journal} {Eur. Phys. J. C}\ }\textbf {\bibinfo {volume} {80}},\
  \bibinfo {pages} {778} (\bibinfo {year} {2020}{\natexlab{b}})}\BibitemShut
  {NoStop}%
\bibitem [{\citenamefont {Kondratenko}\ \emph {et~al.}(2017)\citenamefont
  {Kondratenko}, \citenamefont {Kondratenko}, \citenamefont {Filatov},
  \citenamefont {Derbenev}, \citenamefont {Lin}, \citenamefont {Morozov},\ and\
  \citenamefont {Zhang}}]{Kondratenko2017}%
  \BibitemOpen
  \bibfield  {author} {\bibinfo {author} {\bibfnamefont {A.~M.}\ \bibnamefont
  {Kondratenko}}, \bibinfo {author} {\bibfnamefont {M.~A.}\ \bibnamefont
  {Kondratenko}}, \bibinfo {author} {\bibfnamefont {Y.~N.}\ \bibnamefont
  {Filatov}}, \bibinfo {author} {\bibfnamefont {Y.~S.}\ \bibnamefont
  {Derbenev}}, \bibinfo {author} {\bibfnamefont {F.}~\bibnamefont {Lin}},
  \bibinfo {author} {\bibfnamefont {V.~S.}\ \bibnamefont {Morozov}},\ and\
  \bibinfo {author} {\bibfnamefont {Y.}~\bibnamefont {Zhang}},\ }\href
  {https://doi.org/10.1088/1742-6596/874/1/012011} {\bibfield  {journal}
  {\bibinfo  {journal} {Journal of Physics: Conference Series}\ }\textbf
  {\bibinfo {volume} {874}},\ \bibinfo {pages} {012011} (\bibinfo {year}
  {2017})}\BibitemShut {NoStop}%
\bibitem [{\citenamefont {Pretz}\ \emph {et~al.}(2020)\citenamefont {Pretz},
  \citenamefont {Chang}, \citenamefont {Hejny}, \citenamefont {Karanth},
  \citenamefont {Park}, \citenamefont {Semertzidis}, \citenamefont
  {Stephenson},\ and\ \citenamefont {Str\"oher}}]{Pretz2020}%
  \BibitemOpen
  \bibfield  {author} {\bibinfo {author} {\bibfnamefont {J.}~\bibnamefont
  {Pretz}}, \bibinfo {author} {\bibfnamefont {S.~P.}\ \bibnamefont {Chang}},
  \bibinfo {author} {\bibfnamefont {V.}~\bibnamefont {Hejny}}, \bibinfo
  {author} {\bibfnamefont {S.}~\bibnamefont {Karanth}}, \bibinfo {author}
  {\bibfnamefont {S.}~\bibnamefont {Park}}, \bibinfo {author} {\bibfnamefont
  {Y.}~\bibnamefont {Semertzidis}}, \bibinfo {author} {\bibfnamefont
  {E.}~\bibnamefont {Stephenson}},\ and\ \bibinfo {author} {\bibfnamefont
  {H.}~\bibnamefont {Str\"oher}},\ }\href
  {https://doi.org/10.1140/epjc/s10052-020-7664-9} {\bibfield  {journal}
  {\bibinfo  {journal} {The European Physical Journal C}\ }\textbf {\bibinfo
  {volume} {80}},\ \bibinfo {pages} {107} (\bibinfo {year} {2020})}\BibitemShut
  {NoStop}%
\bibitem [{\citenamefont {Kim}\ and\ \citenamefont
  {Semertzidis}(2021)}]{PhysRevD.104.096006}%
  \BibitemOpen
  \bibfield  {author} {\bibinfo {author} {\bibfnamefont {O.}~\bibnamefont
  {Kim}}\ and\ \bibinfo {author} {\bibfnamefont {Y.~K.}\ \bibnamefont
  {Semertzidis}},\ }\href {https://doi.org/10.1103/PhysRevD.104.096006}
  {\bibfield  {journal} {\bibinfo  {journal} {Phys. Rev. D}\ }\textbf {\bibinfo
  {volume} {104}},\ \bibinfo {pages} {096006} (\bibinfo {year}
  {2021})}\BibitemShut {NoStop}%
\bibitem [{\citenamefont {Adderley}\ \emph {et~al.}(2010)\citenamefont
  {Adderley}, \citenamefont {Clark}, \citenamefont {Grames}, \citenamefont
  {Hansknecht}, \citenamefont {Surles-Law}, \citenamefont {Machie},
  \citenamefont {Poelker}, \citenamefont {Stutzman},\ and\ \citenamefont
  {Suleiman}}]{PhysRevSTAB.13.010101}%
  \BibitemOpen
  \bibfield  {author} {\bibinfo {author} {\bibfnamefont {P.~A.}\ \bibnamefont
  {Adderley}}, \bibinfo {author} {\bibfnamefont {J.}~\bibnamefont {Clark}},
  \bibinfo {author} {\bibfnamefont {J.}~\bibnamefont {Grames}}, \bibinfo
  {author} {\bibfnamefont {J.}~\bibnamefont {Hansknecht}}, \bibinfo {author}
  {\bibfnamefont {K.}~\bibnamefont {Surles-Law}}, \bibinfo {author}
  {\bibfnamefont {D.}~\bibnamefont {Machie}}, \bibinfo {author} {\bibfnamefont
  {M.}~\bibnamefont {Poelker}}, \bibinfo {author} {\bibfnamefont {M.~L.}\
  \bibnamefont {Stutzman}},\ and\ \bibinfo {author} {\bibfnamefont
  {R.}~\bibnamefont {Suleiman}},\ }\href
  {https://doi.org/10.1103/PhysRevSTAB.13.010101} {\bibfield  {journal}
  {\bibinfo  {journal} {Phys. Rev. ST Accel. Beams}\ }\textbf {\bibinfo
  {volume} {13}},\ \bibinfo {pages} {010101} (\bibinfo {year}
  {2010})}\BibitemShut {NoStop}%
\bibitem [{\citenamefont {Adderley}\ \emph {et~al.}(2023)\citenamefont
  {Adderley}, \citenamefont {Bullard}, \citenamefont {Chao}, \citenamefont
  {Garcia}, \citenamefont {Grames}, \citenamefont {Hansknecht}, \citenamefont
  {Hofler}, \citenamefont {Kazimi}, \citenamefont {Musson}, \citenamefont
  {Palatchi}, \citenamefont {Paschke}, \citenamefont {Poelker}, \citenamefont
  {Smith}, \citenamefont {Stutzman}, \citenamefont {Suleiman},\ and\
  \citenamefont {Wang}}]{ADDERLEY2023167710}%
  \BibitemOpen
  \bibfield  {author} {\bibinfo {author} {\bibfnamefont {P.}~\bibnamefont
  {Adderley}}, \bibinfo {author} {\bibfnamefont {D.}~\bibnamefont {Bullard}},
  \bibinfo {author} {\bibfnamefont {Y.}~\bibnamefont {Chao}}, \bibinfo {author}
  {\bibfnamefont {C.}~\bibnamefont {Garcia}}, \bibinfo {author} {\bibfnamefont
  {J.}~\bibnamefont {Grames}}, \bibinfo {author} {\bibfnamefont
  {J.}~\bibnamefont {Hansknecht}}, \bibinfo {author} {\bibfnamefont
  {A.}~\bibnamefont {Hofler}}, \bibinfo {author} {\bibfnamefont
  {R.}~\bibnamefont {Kazimi}}, \bibinfo {author} {\bibfnamefont
  {J.}~\bibnamefont {Musson}}, \bibinfo {author} {\bibfnamefont
  {C.}~\bibnamefont {Palatchi}}, \bibinfo {author} {\bibfnamefont
  {K.}~\bibnamefont {Paschke}}, \bibinfo {author} {\bibfnamefont
  {M.}~\bibnamefont {Poelker}}, \bibinfo {author} {\bibfnamefont
  {G.}~\bibnamefont {Smith}}, \bibinfo {author} {\bibfnamefont
  {M.}~\bibnamefont {Stutzman}}, \bibinfo {author} {\bibfnamefont
  {R.}~\bibnamefont {Suleiman}},\ and\ \bibinfo {author} {\bibfnamefont
  {Y.}~\bibnamefont {Wang}},\ }\href
  {https://doi.org/10.1016/j.nima.2022.167710} {\bibfield  {journal} {\bibinfo
  {journal} {Nuclear Instruments and Methods in Physics Research Section A:
  Accelerators, Spectrometers, Detectors and Associated Equipment}\ }\textbf
  {\bibinfo {volume} {1046}},\ \bibinfo {pages} {167710} (\bibinfo {year}
  {2023})}\BibitemShut {NoStop}%
\bibitem [{\citenamefont {Grames}\ \emph {et~al.}(2011)\citenamefont {Grames},
  \citenamefont {Adderley}, \citenamefont {Benesch}, \citenamefont {Clark},
  \citenamefont {Hansknecht}, \citenamefont {Kazimi}, \citenamefont {Machie},
  \citenamefont {Poelker}, \citenamefont {Stutzman}, \citenamefont {Suleiman},\
  and\ \citenamefont {Zhang}}]{Adderley:2011ri}%
  \BibitemOpen
  \bibfield  {author} {\bibinfo {author} {\bibfnamefont {J.}~\bibnamefont
  {Grames}}, \bibinfo {author} {\bibfnamefont {P.}~\bibnamefont {Adderley}},
  \bibinfo {author} {\bibfnamefont {J.}~\bibnamefont {Benesch}}, \bibinfo
  {author} {\bibfnamefont {J.}~\bibnamefont {Clark}}, \bibinfo {author}
  {\bibfnamefont {J.}~\bibnamefont {Hansknecht}}, \bibinfo {author}
  {\bibfnamefont {R.}~\bibnamefont {Kazimi}}, \bibinfo {author} {\bibfnamefont
  {D.}~\bibnamefont {Machie}}, \bibinfo {author} {\bibfnamefont
  {M.}~\bibnamefont {Poelker}}, \bibinfo {author} {\bibfnamefont
  {M.}~\bibnamefont {Stutzman}}, \bibinfo {author} {\bibfnamefont
  {R.}~\bibnamefont {Suleiman}},\ and\ \bibinfo {author} {\bibfnamefont
  {Y.}~\bibnamefont {Zhang}},\ }\href
  {https://accelconf.web.cern.ch/pac2011/papers/tup025.pdf} {\bibfield
  {journal} {\bibinfo  {journal} {Proceedings of 2011 Particle Accelerator
  Conference, New York, NY, USA}\ ,\ \bibinfo {pages} {862}} (\bibinfo {year}
  {2011})}\BibitemShut {NoStop}%
\bibitem [{\citenamefont {Gay}\ and\ \citenamefont
  {Dunning}(1992)}]{doi:10.1063/1.1143371}%
  \BibitemOpen
  \bibfield  {author} {\bibinfo {author} {\bibfnamefont {T.~J.}\ \bibnamefont
  {Gay}}\ and\ \bibinfo {author} {\bibfnamefont {F.~B.}\ \bibnamefont
  {Dunning}},\ }\href {https://doi.org/10.1063/1.1143371} {\bibfield  {journal}
  {\bibinfo  {journal} {Review of Scientific Instruments}\ }\textbf {\bibinfo
  {volume} {63}},\ \bibinfo {pages} {1635} (\bibinfo {year}
  {1992})}\BibitemShut {NoStop}%
\bibitem [{\citenamefont {Tioukine}\ \emph {et~al.}(2011)\citenamefont
  {Tioukine}, \citenamefont {Aulenbacher},\ and\ \citenamefont
  {Riehn}}]{doi:10.1063/1.3556593}%
  \BibitemOpen
  \bibfield  {author} {\bibinfo {author} {\bibfnamefont {V.}~\bibnamefont
  {Tioukine}}, \bibinfo {author} {\bibfnamefont {K.}~\bibnamefont
  {Aulenbacher}},\ and\ \bibinfo {author} {\bibfnamefont {E.}~\bibnamefont
  {Riehn}},\ }\href {https://doi.org/10.1063/1.3556593} {\bibfield  {journal}
  {\bibinfo  {journal} {Review of Scientific Instruments}\ }\textbf {\bibinfo
  {volume} {82}},\ \bibinfo {pages} {033303} (\bibinfo {year}
  {2011})}\BibitemShut {NoStop}%
\bibitem [{\citenamefont {Aulenbacher}\ \emph {et~al.}(2018)\citenamefont
  {Aulenbacher}, \citenamefont {Chudakov}, \citenamefont {Gaskell},
  \citenamefont {Grames},\ and\ \citenamefont
  {Paschke}}]{doi:10.1142/S0218301318300047}%
  \BibitemOpen
  \bibfield  {author} {\bibinfo {author} {\bibfnamefont {K.}~\bibnamefont
  {Aulenbacher}}, \bibinfo {author} {\bibfnamefont {E.}~\bibnamefont
  {Chudakov}}, \bibinfo {author} {\bibfnamefont {D.}~\bibnamefont {Gaskell}},
  \bibinfo {author} {\bibfnamefont {J.}~\bibnamefont {Grames}},\ and\ \bibinfo
  {author} {\bibfnamefont {K.~D.}\ \bibnamefont {Paschke}},\ }\href
  {https://doi.org/10.1142/S0218301318300047} {\bibfield  {journal} {\bibinfo
  {journal} {International Journal of Modern Physics E}\ }\textbf {\bibinfo
  {volume} {27}},\ \bibinfo {pages} {1830004} (\bibinfo {year}
  {2018})}\BibitemShut {NoStop}%
\bibitem [{\citenamefont {Grames}\ \emph {et~al.}(2020)\citenamefont {Grames},
  \citenamefont {Sinclair}, \citenamefont {Poelker}, \citenamefont {Roca-Maza},
  \citenamefont {Stutzman}, \citenamefont {Suleiman}, \citenamefont {Mamun},
  \citenamefont {McHugh}, \citenamefont {Moser}, \citenamefont {Hansknecht},
  \citenamefont {Moffit},\ and\ \citenamefont {Gay}}]{PhysRevC.102.015501}%
  \BibitemOpen
  \bibfield  {author} {\bibinfo {author} {\bibfnamefont {J.~M.}\ \bibnamefont
  {Grames}}, \bibinfo {author} {\bibfnamefont {C.~K.}\ \bibnamefont
  {Sinclair}}, \bibinfo {author} {\bibfnamefont {M.}~\bibnamefont {Poelker}},
  \bibinfo {author} {\bibfnamefont {X.}~\bibnamefont {Roca-Maza}}, \bibinfo
  {author} {\bibfnamefont {M.~L.}\ \bibnamefont {Stutzman}}, \bibinfo {author}
  {\bibfnamefont {R.}~\bibnamefont {Suleiman}}, \bibinfo {author}
  {\bibfnamefont {M.~A.}\ \bibnamefont {Mamun}}, \bibinfo {author}
  {\bibfnamefont {M.}~\bibnamefont {McHugh}}, \bibinfo {author} {\bibfnamefont
  {D.}~\bibnamefont {Moser}}, \bibinfo {author} {\bibfnamefont
  {J.}~\bibnamefont {Hansknecht}}, \bibinfo {author} {\bibfnamefont
  {B.}~\bibnamefont {Moffit}},\ and\ \bibinfo {author} {\bibfnamefont {T.~J.}\
  \bibnamefont {Gay}},\ }\href {https://doi.org/10.1103/PhysRevC.102.015501}
  {\bibfield  {journal} {\bibinfo  {journal} {Phys. Rev. C}\ }\textbf {\bibinfo
  {volume} {102}},\ \bibinfo {pages} {015501} (\bibinfo {year}
  {2020})}\BibitemShut {NoStop}%
\bibitem [{\citenamefont {Roca-Maza}(2017)}]{Roca_Maza_2017}%
  \BibitemOpen
  \bibfield  {author} {\bibinfo {author} {\bibfnamefont {X.}~\bibnamefont
  {Roca-Maza}},\ }\href {https://doi.org/10.1209/0295-5075/120/33002}
  {\bibfield  {journal} {\bibinfo  {journal} {{EPL} (Europhysics Letters)}\
  }\textbf {\bibinfo {volume} {120}},\ \bibinfo {pages} {33002} (\bibinfo
  {year} {2017})}\BibitemShut {NoStop}%
\bibitem [{\citenamefont {Roca-Maza}(2021)}]{Roca_Maza}%
  \BibitemOpen
  \bibfield  {author} {\bibinfo {author} {\bibfnamefont {X.}~\bibnamefont
  {Roca-Maza}},\ }\href@noop {} {}\bibinfo {howpublished} {private
  communication} (\bibinfo {year} {2021})\BibitemShut {NoStop}%
\bibitem [{\citenamefont {Brantjes}\ \emph {et~al.}(2012)\citenamefont
  {Brantjes}, \citenamefont {Dzordzhadze}, \citenamefont {Gebel}, \citenamefont
  {Gonnella}, \citenamefont {Gray}, \citenamefont {{van der Hoek}},
  \citenamefont {Imig}, \citenamefont {Kruithof}, \citenamefont {Lazarus},
  \citenamefont {Lehrach}, \citenamefont {Lorentz}, \citenamefont {Messi},
  \citenamefont {Moricciani}, \citenamefont {Morse}, \citenamefont {Noid},
  \citenamefont {Onderwater}, \citenamefont {Özben}, \citenamefont {Prasuhn},
  \citenamefont {{Levi Sandri}}, \citenamefont {Semertzidis}, \citenamefont
  {{da Silva e Silva}}, \citenamefont {Stephenson}, \citenamefont {Stockhorst},
  \citenamefont {Venanzoni},\ and\ \citenamefont {Versolato}}]{BRANTJES201249}%
  \BibitemOpen
  \bibfield  {author} {\bibinfo {author} {\bibfnamefont {N.~P.~M.}\
  \bibnamefont {Brantjes}}, \bibinfo {author} {\bibfnamefont {V.}~\bibnamefont
  {Dzordzhadze}}, \bibinfo {author} {\bibfnamefont {R.}~\bibnamefont {Gebel}},
  \bibinfo {author} {\bibfnamefont {F.}~\bibnamefont {Gonnella}}, \bibinfo
  {author} {\bibfnamefont {F.}~\bibnamefont {Gray}}, \bibinfo {author}
  {\bibfnamefont {D.}~\bibnamefont {{van der Hoek}}}, \bibinfo {author}
  {\bibfnamefont {A.}~\bibnamefont {Imig}}, \bibinfo {author} {\bibfnamefont
  {W.}~\bibnamefont {Kruithof}}, \bibinfo {author} {\bibfnamefont
  {D.}~\bibnamefont {Lazarus}}, \bibinfo {author} {\bibfnamefont
  {A.}~\bibnamefont {Lehrach}}, \bibinfo {author} {\bibfnamefont
  {B.}~\bibnamefont {Lorentz}}, \bibinfo {author} {\bibfnamefont
  {R.}~\bibnamefont {Messi}}, \bibinfo {author} {\bibfnamefont
  {D.}~\bibnamefont {Moricciani}}, \bibinfo {author} {\bibfnamefont
  {W.}~\bibnamefont {Morse}}, \bibinfo {author} {\bibfnamefont
  {G.}~\bibnamefont {Noid}}, \bibinfo {author} {\bibfnamefont {C.}~\bibnamefont
  {Onderwater}}, \bibinfo {author} {\bibfnamefont {C.}~\bibnamefont {Özben}},
  \bibinfo {author} {\bibfnamefont {D.}~\bibnamefont {Prasuhn}}, \bibinfo
  {author} {\bibfnamefont {P.}~\bibnamefont {{Levi Sandri}}}, \bibinfo {author}
  {\bibfnamefont {Y.}~\bibnamefont {Semertzidis}}, \bibinfo {author}
  {\bibfnamefont {M.}~\bibnamefont {{da Silva e Silva}}}, \bibinfo {author}
  {\bibfnamefont {E.}~\bibnamefont {Stephenson}}, \bibinfo {author}
  {\bibfnamefont {H.}~\bibnamefont {Stockhorst}}, \bibinfo {author}
  {\bibfnamefont {G.}~\bibnamefont {Venanzoni}},\ and\ \bibinfo {author}
  {\bibfnamefont {O.}~\bibnamefont {Versolato}},\ }\href
  {https://doi.org/10.1016/j.nima.2011.09.055} {\bibfield  {journal} {\bibinfo
  {journal} {Nuclear Instruments and Methods in Physics Research Section A:
  Accelerators, Spectrometers, Detectors and Associated Equipment}\ }\textbf
  {\bibinfo {volume} {664}},\ \bibinfo {pages} {49} (\bibinfo {year}
  {2012})}\BibitemShut {NoStop}%
\bibitem [{\citenamefont {Carli}\ and\ \citenamefont
  {Haj~Tahar}(2022)}]{PhysRevAccelBeams.25.064001}%
  \BibitemOpen
  \bibfield  {author} {\bibinfo {author} {\bibfnamefont {C.}~\bibnamefont
  {Carli}}\ and\ \bibinfo {author} {\bibfnamefont {M.}~\bibnamefont
  {Haj~Tahar}},\ }\href {https://doi.org/10.1103/PhysRevAccelBeams.25.064001}
  {\bibfield  {journal} {\bibinfo  {journal} {Phys. Rev. Accel. Beams}\
  }\textbf {\bibinfo {volume} {25}},\ \bibinfo {pages} {064001} (\bibinfo
  {year} {2022})}\BibitemShut {NoStop}%
\bibitem [{\citenamefont {Lin}\ \emph {et~al.}(2018)\citenamefont {Lin},
  \citenamefont {Grames}, \citenamefont {Guo}, \citenamefont {Morozov},\ and\
  \citenamefont {Zhang}}]{FangleiLin}%
  \BibitemOpen
  \bibfield  {author} {\bibinfo {author} {\bibfnamefont {F.}~\bibnamefont
  {Lin}}, \bibinfo {author} {\bibfnamefont {J.}~\bibnamefont {Grames}},
  \bibinfo {author} {\bibfnamefont {J.}~\bibnamefont {Guo}}, \bibinfo {author}
  {\bibfnamefont {V.}~\bibnamefont {Morozov}},\ and\ \bibinfo {author}
  {\bibfnamefont {Y.}~\bibnamefont {Zhang}},\ }\href
  {https://doi.org/10.1063/1.5040224} {\bibfield  {journal} {\bibinfo
  {journal} {AIP Conference Proceedings}\ }\textbf {\bibinfo {volume} {1970}},\
  \bibinfo {pages} {050005} (\bibinfo {year} {2018})}\BibitemShut {NoStop}%
\bibitem [{\citenamefont {Derbenev}\ \emph {et~al.}(2021)\citenamefont
  {Derbenev}, \citenamefont {Filatov}, \citenamefont {Kondratenko},
  \citenamefont {Kondratenko},\ and\ \citenamefont {Morozov}}]{sym13030398}%
  \BibitemOpen
  \bibfield  {author} {\bibinfo {author} {\bibfnamefont {Y.~S.}\ \bibnamefont
  {Derbenev}}, \bibinfo {author} {\bibfnamefont {Y.~N.}\ \bibnamefont
  {Filatov}}, \bibinfo {author} {\bibfnamefont {A.~M.}\ \bibnamefont
  {Kondratenko}}, \bibinfo {author} {\bibfnamefont {M.~A.}\ \bibnamefont
  {Kondratenko}},\ and\ \bibinfo {author} {\bibfnamefont {V.~S.}\ \bibnamefont
  {Morozov}},\ }\href {https://doi.org/10.3390/sym13030398} {\bibfield
  {journal} {\bibinfo  {journal} {Symmetry}\ }\textbf {\bibinfo {volume}
  {13}},\ \bibinfo {pages} {398} (\bibinfo {year} {2021})}\BibitemShut
  {NoStop}%
\end{thebibliography}%

\end{document}